\documentclass{aa}
\usepackage[colorlinks=true,citecolor=blue,anchorcolor=blue,filecolor=blue,linkcolor=blue]{hyperref}
\usepackage{graphicx}
\usepackage{amssymb,amsmath}
\usepackage{latexsym}
\usepackage{mathtools}
\usepackage{epsfig}
\usepackage{natbib}
\usepackage{txfonts}
\usepackage{mathrsfs}
\begin{document}
   \title{Tilting Uranus via the migration of an ancient satellite}
   %\subtitle{}
   %\titlerunning{}
%
   \author{Melaine Saillenfest\inst{1}
          \and
          Zeeve Rogoszinski\inst{2}
          \and
          Giacomo Lari\inst{3}
          \and
          Kevin Bailli{\'e}\inst{1}
          \and
          Gwena{\"e}l Bou{\'e}\inst{1}
          \and
          Aur{\'e}lien Crida\inst{4}
          \and
          Val{\'e}ry Lainey\inst{1}
          }
   \authorrunning{Saillenfest et al.}
   \institute{IMCCE, Observatoire de Paris, PSL Research University, CNRS, Sorbonne Universit\'e, Universit\'e de Lille, 75014 Paris, France\\
             \email{melaine.saillenfest@obspm.fr}
             \and
             Astronomy Department, University of Maryland College Park, MD 20742, USA
             \and
             Department of Mathematics, University of Pisa, Largo Bruno Pontecorvo 5, 56127 Pisa, Italy
             \and
             Université Côte d'Azur, Observatoire de la Côte d'Azur, CNRS, laboratoire Lagrange (UMR7293), Boulevard de l'observatoire, 06304 Nice cedex 4, France
             }
   \date{Received 2022-05-05 / Accepted 2022-09-21}

%%%%%%%%%%%%%%%%%%%%%%%%%%%%%%%%%%%%%%%%%%%%%%%%%%%%%%%%%%%%%%%%%%%%%%%%%%%%%%%%%%%%%%%%%%%

  \abstract
  % context heading (optional)
  {The $98^\circ$-obliquity of Uranus is commonly attributed to giant impacts that occurred at the end of the planetary formation. This picture, however, is not devoid of weaknesses.}
  % aims heading (mandatory)
  {On a billion-year timescale, the tidal migration of the satellites of Jupiter and Saturn has been shown to strongly affect their spin-axis dynamics. We aim to revisit the scenario of tilting Uranus in light of this mechanism.}
  % methods heading (mandatory)
  {We analyse the precession spectrum of Uranus and identify the candidate secular spin-orbit resonances that could be responsible for the tilting. We determine the properties of the hypothetical ancient satellite required for a capture and explore the dynamics numerically.}
  % results heading (mandatory)
  {If it migrates over $10$~Uranus' radii, a single satellite with minimum mass $4\times 10^{-4}$~Uranus' mass is able to tilt Uranus from a small obliquity and make it converge towards $90^\circ$. In order to achieve the tilting in less than the age of the Solar System, the mean drift rate of the satellite must be comparable to the Moon's current orbital expansion. Under these conditions, simulations show that Uranus is readily tilted over $80^\circ$. Beyond this point, the satellite is strongly destabilised and triggers a phase of chaotic motion for the planet's spin axis. The chaotic phase ends when the satellite collides into the planet, ultimately freezing the planet's obliquity in either a prograde, or plainly retrograde state (as Uranus today). Spin states resembling that of Uranus can be obtained with probabilities as large as $80\%$, but a bigger satellite is favoured, with mass $1.7\times 10^{-3}$~Uranus' mass or more. Yet, a smaller ancient satellite is not categorically ruled out, and there is room for improving this basic scenario in future studies. Interactions among several pre-existing satellites is a promising possibility.}
  % conclusions heading (optional)
  {The conditions required for the tilting seem broadly realistic, but it remains to be determined whether Uranus could have hosted a big primordial satellite subject to substantial tidal migration. The efficiency of tidal energy dissipation within Uranus is required to be much higher than traditionally assumed, more in line with that measured for the migration of Titan. Hints about these issues would be given by a measure of the expansion rate of Uranus' main satellites.}


  \maketitle

\section{Introduction}\label{sec:intro}

The spin axis of Uranus is inclined by about $98^\circ$ with respect to its orbit normal. This large inclination is particularly puzzling, as all other giant planets in the Solar System have obliquity values smaller than $30^\circ$. The most widely accepted explanation for this discrepancy is a series of impacts with large planetesimals that probably occurred at the end of the accretion phase (see e.g. \citealp{Morbidelli-etal_2012,Izidoro-etal_2015,Salmon-Canup_2022}). In this picture, the two ice giants Uranus and Neptune would have acquired random spin-axis orientations, whereas runaway gas accretion would have driven the obliquities of Jupiter and Saturn towards~$0^\circ$ \citep{Ward-Hamilton_2004,Hamilton-Ward_2004}. Differing impact histories may also explain why Uranus and Neptune have qualitatively different satellite systems and show dissimilar heat fluxes \citep{Reinhardt-etal_2020}. However, as argued by \cite{Rogoszinski-Hamilton_2020}, this scenario suffers from several weaknesses. Uranus and Neptune have strikingly similar masses, radii, and spin rates; they also feature similar atmospheric dynamics \citep{Aurnou-etal_2007}, and both have non-dipolar magnetic fields which are likely produced through the same internal convection processes (see e.g. \citealp{Bailey-Stevenson_2021}, for a review). If the final formation stages of Uranus and Neptune had been dominated by giant impacts, one would expect them to present a much larger diversity of properties (see e.g. \citealp{Chau-etal_2021}), perhaps in the spirit of the terrestrial planets of the Solar System. Even though the rotation state of both Uranus and Neptune can be reproduced individually by a finely tuned collision (see e.g. \citealp{Slattery-etal_1992,Reinhardt-etal_2020}, and the insightful discussions of \citealp{Rufu-Canup_2022}), similar spin rates, above all, are not what one would expect as the outcome of random collisions. Instead, the strong similarities between Uranus and Neptune may indicate that their final properties result from smoother processes, and that gas accretion supplied enough angular momentum to dominate their rotation states, similarly to Jupiter and Saturn. This idea is supported by the fact that the spin rates of all four giant planets are a fraction of their critical rotation speeds in a consistent way with formation theories \citep{Machida-etal_2008,Ward-Canup_2010,Lovelace-etal_2011,Batygin_2018,Bryan-etal_2018,Dong-etal_2021,Dittmann_2021}. In this unified picture, we would expect the obliquities of Uranus and Neptune to have been small after their formations, and to have evolved to their observed values (about $98^\circ$ and $30^\circ$) at later stages.

Today, the orbits and spin axes of Uranus and Neptune are stable and far from any kind of strong perturbation. Therefore, it seems natural to assume that their tilting occurred early, when the Solar System was still evolving. It is now believed that the last large-scale changes in the Solar System happened during a phase of instability that followed the dissipation of the protoplanetary disc, as the giant planets were migrating in a swarm of planetesimals \citep{Tsiganis-etal_2005}. It was initially proposed that this event could be related to a phase of resurging collisions recorded in the lunar craters (called the Late heavy bombardment; see \citealp{Gomes-etal_2005}). Numerous refinements were then brought to the instability model in order to properly reproduce the structure of the Solar System as we observe it today. The most recent advances favour a very early migration and instability (within $10$~Myrs, or even immediately after dispersal of the gas disc), and no relation with the Late heavy bombardment (see \citealp{Clement-etal_2018,Clement-etal_2021,Nesvorny-etal_2018,Morgan-etal_2021}).

The instability itself did not exactly place the giant planets on their current orbits. After the instability, the scattering of the remaining planetesimals produced a last grasp of migration, during which Uranus and Neptune still migrated over a few astronomical units. In this much depleted environment, the migration of Uranus and Neptune was slower, in accordance with observations: in order to reproduce the observed Kuiper belt, Neptune probably migrated during this last stage with an $e$-folding timescale larger than $10$~Myrs and possibly extending to about $100$~Myrs at the very end of the migration (see \citealp{Nesvorny_2015}, or \citealp{Nesvorny_2018} for a review). Yet, as noted by \cite{Ida-etal_2000}, Neptune's migration cannot have been arbitrarily slow, otherwise mean-motion resonances in the Kuiper belt would be overcrowded. Putting everything together, Uranus and Neptune most likely finished their migration from several tens to a few hundreds of million years at most after the dispersal of the gas disc. This timespan puts strong constraints on their tilting scenarios if they are required to happen during the planetary migration.

Apart from collisions, possible tilting mechanisms involve some kind of spin-orbit coupling \citep{Boue-Laskar_2010,Quillen-etal_2018,Millholland-Batygin_2019,Rogoszinski-Hamilton_2020,Rogoszinski-Hamilton_2021}. It means that at some point, the spin-axis precession motion of the planet enters into resonance with some harmonics stemming from its orbital dynamics. If this happens while the orbits of the planets are still being reshaped, then the resonance properties change over time. Previous authors found that during planetary migration, there exist several ways to displace the resonant equilibrium configuration of the spin axis from a small obliquity $\varepsilon\approx 0^\circ$ to over $90^\circ$ (and even up to $180^\circ$ in the case described by \citealp{Quillen-etal_2018}). However, because Uranus and Neptune are today so far from the Sun, their spin-axis precession timescales are extremely large and count in hundreds of million years. In order to yield reasonable tilting probabilities, the displacement of the equilibrium must be very slow (i.e. adiabatic) compared to the libration inside the resonance, which itself is slower than the spin-axis precession motion. Hence, for Uranus placed near its current location, authors invariably obtained that the timescale for a tilting up to $90^\circ$ or more greatly exceeds the expected duration of the planetary migration (see e.g. \citealp{Quillen-etal_2018}).

In order to speed up the spin-axis precession of Uranus and catch a strong resonance, we can imagine that Uranus formed much closer to the Sun than it is now \citep{Rogoszinski-Hamilton_2021}, or that it once had a big satellite \citep{Boue-Laskar_2010} or a massive disc around it \citep{Harris-Ward_1982,Rogoszinski-Hamilton_2020}. Indeed, close-in planets precess faster, and satellites or circumplanetary discs have the ability to magnify the spin-axis precession rate of their host planet (see e.g. \citealp{Tremaine_1991,Boue-Laskar_2006,Millholland-Batygin_2019}). However, authors found that even with these refinements, the timescale required for a purely adiabatic tilting of Uranus hardly fits in the limited amount of time offered by planetary migration (see e.g. \citealp{Rogoszinski-Hamilton_2021}). Moreover, the late planetary migration has certainly not been perfectly smooth, but `grainy' (as a result of the scattering of large planetesimals; see e.g. \citealp{Nesvorny-Vokrouhlicky_2016}). The level of smoothness of the planetary migration further complicates the adiabatic tilting process, as a sudden change in the semi-major axis of a given planet may release Uranus out of resonance, in which case the tilting would stop.

Yet, the tilting of Uranus may not have been produced through an adiabatic process: instead of moving the stable equilibrium point, the quickest way to incline a planet via a resonance is to do so over a single large-amplitude libration cycle (a `resonance kick'). In this case, the resonance itself must be extremely wide, which puts new constraints on the orbital dynamics of the planets. In Appendix~\ref{asec:migtilt}, we show that the orbital inclination $I$ of the planet with respect to the invariable plane must verify $\tan I \geqslant 3-2\sqrt{2}$ in order to allow for trajectories going from $\varepsilon\approx 0^\circ$ to over $90^\circ$ during a single libration inside a secular spin-orbit resonance. Therefore, irrespectively of the secular spin-orbit resonance involved (e.g. $s_7$ or $s_8$, as considered by previous authors), a large orbital inclination of more than about $10^\circ$ is required for the resonant harmonic. This limit quite accurately matches the numerical experiments of \cite{Rogoszinski-Hamilton_2020}. But in order to yield a reasonable probability of reproducing Uranus' current state (instead of a single finely-tuned pathway), the inclination needed is actually much larger than $10^\circ$. In fact, the most successful tilting scenarios found in previous studies involve altogether a massive satellite (or disc), a resonance drift due to planetary migration, and a phase of very large orbital inclination (see \citealp{Boue-Laskar_2010,Rogoszinski-Hamilton_2020}). Unfortunately, as discussed by \cite{Vokrouhlicky-Nesvorny_2015}, large inclination values for Uranus and/or Neptune are an improbable transient feature of Solar System migration models, and none was reported in the extensive simulations of \cite{Nesvorny-Morbidelli_2012}. Moreover, the mass required for the satellite in order to reach a strong secular spin-orbit resonance was found to be implausibly large (e.g. $0.01$ Uranus' mass), and some additional mechanism was needed to remove the satellite after the tilting. Hence, the timescale issue remains a major problem of previous `collisionless' scenarios for the tilting of Uranus.

In the alternative migration scenario recently proposed by \cite{Lu-Laughlin_2022}, the large-inclination harmonic is provided by the hypothetical Planet~9 (see e.g. \citealp{Batygin-etal_2019}). Even though this scenario does not alleviate the need for a massive circumplanetary disc, the large inclination of Planet~9 allows for dramatic resonance kicks that can easily produce the required level of obliquity excitation for Uranus. However, much uncertainty remains about the migration history (and existence) of Planet~9, and it is not clear whether its large inclination actually predates its scattering away, as assumed by the authors. This scenario also suffers from the same caveats as previous works regarding Uranus' disc: a massive equatorial disc needs to be maintained and continuously fed during the whole tilting phase, which may not be realistic (see \citealp{Rogoszinski-Hamilton_2020}); besides, the accretion of matters from the disc onto Uranus would slow down the resonance crossing phenomenon possibly up to beyond the timescale of planetary migration.

As the spin axis of Neptune is much less inclined, its current orientation is easier to explain. Possible scenarios involve a very early tilt during planetary formation \citep{Martin-etal_2020} and/or a resonance crossing during the late planetary migration \citep{Rogoszinski-Hamilton_2020}. If Neptune was surrounded by a massive circumplanetary disc at that time, even a moderate resonance kick would be enough. For this reason, we may consider that the case of Neptune is (at least partially) solved.

Independently of Uranus and Neptune, \cite{Saillenfest-Lari_2021} have recently described a mechanism through which a migrating satellite can tilt an initially small-obliquity planet up to over $\varepsilon=90^\circ$ with almost no modification to its spin rate. Their study was prompted by the works of \cite{Lainey-etal_2009,Lainey-etal_2020}, who discovered that the orbital expansion of satellites around Jupiter and Saturn are much faster than previously thought, indicating that efficient mechanisms of tidal dissipation can take place not only in terrestrial planets, as it was known for the Earth-Moon system, but also in giant planets. Such a fast satellite migration has been found to deeply affect the spin-axis dynamics of Jupiter and Saturn (see \citealp{Lari-etal_2020,Saillenfest-etal_2020,Saillenfest-etal_2021a,Saillenfest-etal_2021b}). In fact, while exploring the future evolution of Saturn and Titan, \cite{Saillenfest-Lari_2021} have shown that if a satellite substantially migrates inwards or outwards, then its influence on the spin-axis precession of its host planet changes over time in a way that is intimately linked to the properties of its equilibrium orbital plane, called `Laplace plane' \citep{Tremaine-etal_2009}. The Laplace plane is well defined everywhere in the parameter space, except in a specific point, called $\mathrm{S}_1$, where it degenerates into a continuum of equilibria. When it comes to the spin-axis dynamics of the planet, $\mathrm{S}_1$ appears as a singularity located at an obliquity $\varepsilon=90^\circ$, towards which all secular spin-orbit resonances converge if the satellite is massive enough. Over the migration of its satellites, a planet has therefore a chance of being captured in resonance and brought to extreme obliquities. Depending on the migration rate of the satellite, this whole process can take place over the whole lifetime of the planetary system. Importantly, $\mathrm{S}_1$ is surrounded by an unstable region which may lead to the destruction of the satellite once the tilting is achieved \citep{Tremaine-etal_2009,Saillenfest-Lari_2021}. New satellites could then be formed from the leftover debris disc \citep{Crida-Charnoz_2012}.

In this context, it appears tempting to revisit the case of Uranus's obliquity and see whether this new satellite migration-destabilisation mechanism could solve at least part of the mystery. Since tidal migration of satellites is an ever-going process, the tilting would not need to be restricted to a short timespan, contrary to previous models, and the resonance involved would not need to be very strong. As a result, this new scenario would not require a particularly massive satellite, nor a particularly high orbital inclination for the planet. This tilting mechanism was actually proven to work for Saturn on its current orbit, and it is expected to work as well for any giant planet of the solar system, should it have the adequate long-range migrating satellite \citep{Saillenfest-Lari_2021}.

In this article, we aim to assess the feasibility of this tilting mechanism for Uranus and to determine the main parameters that would be required: mass of the satellite, migration range and timescale, etc. Due to the multi-scale nature of the mechanism, the current structure of the Solar System actually offers many usable constraints, such as its age, the orbital dynamics of the planets, and the properties of Uranus' current satellite system.

The basic mechanism considered here is composed of two stages. The resonance capture and adiabatic tilting up to the unstable zone is studied in Sect.~\ref{sec:tilting}. We recall the basic mechanism through which a migrating satellite can tilt its host planet (Sect.~\ref{sec:mec}), and conduct a preliminary analysis of the parameters needed in the case of Uranus: mass of the satellite (Sect.~\ref{sec:mass}), migration range (Sect.~\ref{sec:res}), and resonance capture probabilities (Sect.~\ref{sec:proba}). The final destabilisation phase is studied in Sect.~\ref{sec:coupledmod}. We show that the satellite triggers a double synergistic destabilisation involving the planet's spin axis (Sect.~\ref{sec:doubledestab}) and we explore numerically the range of possible outcomes (Sect.~\ref{sec:explor}); we then quantify the probability of reproducing Uranus' current spin state (Sect.~\ref{sec:coupledproba}) and determine the conditions under which the satellite collides into the planet (Sect.~\ref{sec:keyholes}). In Sect.~\ref{sec:discuss}, we discuss whether the parameters that we obtain appear realistic in the broader context of satellite formation theories and tidal migration mechanisms. We conclude in Sect.~\ref{sec:ccl}.

\section{Adiabatic tilting up to the unstable zone}\label{sec:tilting}

\subsection{Tilting mechanism}\label{sec:mec}
As shown by \cite{Saillenfest-Lari_2021}, the influence of a given regular satellite on the spin-axis precession of its host planet critically depends on its non-dimensional `mass parameter' $\eta$, defined by
\begin{equation}\label{eq:etamass}
   \eta = \frac{1}{2}\frac{m}{M}\frac{r_\mathrm{M}^2}{J_2R_\mathrm{eq}^2}\,,
\end{equation}
where $m$ is the mass of the satellite, $M$ is the mass of the planet, $J_2$ is its second zonal gravity coefficient, and $R_\mathrm{eq}$ is a normalising radius (that we take equal to the planet's equatorial radius). The characteristic length $r_\mathrm{M}$ is the distance at which the equilibrium orbital plane of the satellite lies exactly halfway between the equatorial plane and the orbital plane of the planet (the index $\mathrm{M}$ stands for `midpoint'). Its definition is closely related to the Laplace radius\footnote{There is currently no consensus in the literature on the exact definition of the `Laplace radius'. To avoid confusion with previous works, we stick to the definition of \cite{Tremaine-etal_2009} for $r_\mathrm{L}$ and introduce $r_\mathrm{M}$ as a distinct characteristic length.} $r_\mathrm{L}$ introduced by \cite{Tremaine-etal_2009}, as
\begin{equation}\label{eq:rM}
   r_\mathrm{M}^5 = 2r_\mathrm{L}^5 = 2\frac{M}{m_\odot}J_2R_\mathrm{eq}^2a_\odot^3(1-e_\odot^2)^{3/2}\,.
\end{equation}
In this expression, $m_\odot$ is the mass of the star, and $a_\odot$ and $e_\odot$ are the semi-major axis and the eccentricity of the planet on its orbit around the star. The value of $r_\mathrm{M}$ for Uranus is about $53$~$R_\mathrm{eq}$. We neglect for now the influence of inner satellites in the effective value of $J_2$ (see e.g. \citealp{Tremaine-etal_2009}). This means either that Uranus possessed a single satellite at that time, or that other satellites were too small and/or too close to Uranus to substantially contribute. We go back to this point in Sect.~\ref{sec:discuss}.

For a small mass ratio $m/M$, the free spin-axis precession frequency of the planet taking into account the influence of its satellite can be approximated by
\begin{equation}\label{eq:psidotsimp}
   \Omega_0 = p\left(\cos\varepsilon + \eta\frac{a^2}{r_\mathrm{M}^2}\frac{\sin(2\varepsilon-2I_\mathrm{L})}{2\sin(\varepsilon)}\right)\,,
\end{equation}
where $\varepsilon$ is the planet's obliquity, $a$ is the semi-major axis of the satellite, and $I_\mathrm{L}$ is the inclination of the satellite's local Laplace plane with respect to the planet's equator. The leading factor $p$ is the characteristic spin-axis precession rate of the planet (without satellite); it relates to the classic `precession constant' $\alpha$ as $p=\alpha(1-e_\odot^2)^{-3/2}$, and it can be written as
\begin{equation}\label{eq:charprec}
   p = \frac{3}{2}\frac{\mathcal{G}m_\odot}{\omega a_\odot^3(1-e_\odot^2)^{3/2}}\frac{J_2}{\lambda}\,,
\end{equation}
where $\mathcal{G}$ is the gravitational constant, $\omega$ is the spin rate of the planet, and $\lambda$ is its normalised polar moment of inertia. Equation~\eqref{eq:psidotsimp} is valid as long as the satellite oscillates around its local circular Laplace equilibrium on a timescale that is much smaller than the spin-axis precession motion of the planet, which is usually well verified in practice \citep{Saillenfest-Lari_2021}. A good approximation for $I_\mathrm{L}$ is given by the formula
\begin{equation}
   I_\mathrm{L} = \frac{\pi}{2} + \frac{1}{2}\mathrm{atan}2\big[-\sin(2\varepsilon),-r_\mathrm{M}^5/a^5 - \cos(2\varepsilon)\big]\,,
\end{equation}
which becomes exact in the limit $m/M\rightarrow 0$ \citep{Tremaine-etal_2009}. It should be noted that even a satellite with a small mass ratio $m/M$ can have a large mass parameter $\eta$, and therefore a strong influence in Eq.~\eqref{eq:psidotsimp}; see Table~1 of \cite{Saillenfest-Lari_2021} for examples.

Because of orbital perturbations (e.g. mutual gravitational interactions with other planets), the orbital plane and eccentricity of the planet\footnote{Throughout this article, the orbit of the planet-satellite barycentre around the star is generally referred to as `the orbit of the planet' for simplicity. This distinction will be important in Sect.~\ref{sec:coupledmod}, when using a self-consistent coupled model for the planet and its satellite.} are not fixed but vary on a secular timescale. Assuming that the orbit of the planet is long-term stable, its secular motion can be expressed (at least locally) in convergent quasi-periodic series:
\begin{equation}\label{eq:qprep}
   \begin{aligned}
      z = e_\odot\exp(i\varpi_\odot) &= \sum_k E_k\exp(i\theta_k) \,,\\
      \zeta = \sin\frac{I_\odot}{2}\exp(i\Omega_\odot) &= \sum_k S_k\exp(i\phi_k)\,,
   \end{aligned}
\end{equation}
where $\varpi_\odot$, $I_\odot$, and $\Omega_\odot$ are the planet's longitude of pericentre, orbital inclination, and longitude of ascending node measured in an inertial frame. The amplitudes $E_k$ and $S_k$ are real constants and the angles $\theta_k$ and $\phi_k$ evolve linearly over time $t$ with frequencies $\mu_k$ and $\nu_k$:
\begin{equation}\label{eq:munu}
   \theta_k(t) = \mu_k\,t + \theta_k^{(0)}
   \hspace{0.5cm}\text{and}\hspace{0.5cm}
   \phi_k(t) = \nu_k\,t + \phi_k^{(0)}\,,
\end{equation}
where the index $k$ runs over all terms that have non-negligible amplitudes. In Appendix~\ref{asec:QPS}, we give the orbital solution of \cite{Laskar_1990} for Uranus with amplitudes down to $10^{-9}$. This solution can be considered valid (at least in a qualitative point of view) since the end of the planetary migration. In the integrable approximation, the frequency of each term corresponds to a unique combination of the fundamental frequencies of the system, usually noted $g_j$ and $s_j$. Table~\ref{tab:zetashort} shows the combinations of fundamental frequencies identified for the twenty largest terms of Uranus' $\zeta$ series. In contrast to previous works, in which authors studied the effect of large inclination variations for Uranus or Neptune during early stages of their evolution, we stress that the orbit of Uranus is now very steady. Over billions of years, its inclination barely exceeds $1^\circ$ with respect to the invariable plane of the Solar System. Yet, as we show below, dramatic obliquity variations can still occur if we leave the system evolve for a sufficient amount of time.

\begin{table}
   \caption{First twenty terms of Uranus' inclination and longitude of ascending node in the J2000 ecliptic and equinox reference frame.}
   \label{tab:zetashort}
   \vspace{-0.7cm}
   \begin{equation*}
      \begin{array}{rcrrr}
      \hline
      \hline
      k & \text{identification}\tablefootmark{*} & \nu_k\ (''\,\text{yr}^{-1}) & S_k\times 10^9 & \phi_k^{(0)}\ (^\text{o}) \\
      \hline   
          1 &              s_5 &   0.00000 &  13773646 & 107.59 \\
          2 &              s_7 &  -3.00557 &   8871413 & 320.33 \\
          3 &              s_8 &  -0.69189 &    563042 & 203.96 \\
          4 &              s_6 & -26.33023 &    347710 & 307.29 \\
          5 & -g_5 + g_6 + s_6 &  -2.35835 &    299979 & 224.75 \\
          6 & -g_5 + g_7 + s_7 &  -4.16482 &    187859 & 231.66 \\
          7 &  g_5 - g_7 + s_7 &  -1.84625 &    182575 & 224.56 \\
          8 & -g_7 + g_8 + s_8 &  -3.11725 &     59252 & 146.97 \\
          9 &  g_6 - g_7 + s_6 &  -1.19906 &     25881 & 313.99 \\
         10 &       2g_5 - s_7 &  11.50319 &     18941 & 101.01 \\
         11 &  g_5 + g_7 - s_7 &  10.34389 &     11930 &  11.68 \\
         12 &  g_5 - g_6 + s_7 & -26.97744 &     10362 & 225.10 \\
         13 &              s_1 &  -5.61755 &     10270 & 348.70 \\
         14 & -g_5 + g_6 + s_7 &  20.96631 &      7346 & 237.78 \\
         15 &  g_7 - g_8 + s_7 &  -0.58033 &      5474 & 197.32 \\
         16 &     s_1 + \gamma &  -5.50098 &      3662 & 342.89 \\
         17 &  g_5 - g_6 + s_6 & -50.30212 &      2748 &  29.83 \\
         18 &              s_2 &  -7.07963 &      2372 & 273.81 \\
         19 &  g_5 - g_7 + s_8 &   0.46547 &      1575 & 106.88 \\
         20 &       2g_6 - s_6 &  82.77163 &      1514 & 308.95 \\
      \hline
      \end{array}
   \end{equation*}
   \vspace{-0.5cm}
   \tablefoot{Due to the secular resonance $(g_1-g_5)-(s_1-s_2)$, an additional fundamental frequency $\gamma$ appears in term 16 (see \citealp{Laskar_1990}).\\
   \tablefoottext{*}{There are typographical errors in \cite{Laskar_1990} in the identification of the 6th and 7th terms.}
   }
\end{table}

As a result of the motion of the planet's orbital plane, its long-term spin-axis dynamics is shaped by secular spin-orbit resonances, that is, by resonances between the unperturbed spin-axis precession frequency $\Omega_0$ in Eq.~\eqref{eq:psidotsimp} and the forcing frequencies $\mu_k$ and $\nu_k$ appearing in Eq.~\eqref{eq:munu}. As detailed by \cite{Saillenfest-etal_2019}, the largest resonances are those of order $1$ in the amplitudes $\{S_k\}$, for which the resonance angle is $\sigma=\psi+\phi_j$, where $\psi$ is the precession angle of the planet's spin axis and $j$ is a given index in the $\zeta$ series\footnote{For short, the first-order secular spin-orbit resonance with resonant angle $\sigma=\psi+\phi_j$ will be called the `$\nu_j$ resonance', where $j$ is a given index in Table~\ref{tab:zetashort}.}. If the planet is trapped in a secular spin-orbit resonance while its satellite migrates, then its mean obliquity $\varepsilon$ evolves together with $a$ along a level curve of $\Omega_0$, such that the relation $\Omega_0+\nu_j\approx 0$ is maintained.

\begin{figure}
   \includegraphics[width=\columnwidth]{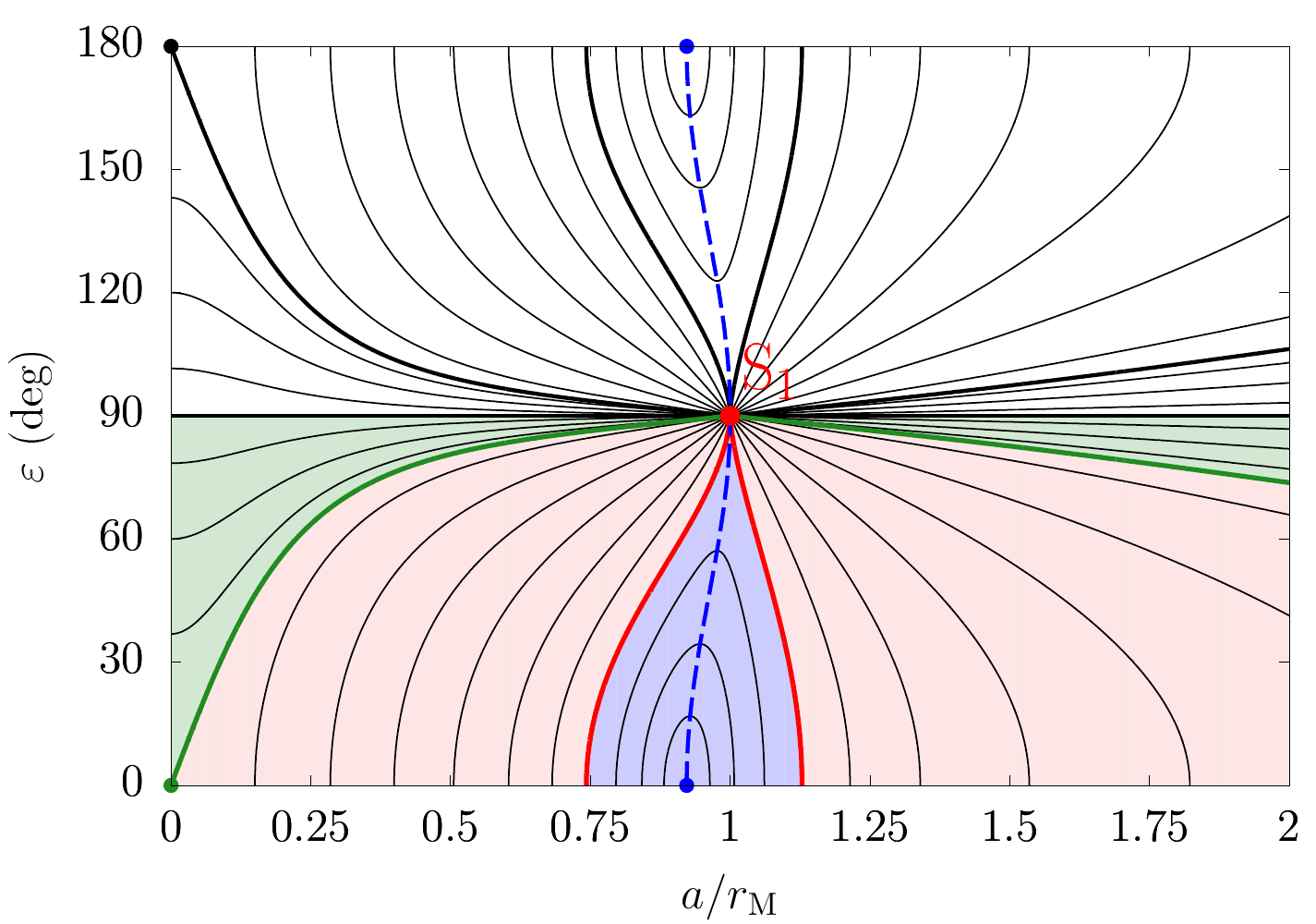}
   \caption{Level curves of the spin-axis precession rate of the planet as a function of its obliquity and the distance of its satellite. The simplified formula in Eq.~\eqref{eq:psidotsimp} is used here with $\eta=20$. Curves in the pink region connect $\varepsilon=0^\circ$ to $\mathrm{S}_1$. Curves in the blue region connect $\varepsilon=0^\circ$ to $\varepsilon=0^\circ$ again. Curves in the green region never go to $\varepsilon=0^\circ$ but they connect to $\mathrm{S}_1$. The mirror level curves exist for $\varepsilon>90^\circ$ with reversed precession motion. The precession rate along the red level is $\Omega_0=p\eta/2$. The precession rate along the dark green level is $\Omega_0=p$; it connects $(a,\varepsilon)=(0,0^\circ)$ to $\mathrm{S}_1$. The dashed blue curve is the ridge line separating the close and far satellite regimes; it has expression $(a/r_\mathrm{M})^5 = [\sqrt{\cos^2(2\varepsilon)+24}-\cos(2\varepsilon)]/6$. See Fig.~17 of \cite{Saillenfest-Lari_2021} for examples with other values of $\eta$.}
   \label{fig:precrate}
\end{figure}

\cite{Saillenfest-Lari_2021} have shown that for a mass parameter $\eta\geqslant 2$, all level curves of $\Omega_0$ with values between $p$ and $p\eta/2$ connect $\varepsilon=0^\circ$ to the singular point $\mathrm{S}_1$, which has coordinates $(a,\varepsilon)=(r_\mathrm{M},90^\circ)$; this property is illustrated in Fig.~\ref{fig:precrate}. In other words, any secular spin-orbit resonance with a forcing term having a negative frequency $\nu_j$ such that
\begin{equation}\label{eq:condtilt}
   p\leqslant|\nu_j|\leqslant p\frac{\eta}{2}
\end{equation}
would allow for a tilting of the planet between $\varepsilon=0^\circ$ and $\varepsilon=90^\circ$. If the resonance is large and/or if there are large neighbouring resonances, the planet may even go beyond $90^\circ$. For retrograde resonances (i.e. positive frequency $\nu_j$), the condition in Eq.~\eqref{eq:condtilt} allows for a tilting between $\varepsilon=180^\circ$ and $\varepsilon=90^\circ$. Equation~\eqref{eq:condtilt} can easily be transposed to higher-order secular spin-orbit resonances: $\nu_j$ should simply be replaced by a combination of frequencies $\nu_k$ and $\mu_k$ (see \citealp{Saillenfest-etal_2019} for the variety of possible resonances). However, in the case of Uranus, resonances of order~2 and higher are very small and only few of them may possibly allow for a capture.

\subsection{Minimum mass for the satellite}\label{sec:mass}
The simplified condition in Eq.~\eqref{eq:condtilt} offers a quick way to assess whether a given satellite can produce large obliquity variations of its host planet, or which are the optimum parameters of a hypothetical satellite in order to produce a dramatic tilting. In the case of Uranus, $p$ is much smaller than any frequency $|\nu_k|$ appearing in its $\zeta$ series; therefore, the condition for a satellite to allow for a capture into resonance and tilting up to $90^\circ$ is simply $\eta \geqslant 2|\nu_\star|/p$, where $\nu_\star$ is the frequency of the closest resonance. This condition directly gives the minimum mass required for the satellite, that we compute below.

The largest uncertainty in Uranus' precession constant comes from its polar moment of inertia $\lambda$. \cite{Helled-etal_2010} also warn readers about the poor accuracy of the spin rate $\omega$ measured by Voyager~2, whose value is usually cited (see e.g. \citealp{Yoder_1995,Archinal-etal_2018}); they provide a new estimate for the solid-body rotation period of Uranus that differs from the Voyager value by several percent. As for $\lambda$, \cite{Hubbard-Marley_1989} give $\lambda=0.22680$ obtained from their interior models using a normalising radius $R_\mathrm{eq}=25\,559$~km. Guided by previous works, we assume $\lambda=0.23$ for simplicity and consider a $10\%$ uncertainty on the product $\omega\lambda$, used to account for model dependency and the uncertainty on $\omega$ (the impact of this choice is commented below). This exploration interval entirely contains the more recent estimates of \cite{Nettelmann-etal_2013} and \cite{Neuenschwander-Helled_2022}, whose values for $\lambda$ range from $0.21840$ to $0.22670$ depending on the model used. We obtain values of the frequency parameter $p$ of Uranus ranging from $0.0075$ to $0.0092''$yr$^{-1}$, with minor contribution to the uncertainty range coming from other parameters.

Among all terms in Uranus' $\zeta$ series computed by \cite{Laskar_1990}, the negative frequency with smallest absolute value is $\nu_{15}=g_7-g_8+s_7$ (see Table~\ref{tab:zetashort} and Appendix~\ref{asec:QPS}). Even though this term is very small and unlikely to produce a resonance capture (see below), we can use it to compute the minimum mass ever that a satellite should have in order to tilt Uranus via a secular spin-orbit resonance after the end of the planetary migration. According to Eq.~\eqref{eq:condtilt} and with our $10\%$ uncertainty range on the value of $\omega\lambda$, we obtain a minimum mass parameter $\eta_\mathrm{min}$ ranging from about $130$ to $150$. From Eq.~\eqref{eq:etamass}, this gives a minimum mass $m_\mathrm{min}$ for the satellite lying between about $3.2\times 10^{-4}$ and $3.8\times 10^{-4}$ the mass of Uranus. Two comments can be made about this estimate: Firstly, we note that it does not strongly depend on the poorly known value of $\omega\lambda$; since we look here for an order-of-magnitude estimate, we adopt its central value for definitiveness and use it in the rest of the article. Secondly, the minimum mass obtained here is not absurdly high as compared to the mass ratios of regular satellites found in the Solar System (see e.g. \citealp{Murray-Dermott_1999}) or obtained in simulations of satellite formation around Uranus and Neptune \citep{Szulagyi-etal_2018}. For comparison, a satellite with mass $m/M\approx 3.5\times 10^{-4}$, that is, $m\approx3\times 10^{22}$\,kg, would be smaller than Jupiter's moon Europa. Yet, it is still about ten times the mass of Titania or Oberon, which are today the most massive satellites of Uranus. Hence, if Uranus has been tilted with the mechanism described here, then it should have involved an ancient satellite which has now disappeared.

The need for a hypothetical past satellite does not seem to be a problem because the satellite involved is likely to be destructed anyway during the final stage of the tilting mechanism (see \citealp{Saillenfest-Lari_2021}). We will come back to this point below. Now that the need for a hypothetical ancient satellite of Uranus is established, we must determine what could have been its physical and orbital properties in order for the tilting scenario to be plausible. Apart from the very small $\nu_{15}$ term, Uranus' $\zeta$ series features several low-frequency terms that may be good candidates for secular spin-orbit resonances. Table~\ref{tab:res} lists the eight closest resonances with the corresponding minimum masses $m_\mathrm{min}$ for the hypothetical satellite computed through Eq.~\eqref{eq:condtilt}. For completeness, we include in this table some terms that would require quite a massive satellite to trigger a resonance, even though this possibility appears more unlikely in the context of satellite formation theories (see the discussions in Sect.~\ref{sec:discuss}).

\begin{table}
   \caption{Minimum mass of the satellite in order for Uranus to reach the few closest resonances and be tilted from $0^\circ$ to $90^\circ$.}
   \label{tab:res}
   \vspace{-0.7cm}
   \begin{equation*}
      \begin{array}{rcrrrr}
         \hline
         \hline
         k  & \text{identification} &  \nu_k & T_\mathrm{lib}^{(\mathrm{sep})} & m_\mathrm{min}/M & m_k/M \\
         & & (''\,\text{yr}^{-1}) & (\text{Myrs}) & (\times 10^{-5}) & (\times 10^{-5}) \\
         \hline
         15 &  g_7 - g_8 + s_7 & -0.58033 & 3021 &  35 &  36 \\
          3 &              s_8 & -0.69189 &  115 &  41 &  44 \\
          9 &  g_6 - g_7 + s_6 & -1.19906 &  519 &  71 &  79 \\
          7 &  g_5 - g_7 + s_7 & -1.84625 &   92 & 110 & 129 \\
          5 & -g_5 + g_6 + s_6 & -2.35835 &   51 & 141 & 172 \\
          2 &              s_7 & -3.00557 &    4 & 179 & 233 \\
          8 & -g_7 + g_8 + s_8 & -3.11725 &  115 & 186 & 244 \\
          6 & -g_5 + g_7 + s_7 & -4.16482 &   40 & 248 & 368 \\
         \hline
      \end{array}
   \end{equation*}
   \vspace{-0.5cm}
   \tablefoot{$k$ is the index of the resonant term in Uranus' $\zeta$ series ordered by decreasing amplitude (see Table~\ref{tab:zetashort}); $T_\mathrm{lib}^{(\mathrm{sep})}$ is the libration period inside the resonance when the separatrix appears; $m_\mathrm{min}$ is the minimum mass of the satellite obtained analytically through Eq.~\eqref{eq:condtilt}; $m_k$ is the minimum mass of the satellite obtained semi-analytically using a more realistic model. The terms are sorted here by increasing value of the frequency $|\nu_k|$ (i.e. by increasing value of $m_\mathrm{min}$).}
\end{table}

\subsection{Migration range and velocity}\label{sec:res}

\begin{figure*}
   \includegraphics[width=\textwidth]{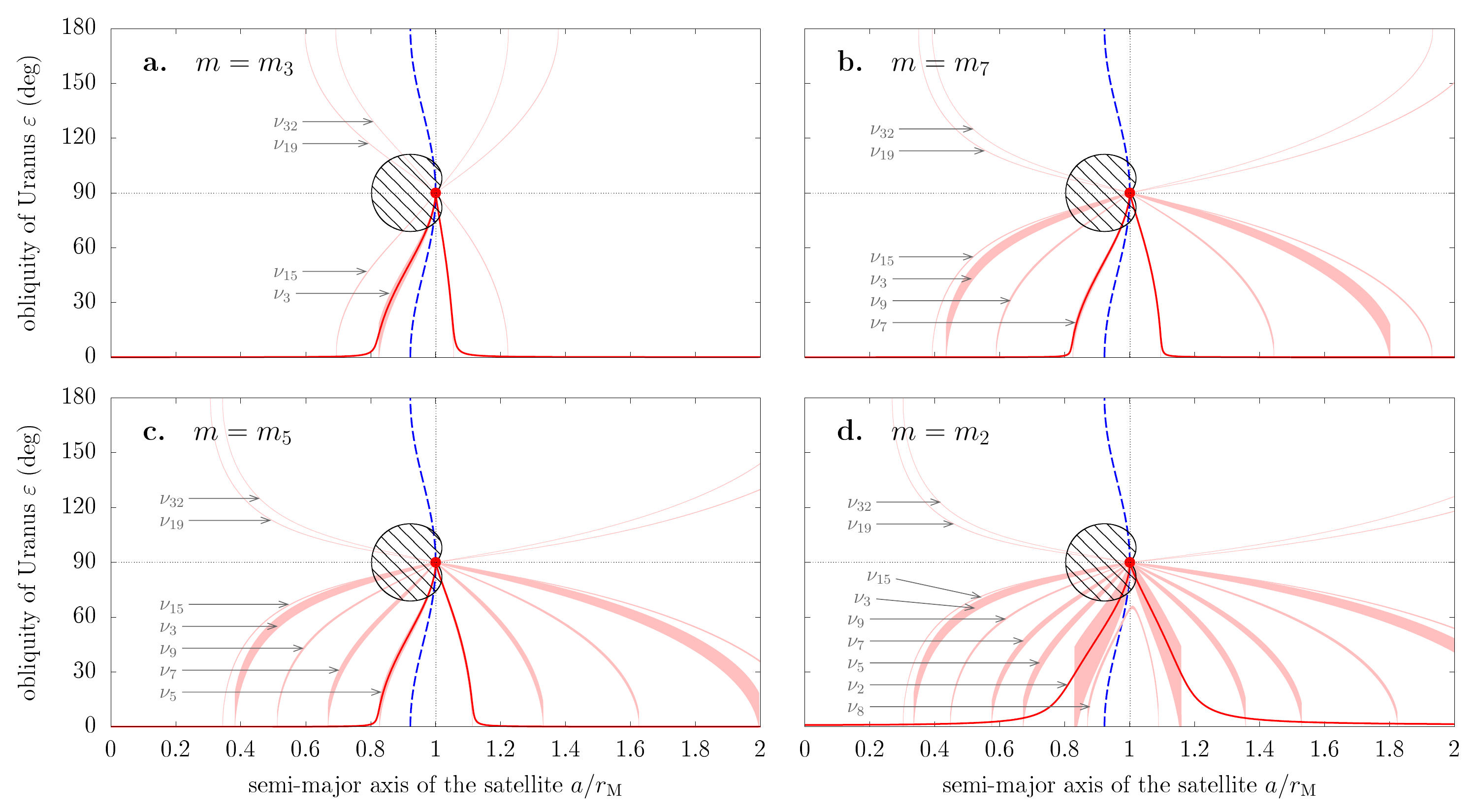}
   \caption{Locations and widths of first-order secular spin-orbit resonances for different masses of the satellite. In each panel, the labelled mass $m_j$ is the minimum satellite mass in order to allow for Uranus to reach the $\nu_j$ resonance and be tilted from $0^\circ$ to $90^\circ$ (see Table~\ref{tab:res}). The extent of all resonances is shown in pink and the centre of the $\nu_j$ resonance is highlighted with a red curve. The approximate ridge line separating the close and far satellite regimes is shown by the dashed blue line (same as in Fig.~\ref{fig:precrate}). The black striped area is the region $\mathrm{E}_1$ where the satellite's classic Laplace plane is unstable. The red dot is the singular point $\mathrm{S}_1$. Resonances are labelled by the frequency $\nu_j$ of the forcing term (arrows).}
   \label{fig:reswidths}
\end{figure*}

In order to allow for a resonance capture, the migration of the satellite must be slow compared to the oscillations of the resonance angle $\sigma=\psi+\phi_j$, so that the parameter change is close to the adiabatic regime (see e.g. \citealp{Su-Lai_2020,Saillenfest-etal_2020,Saillenfest-etal_2021b}). Among the terms listed in Table~\ref{tab:res}, this condition may not be verified for all realistic sets of parameters. For each resonance, the oscillation period near the resonance centre (called Cassini state~2) can be computed by modelling the influence of the satellite as an enhanced precession constant $\alpha'$ for Uranus (see e.g. \citealp{French-etal_1993,Ward-Hamilton_2004,Saillenfest-Lari_2021}) and by applying the analytical formulas of \cite{Saillenfest-etal_2019} or \cite{Su-Lai_2020}. As described by \cite{Boue-Laskar_2006}, the spin-axis precession rate of a planet increases over time if it hosts an outward-migrating close-in satellite, or an inward-migrating far-away satellite. When the satellite reaches some threshold distance, a separatrix appears around Cassini state~2, after which point the planet may be captured in the secular spin-orbit resonance considered. The libration period $T_\mathrm{lib}^\mathrm{(sep)}$ when the separatrix appears is given in Table~\ref{tab:res} for each candidate resonance. An adiabatic resonance encounter could possibly have taken place for Uranus if $T_\mathrm{lib}^\mathrm{(sep)}$ is much smaller than the age of the Solar System. In this regard, the $\nu_9$ and $\nu_{15}$ resonances are very unlikely candidates, as adiabaticity would impose the satellite to hardly migrate at all over the Solar System lifetime, in which case the tilting of Uranus is impossible. Indeed, these two terms have the smallest amplitudes in our list, and the resonance width and oscillation frequency scale as the square root of the amplitude of the term (see e.g. \citealp{Atobe-etal_2004,Li-Batygin_2014}). The situation is different for the other candidate resonances in Table~\ref{tab:res}: we see that Uranus would have enough time to oscillate at least several dozens of time inside the resonance over the Solar System age, potentially allowing for an adiabatic capture and tilting to extreme obliquities. We quantify this point below.

Equation~\eqref{eq:psidotsimp} provides a qualitative understanding of how the spin-axis precession rate varies over the space of parameters, and we used it to obtain a quick estimate for the minimum mass of the satellite $m_\mathrm{min}$ required for each candidate resonance. However, one may recall that Eq.~\eqref{eq:psidotsimp} is approximate and strictly valid only for small satellites. As argued by \cite{Saillenfest-Lari_2021}, the complete formula of \cite{French-etal_1993} is more precise and gives satisfactory results up to larger satellite masses; it can then be used to compute the locations of the resonances and their widths for any distance of the satellite. In this case, there is no closed-form expression for the minimum mass of the satellite but we can compute it in the following way: For a given mass $m$, the formula of \cite{French-etal_1993} is used to obtain an effective precession constant $\alpha'$. This precession constant is then injected into the equations of \cite{Saillenfest-etal_2019} in order to compute the location and width of all resonances in the plane $(a,\varepsilon)$. Since $\alpha'$ depends on the obliquity, the equations are implicit and must be solved numerically. We repeat this procedure for different masses (e.g. using the bisection method) until we obtain the minimum mass needed for a given resonance to exist somewhere in the plane $(a,\varepsilon)$ and go from $\varepsilon=0^\circ$ to $90^\circ$. The minimum masses obtained in this way are listed in Table~\ref{tab:res}; we call them $m_k$, where $k$ is the index of the resonant term in Uranus' $\zeta$ series. When $m_k$ is large, we see that our simplified analytical expression systematically underestimates its value (i.e. $m_\mathrm{min}\lesssim m_k$), even though it gives the correct order of magnitude.

Figure~\ref{fig:reswidths} shows the location and width of all resonances reachable by the system for different masses of the satellite. It also features the region $\mathrm{E}_1$ where the classic Laplace plane of the satellite is unstable (black striped area; see \citealp{Tremaine-etal_2009} and \citealp{Saillenfest-Lari_2021}). If the system enters the region $\mathrm{E}_1$, the satellite first transfers to an eccentric equilibrium, and then it suffers from wild orbital changes, that are the most extreme in the vicinity of the singular point $\mathrm{S}_1$. The behaviour of the system in this region is investigated in Sect.~\ref{sec:coupledmod}. As expected, resonances are more numerous for a more massive satellite, because their respective minimum mass criteria are verified simultaneously. When we increase the mass of the satellite, resonances appear near $a=r_\mathrm{M}$ (blue zone in Fig.~\ref{fig:precrate}), then touch the singular point $\mathrm{S}_1$ (red curve in Fig.~\ref{fig:precrate}), at which point we define the mass $m_k$. Then, if we increase the mass further, resonances expand towards smaller and larger semi-major axes (pink region in Fig.~\ref{fig:precrate}). Figure~\ref{fig:reswidths} shows that the $\nu_9$ and $\nu_{15}$ resonances are indeed very thin, and that Uranus also possesses two extremely weak retrograde resonances ($\nu_{19}$ and $\nu_{32}$). Since $m_2$ is only slightly smaller than $m_8$ (see Table~\ref{tab:res}), the $\nu_8$ resonance appears in panel~d but does not connect to $\mathrm{S}_1$ (as in the blue zone in Fig.~\ref{fig:precrate}). From these examples, we deduce that in order to tilt Uranus from $0^\circ$ to $90^\circ$, its hypothetical ancient satellite must migrate at least over $\Delta a\approx 0.2\,r_\mathrm{M}$, either inwards or outwards. This semi-major axis range represents a physical distance of about $10$ Uranus' radii. In order for the satellite to go through this distance in less than the age of the Solar System, its mean migration velocity must be larger than about $6$~cm\,yr$^{-1}$. Compared to the measured expansion rates of the Moon from the Earth ($4$~cm\,yr$^{-1}$; see e.g. \citealp{Williams-Boggs_2016}), Ganymede from Jupiter ($11$~cm\,yr$^{-1}$; see \citealp{Lainey-etal_2009}) and Titan from Saturn ($11$~cm\,yr$^{-1}$; see \citealp{Lainey-etal_2020}), this migration velocity appears quite realistic. Hence, even though the specific dissipation mechanisms acting in the interior of Uranus are essentially unknown yet, at least we are assured that both terrestrial and gaseous planets are able to generate the appropriate range of satellite migration velocities.

\begin{figure}
   \centering
   \includegraphics[width=\columnwidth]{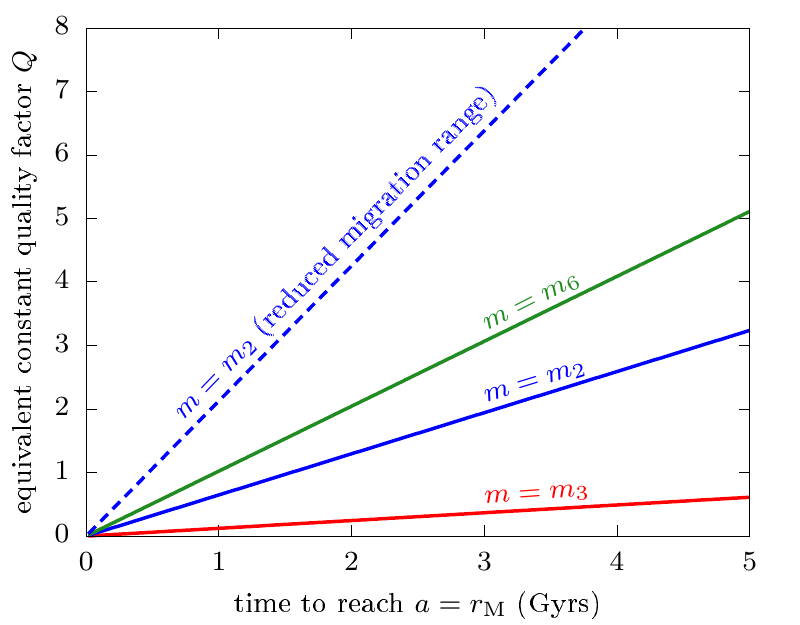}
   \caption{Quality factor of tidal energy dissipation within Uranus in order for a satellite to migrate from $a=0.8\,r_\mathrm{M}$ to $a=r_\mathrm{M}$. The ratio $k_2/Q$ is computed from classical formulas and plotted here assuming a Love number $k_2=0.1$ for Uranus \citep{Gavrilov-Zharkov_1977}. The mass of the satellite is labelled above the curves. For comparison, the dotted blue line shows the case of a satellite migration from $a=0.96\,r_\mathrm{M}$ to $a=r_\mathrm{M}$ (i.e. a migration range reduced to $\Delta a=2$~$R_\mathrm{eq}$), as obtained in the numerical experiments of Sect.~\ref{sec:coupledproba}.}
   \label{fig:k2sQ}
\end{figure}

It should be noted, however, that the satellite migration needs to be sustained at least up to a distance of $a\gtrsim 0.8\,r_\mathrm{M}$ in order for the tilting mechanism to fully operate (see Fig.~\ref{fig:reswidths}). Since the characteristic radius $r_\mathrm{M}$ is quite large for Uranus ($r_\mathrm{M}\approx 53$~$R_\mathrm{eq}$), this may require the satellite to be fairly big to raise a substantial tidal bulge inside Uranus even at this distance. Assuming that the satellite migrates from $a=0.8\,r_\mathrm{M}$ to $a=r_\mathrm{M}$, the equivalent constant parameter $k_2/Q$ of energy dissipation within Uranus can be estimated from classical theories (see e.g. \citealp{Goldreich-Soter_1966,Efroimsky-Lainey_2007}). As shown by Fig.~\ref{fig:k2sQ}, the mean energy dissipation within Uranus is required to be much higher than assumed in historical works, even if the whole process takes as long as the age of the Solar System (for comparison, \citealp{Goldreich-Soter_1966} gave $Q\gtrsim 10^5$). This new scenario for the tilting of Uranus must therefore be viewed in the context of the fast satellite migration predicted by \cite{Fuller-etal_2016} and measured by \cite{Lainey-etal_2020}. Within this change of paradigm, `effective' quality factors $Q$ of order unity are realistic for distant satellites around gaseous planets. For instance, \cite{Fuller-etal_2016} predicted $Q\approx 20$ for Titan, and $Q\approx 1$ for Callisto if a similar resonance-locking mechanism is currently at play in the Jupiter-Callisto system as well. Effective quality factors can also be smaller than unity, because they are not directly linked to a physical tidal lag as they would in classical theories (see \citealp{Fuller-etal_2016}). The definition of an equivalent constant $Q$ here is therefore only used for comparison purpose.

Alternatively, a chain of mean-motion resonances among several satellites might offer another way to produce a fast satellite migration up to large distances (as it is the case for Ganymede; see e.g. \citealp{Lari-etal_2020}). But again, these aspects intimately depend on the specific dissipation mechanism at play and we can hardly give any more quantitative argument at this stage.

\subsection{Tilting efficiency}\label{sec:proba}
The few closest resonances reachable by Uranus and its hypothetical satellite have diverse properties, and apart from general considerations about the level of adiabaticity required, it is not obvious which resonances are more prone to tilting Uranus than others. For this reason, we turn to a numerical exploration of the capture mechanism.

Our setting is similar to the one used by \cite{Saillenfest-Lari_2021} for Saturn: we implement the classic equations of motion for the secular spin-axis dynamics (see e.g. \citealp{Laskar-Robutel_1993,NerondeSurgy-Laskar_1997}), in which we modify the precession constant $\alpha$ according to the formula of \cite{French-etal_1993}. The satellite is assumed to lie on its local Laplace plane at all time and to adiabatically follow its drift as the obliquity is changing. This approximation is valid as long as the orbital timescale of the satellite is much shorter than the motion of the planet's spin axis, which is well verified in practice. However, it assumes that the satellite's orbital equilibrium is stable, which is not true in region $\mathrm{E}_1$. Modelling the satellite's influence as a modified $\alpha$ also requires that the variations of $\alpha$ are slow compared to the spin-axis dynamics, which is not the case in the vicinity of $\mathrm{S}_1$. For these reasons, this approach is very efficient for the stage of capture in secular spin-orbit resonance and adiabatic tilting, but it fails when the system reaches the unstable zone. We use it here to conduct a statistical analysis of the capture process and to perform millions of numerical integrations in a reasonable amount of time. As for the destabilisation stage, it is studied in Sect.~\ref{sec:coupledmod} below using a self-consistent model.

\begin{figure*}
   \centering
   \includegraphics[width=0.85\textwidth]{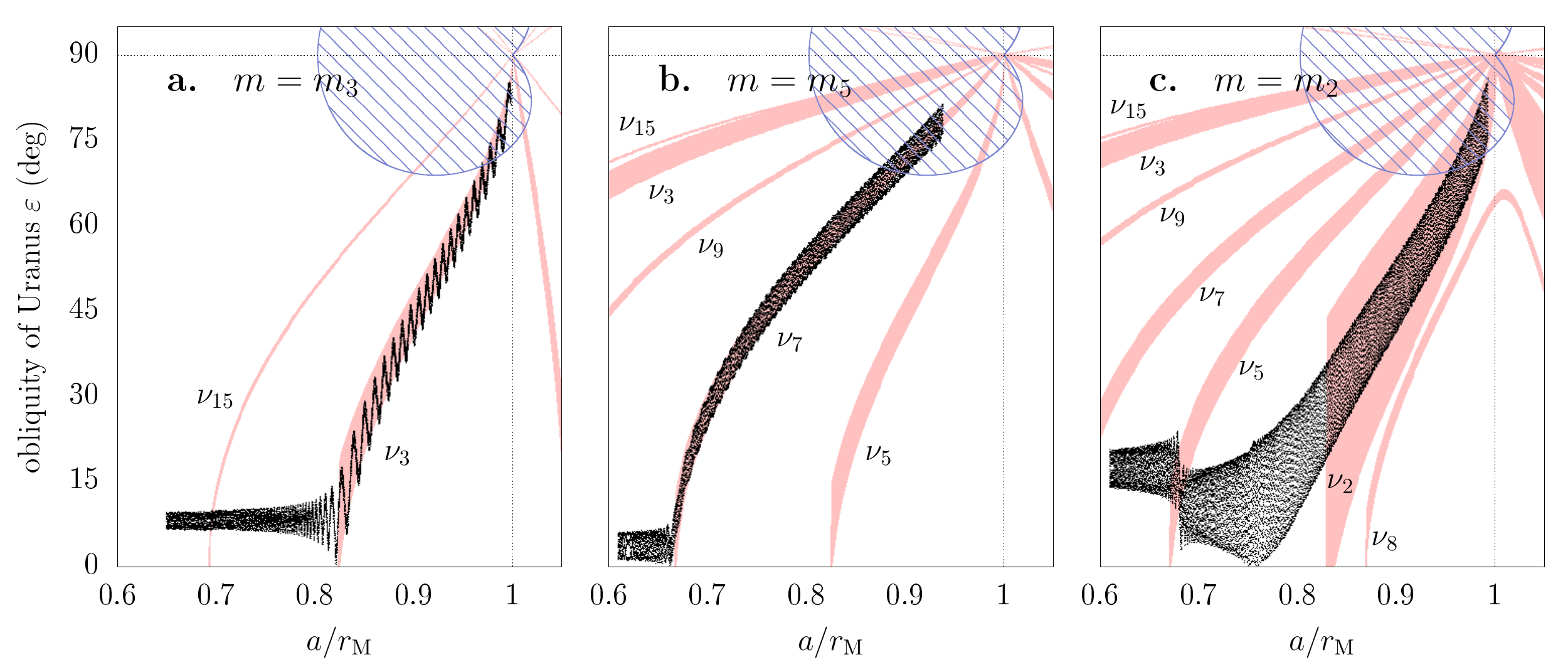}
   \caption{Examples of capture in secular spin-orbit resonance and tilting to large obliquities. The satellite's mass and the resonances are labelled as in Fig.~\ref{fig:reswidths}. Numerical trajectories go from left to right (black dots) over a timespan of billions of years. The migration timescale $\tau$ of the satellite is $9$~Gyrs for panel~a and $6$~Gyrs for panels~b and c. Integrations are stopped in the hatched region, where the numerical model used here fails.}
   \label{fig:exsuccess}
\end{figure*}

\begin{figure*}
   \centering
   \includegraphics[width=0.85\textwidth]{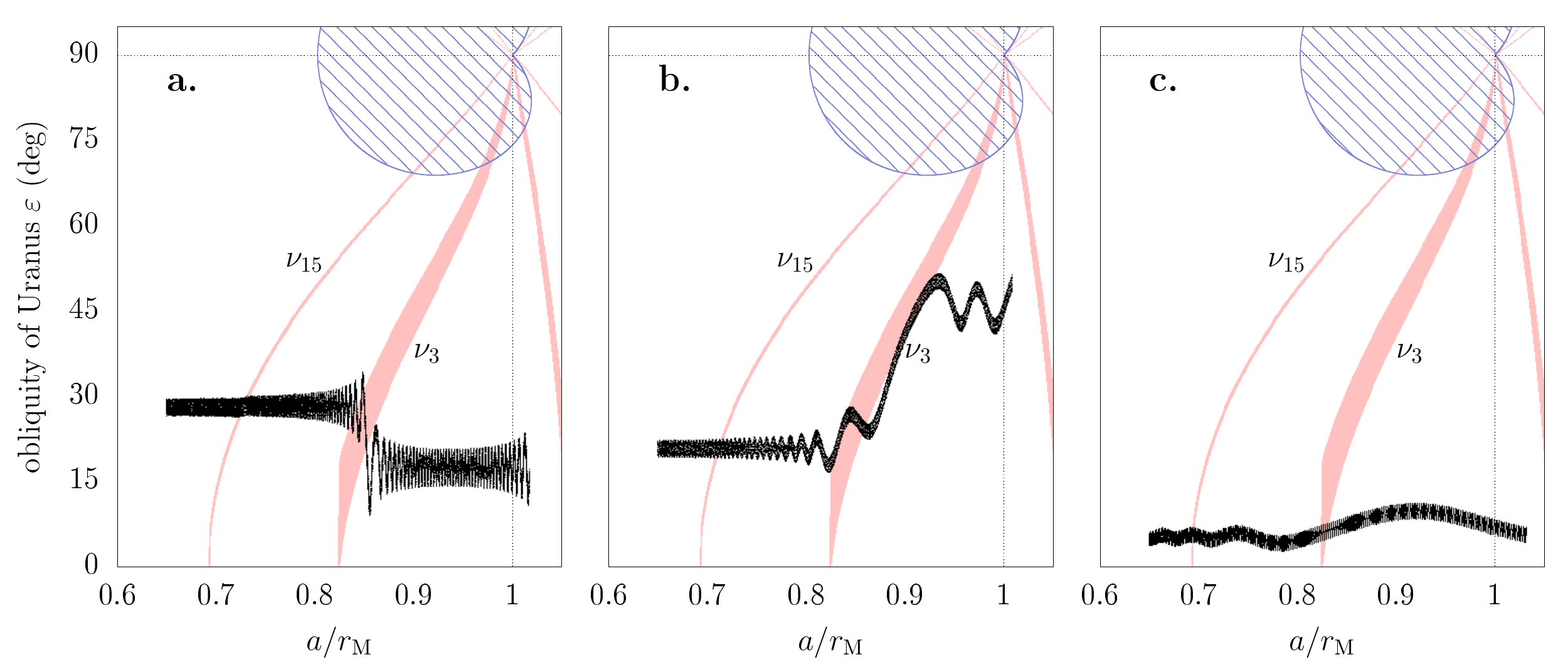}
   \caption{Examples of resonant interactions without tilting all the way to the unstable region. Symbols are the same as in Fig.~\ref{fig:exsuccess} and the satellite has mass $m_3$ is all panels. The migration timescale $\tau$ of the satellite is $9$, $2$, and $0.2$~Gyrs for panels~a, b, and c, respectively.}
   \label{fig:exfailure}
\end{figure*}

The tilting mechanism is driven by the migration of the satellite, which runs on a billion-year timescale. Tidal migration of satellites is a complex phenomenon that deeply depends on the internal structure of their host planet, and there is much ongoing research on this topic (see e.g. \citealp{AuclairDesrotour-etal_2014,Fuller-etal_2016,Lin-Ogilvie_2021}). Satellite migration around terrestrial or gaseous planets are known to involve very different physical mechanisms, and little is known about the diversity of satellite evolutions that can be produced around ice giants like Uranus and Neptune. Instead of relying on one migration scenario, we explore a large range of possibilities for the variations of the satellite's semi-major axis $a$ over time $t$. For definitiveness, we adopt a power law of the type
\begin{equation}\label{eq:asat}
   a(t) = a_0\left(\frac{t_0+t}{t_0}\right)^{t_0/\tau} \,,
\end{equation}
where $a_0$ is the initial location of the satellite, and $t_0$ and $\tau$ are parameters that have the dimension of time. In the following, we adopt $t_0=4.5$~Gyrs and vary the parameter $\tau$, that we call the `migration timescale'. We stress that Eq.~\eqref{eq:asat} is purely ad hoc, and that one could have chosen a linear law instead. The study of more complex migration laws (e.g. involving fast and slow regimes) is beyond the scope of this article; however, if the migration is slow enough to remain in the adiabatic regime, only the total migration range matters, and not its precise rate.

At this point, it should be recalled that, even if we do not model it in detail, the tidal migration of a satellite occurs because of a transfer of angular momentum between the planet's rotation and the satellite's orbit. The outward migration of a satellite is therefore accompanied by a decrease in the planet's rotation rate $\omega$ (along with a change in its equilibrium shape). For the largest masses $m_k$ listed in Table~\ref{tab:res}, and assuming that the increase in the satellite's angular momentum is fully compensated by a change in $\omega$, the migration of a satellite from $a\approx 0.8\,r_\mathrm{M}$ to $1\,r_\mathrm{M}$ would produce a change in $\omega$ of several percent. Such a variation is of the same order of magnitude as the error on the value of $\omega$ measured by Voyager~2 \citep{Helled-etal_2010}, and much less than our assumed uncertainty on the value of $\lambda$ (see Sect.~\ref{sec:mass}). Moreover, as the planet slowly cools down on a gigayear timescale after its formation, it is expected to contract slightly (see \citealp{Bodenheimer-Pollack_1986,DAngelo-etal_2021}), with the result of increasing its spin rate. This opposite effect would further reduce the tidal spin down due to satellites. For these reasons, we neglect the variations of $\omega$ and $J_2$ in our dynamical analysis. We also consider that their changes are small enough so as to present no contradiction with our basic assumption that the spin rate of Uranus is primordial (see Sect.~\ref{sec:intro}).

We are looking for trajectories that go deep inside the unstable region for the satellite. This way, the obliquity not only grows large, but we are also assured that the satellite is destabilised and potentially removed, as required to reproduce the current state of Uranus. Examples of such trajectories are given in Fig.~\ref{fig:exsuccess}. Panel~a shows a capture in the $\nu_3$ resonance; in this example, the satellite has the minimum mass allowing Uranus to be captured in this resonance and tilted all the way to $90^\circ$. The thin $\nu_{15}$ resonance is crossed as well but it hardly affects the trajectory at all. Panel~b shows a case where the satellite has mass $m_5$ but is captured in the $\nu_7$ resonance. Indeed, for such a large mass, resonances are numerous and there are many possibilities to tilt the planet. As another example, panel~c shows a case where the resonance $\nu_5$ is crossed without capture but the planet is then captured and tilted in the $\nu_2$ resonance.

\begin{figure*}
   \centering
   \includegraphics[width=0.9\textwidth]{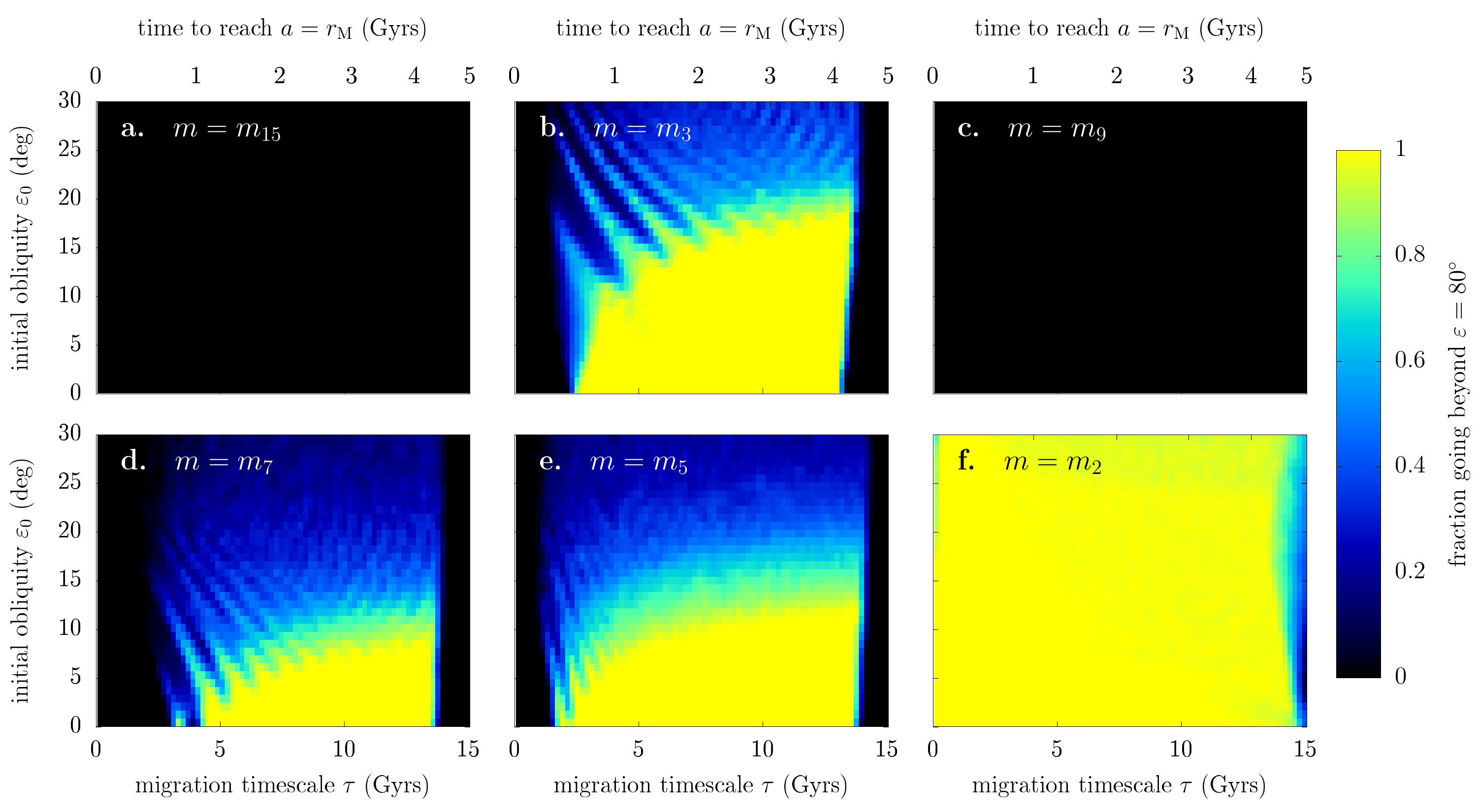}
   \caption{Probability of tilting Uranus as a function of its initial obliquity and the migration velocity of its satellite. On each panel, parameters are sampled on a regular grid, and for each set of parameters, $240$ numerical integrations are performed with random initial spin-axis phases $\psi\in [0,2\pi)$. The colour map shows the fraction of the $240$ trajectories that go beyond $\varepsilon=80^\circ$ during the integration. The satellite is initialised at $a_0=0.8\,r_\mathrm{M}$ and propagated with the migration law in Eq.~\eqref{eq:asat}. Numerical integrations are stopped after $4$~Gyrs or when the satellite goes beyond the limit of the unstable zone. The mass of the satellite is labelled on each panel using the notation of Table~\ref{tab:res}. The top horizontal axis shows the time required for the satellite to reach $a=r_\mathrm{M}$. It can be converted into an effective tidal quality factor $Q$ for Uranus (see Fig.~\ref{fig:k2sQ}).}
   \label{fig:resstats}
\end{figure*}

However, capture and extreme tilting are not guaranteed even if the satellite has the required mass. There are four cases in which the mechanism can fail to incline the planet enough to reach the unstable zone. The first case is when the satellite does not migrate enough over the Solar System age so as to reach $a\approx r_\mathrm{M}$. This can happen if it formed too close to the planet (i.e. too far from the Laplace radius) or if it migrates too slowly. The three other cases are illustrated in Fig.~\ref{fig:exfailure}. In panel~a, the planet has a large initial obliquity, which means that it must inevitably cross the resonance separatrix instead of entering smoothly from below (see e.g. \citealp{Saillenfest-etal_2020}). The outcome of separatrix crossings is sensitive to the initial precession phase, and the planet has therefore a given probability of being captured or not (see e.g. \citealp{Henrard-Murigande_1987,Ward-Hamilton_2004,Hamilton-Ward_2004,Su-Lai_2020}). In our case, the capture probability decreases for growing initial obliquity, and panel~a is a typical example of failed capture. The numerical experiments of \cite{Saillenfest-etal_2021b} show that a fast tidal migration of the satellite tends to smooth the probability profile and to reduce the interval of obliquity leading to a $100\%$-sure capture. In panel~b, the planet is captured but evolves very close to the separatrix; as the resonance width decreases, the trajectory is then pushed out of the resonance before having reached a large obliquity. In panel~c, the migration of the satellite is so fast that the resonance is crossed without altering much the obliquity. This is a typical example of non-adiabatic crossing, in which the resonance angle has not enough time to oscillate before the trajectory is already gone.

In order to quantify the ability of each candidate resonance to tilt Uranus and destroy the satellite, we turn to Monte Carlo experiments. In these experiments, the satellite is given the minimum mass $m_k$ required for a given resonance to exist and go to $90^\circ$ (see Table~\ref{tab:res}). The satellite is initialised at $a_0 = 0.8\,r_\mathrm{M}$, so as to focus on the effects of resonance $\nu_k$ and minimise the migration range required for the tilting (see Fig.~\ref{fig:reswidths}). Over a grid of initial obliquity $\varepsilon_0$ and migration timescale $\tau$, we sample random values of initial spin-axis precession angle $\psi\in[0,2\pi)$ and integrate all trajectories numerically until the satellite goes beyond $a^5/r_\mathrm{M}^5=(10\sqrt{22}-4)/39$, which is the rightmost limit of the unstable zone, and for a maximum timespan of $4$~Gyrs. This integration timespan is chosen so that the entire tilting mechanism can take place between the formation of Uranus and today, with possibly some time left for the clearing of the debris disc after the satellite's destruction (see Sects.~\ref{sec:coupledmod} and \ref{sec:discuss} below). A statistic is drawn from each sample of initial precession angles. We consider that the system goes deep enough inside the unstable region if the obliquity grows over $\varepsilon = 80^\circ$. This threshold does not mean that the obliquity cannot grow more than $80^\circ$, but the model used here fails in this regime, so we prefer to use a conservative limit. The behaviour of the system after destabilisation will be studied in Sect.~\ref{sec:coupledmod}.

Figure~\ref{fig:resstats} shows the result of our experiments for the few closest resonances reachable by Uranus. Even though near-zero initial obliquities are expected from formation models, a variety of different mechanisms may have produced small primordial obliquity excitations (see e.g. \citealp{Millholland-Batygin_2019,Martin-etal_2020,Rogoszinski-Hamilton_2020,Rogoszinski-Hamilton_2021}); for this reason, we explore a wide range of values for $\varepsilon_0$. Panel~b for $m=m_3$ is a typical illustration of the different possible behaviours: for $\tau\gtrsim 13$~Gyrs the migration is too slow to tilt the planet in only $4$~Gyrs, so the success ratio is zero. For $\tau\lesssim 2$~Gyrs the migration is too fast for an adiabatic capture in the $\nu_3$ resonance (as in Fig.~\ref{fig:exfailure}c). For $\varepsilon_0\gtrsim 20^\circ$ the trajectories cross the separatrix, leading to an inefficient probabilistic capture (as in Fig.~\ref{fig:exfailure}a,b). This leaves us with an optimal parameter region with initial obliquity $\varepsilon_0\lesssim 20^\circ$ and migration timescale $2\lesssim\tau\lesssim 13$~Gyrs, which shows a success probability close to~$1$. The same structure is observed in panels~d and e for resonances $\nu_7$ and $\nu_5$. In panels~a and c, on the contrary, the resonances $\nu_{15}$ and $\nu_9$ are so thin that the resonance encounter is strongly non-adiabatic in the whole parameter region explored and no trajectory can possibly achieve the tilting of Uranus in the required time interval. In panel~f, the resonance $\nu_2$ is so large that it produces a $100\%$-sure capture up to large initial obliquities and down to very small migration timescales; moreover, due to the large amplitude of obliquity oscillations (see Fig.~\ref{fig:exsuccess}c), trajectories can go beyond $\varepsilon=80^\circ$ much before the centre of the resonance does so, which enlarges the yellow region towards the right side of the graph.

From this preliminary analysis, we conclude that a promising resonance candidate is $\nu_3=s_8$ (i.e. a resonance with the nodal orbital precession mode of Neptune): it requires a minimum mass for the satellite of only $m/M\approx 4\times 10^{-4}$ and has a high tilting efficiency for a large range of initial obliquities ($0^\circ\leqslant\varepsilon_0\lesssim 20^\circ$). For larger masses $m/M\gtrsim 1$ to $2\times 10^{-3}$, the nodal precession spectrum of Uranus allows for several other possibilities: the resonances with $\nu_7=g_5 - g_7 + s_7$ and $\nu_5=-g_5 + g_6 + s_6$ result to be quite good candidates, but in a more limited range of initial obliquities ($\varepsilon_0\lesssim 10^\circ$ and $15^\circ$, respectively). As already argued by previous authors \citep{Boue-Laskar_2010,Rogoszinski-Hamilton_2020}, the most powerful resonance in terms of tilting efficiency is $\nu_2=s_7$ (i.e. a resonance with the nodal orbital precession mode of Uranus itself). Indeed, it shows a $100\%$ capture ratio up to large initial obliquities for Uranus ($\varepsilon_0\lesssim 45^\circ$) and for any conceivable migration rate for the satellite. However, it requires quite a massive satellite with $m/M\gtrsim 2.3\times 10^{-3}$: even though this value is not absurdly large, it may appear less realistic according to satellite formation theories (see the discussions in Sect.~\ref{sec:discuss}).

We are now assured that Uranus can be efficiently tilted to large obliquities through the migration of a hypothetical ancient satellite. The next step is to understand the behaviour of the system once it enters the unstable region. This is the purpose of the next section.

\section{Coupled destabilisation of the planet's spin and satellite's orbit}\label{sec:coupledmod}

We saw that several secular spin-orbit resonances allow for a $100\%$-sure capture probability and tilting to $\varepsilon\gtrsim 80^\circ$ in large regions of the parameter space. Such trajectories go deep into the unstable region (hatched zone in Figs.~\ref{fig:reswidths}, \ref{fig:exsuccess}, \ref{fig:exfailure}); this is a necessary condition, but is not sufficient to reproduce the current state of Uranus. Indeed, assuming that Uranus was indeed tilted by the mechanism presented here, its $98^\circ$-obliquity implies that the system not only reached the unstable region, but that some additional dynamical mechanism pushed it much further up, beyond the $90^\circ$ critical point.

\begin{figure}
   \centering
   \includegraphics[width=\columnwidth]{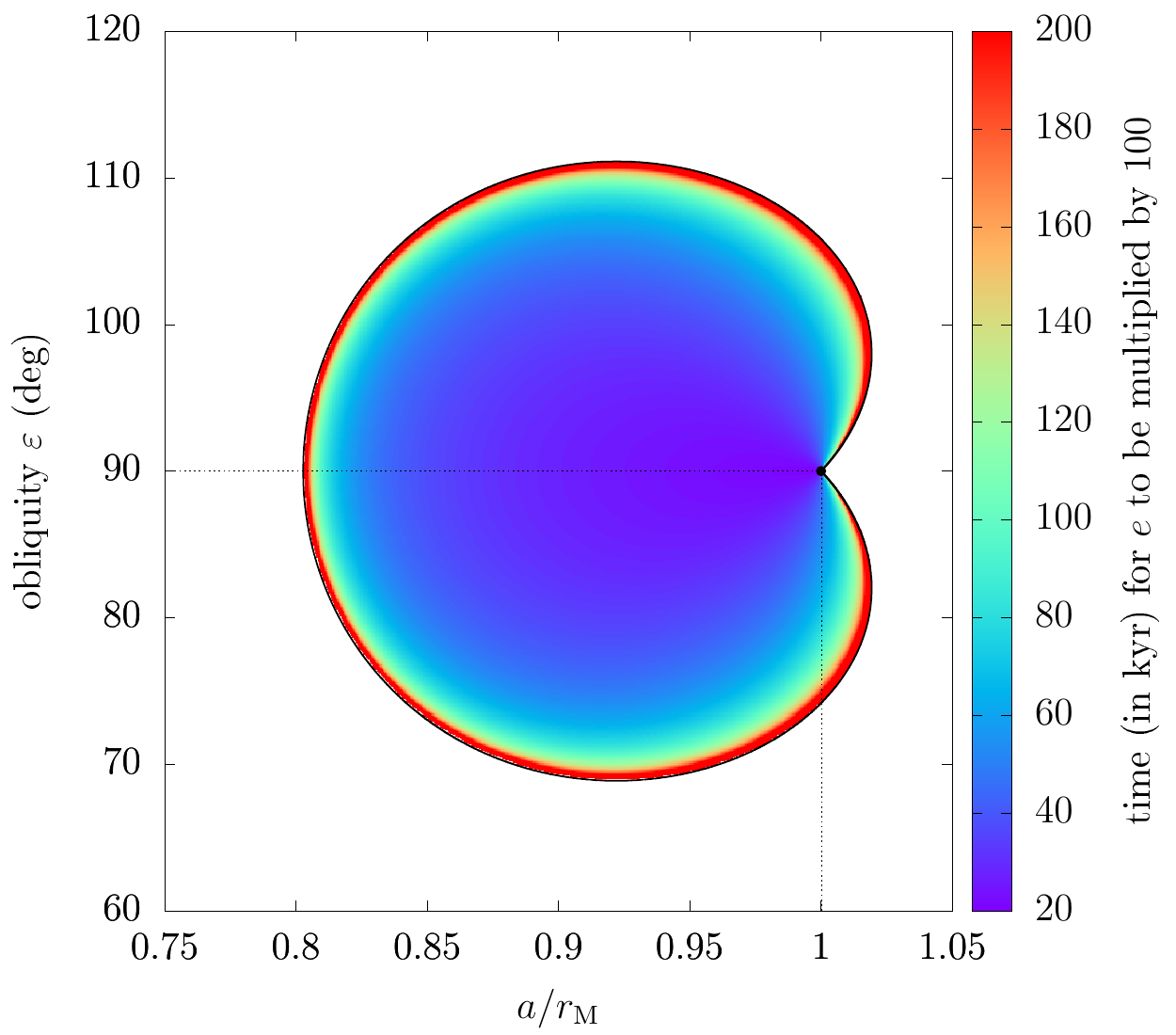}
   \caption{Timescale of the satellite's eccentricity increase when the system becomes unstable. The colour scale shows the time needed for the eccentricity to be multiplied by a factor $100$, as computed from the equations linearised at the unstable equilibrium point (see \citealp{Tremaine-etal_2009,Saillenfest-Lari_2021}).}
   \label{fig:ecc_increase}
\end{figure}

The simplified models used in Sect.~\ref{sec:tilting} are not valid when the satellite becomes unstable. In order to investigate what happens to the system at the end of the tilting mechanism, we must use models that include the self-consistent interactions between the planet's spin and satellite's orbit. For this purpose, the secular model described by \cite{Correia-etal_2011} for the hierarchical planetary case can be transposed to the interactions of a planet and its satellite subject to the attraction of their host star. Due to the highly hierarchical nature of the problem, it is enough to expand all cross interactions to quadrupolar order. The Hamiltonian function is then averaged over the fast orbital and rotational angles. Equations are written in a vectorial form and integrated numerically. In our case, the dynamical variables are the rotational angular momentum of the planet $\mathbf{G}$, the orbital angular momentum of the satellite $\mathbf{G}_1$, and the eccentricity vector of the satellite $\mathbf{e}_1$. The orbital angular momentum of the planet-satellite barycentre around the star and its eccentricity vector (written $\mathbf{G}_2$ and $\mathbf{e}_2$ using the notations of \citealp{Correia-etal_2011}) are taken as quasi-periodic functions of time, varying according to the full orbital solution of \cite{Laskar_1990}. This way, all planets of the Solar System are included in the orbital motion of the planet-satellite barycentre (but we neglect their direct attraction on the satellite and on the planet's equatorial bulge). A full description of this model is given in Appendix~\ref{asec:selfcons}. Tidal dissipation in the planet is mimicked by applying a slow variation to the satellite's semi-major axis as in Eq.~\eqref{eq:asat}. In this first exploration of the dynamics, we do not include tidal dissipation within the satellite and do not track its rotational dynamics. This amounts to assuming that the energy dissipated within the satellite always remains negligible with respect to that dissipated in the planet, and that the eccentricity damping of the satellite is much slower than its dynamical evolution. The latter point can be verified a priori: for realistic dissipative parameters (see e.g. \citealp{Murray-Dermott_1999}), the eccentricity damping timescale of the satellite at a distance $a\gtrsim 0.8r_\mathrm{M}$ would count in gigayears. In contrast, Fig.~\ref{fig:ecc_increase} shows that as soon as the system enters (even moderately deep) into the unstable region, the increase in the satellite's eccentricity unfolds in less than a few hundred thousand years\footnote{The destabilisation process is complicated by the existence of the stable eccentric equilibrium (see \citealp{Tremaine-etal_2009,Saillenfest-Lari_2021}). Yet, we have checked that when the destabilisation is triggered, the timescale of the eccentricity increase in our simulations is indeed consistent with that shown in Fig.~\ref{fig:ecc_increase}.}.

\subsection{Destabilisation mechanism}\label{sec:doubledestab}

\begin{figure}
   \centering
   \includegraphics[width=0.85\columnwidth]{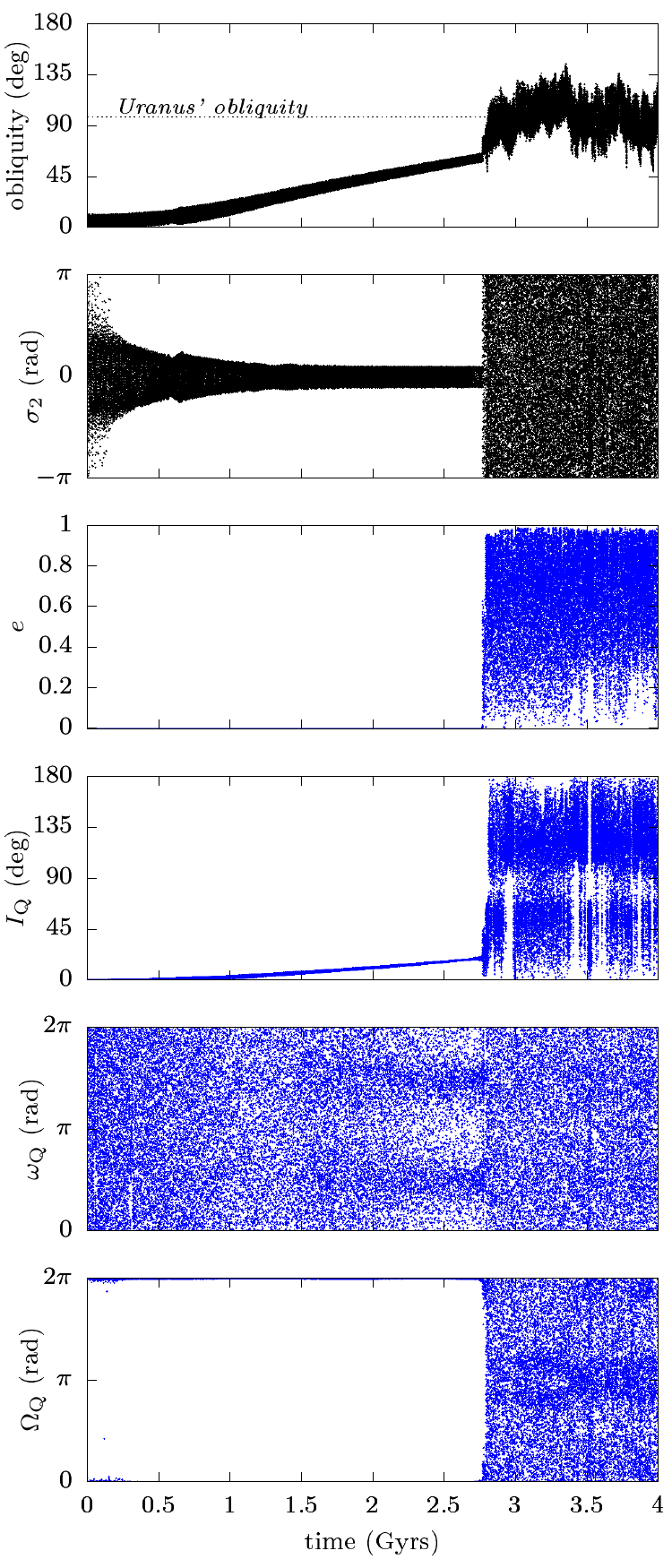}
   \caption{Example of simulation using the fully coupled secular model. Black and blue are used for quantities related to the planet's spin axis and satellite's orbit, respectively. The secular spin-orbit resonance angle is $\sigma_2=\psi+\phi_2$. The orbital elements of the satellite are measured with respect to the equator and equinox of date. The mass of the satellite is $m/M=2.2\times 10^{-3}$. The satellite is initialised close to the local Laplace equilibrium at $a=37$~$R_\mathrm{eq}$ and made migrating outwards with timescale $\tau=6.5$~Gyrs.}
   \label{fig:fullycoupled_ex1}
\end{figure}

Figure~\ref{fig:fullycoupled_ex1} shows a typical example of simulation obtained with the coupled model. The migration timescale $\tau$ is chosen to be in the middle of the parameter region producing a $100\%$-sure resonance capture (see Fig.~\ref{fig:resstats}). The first portion of the trajectory is identical to what was predicted in Sect.~\ref{sec:tilting} using the simplified dynamical model (see e.g. Fig.~\ref{fig:exsuccess}c for a similar satellite mass). Indeed, the oscillations of the satellite around the Laplace equilibrium are very fast compared to the motion of the planet's spin axis; the previous assumption that the satellite instantly follows the Laplace plane is therefore a very good approximation. The smooth increase in the satellite's orbital inclination while its longitude of node closely oscillates around zero is a direct manifestation of this adiabatic dynamics. Once the system reaches the border of the unstable zone, however, the dynamics brutally change: Figure~\ref{fig:fullycoupled_ex1} shows that the satellite's eccentricity increases very quickly, which releases the planet from the secular spin-orbit resonance ($\sigma_2$ circulates again). From this point on, the satellite's orbit starts to evolve chaotically and it can reach almost any eccentricity and inclination value, while switching sporadically between being prograde and retrograde with respect to the planet's spin. Interestingly, these strong chaotic orbital variations affect the planet's obliquity, which evolves pretty much as a random walk. At this stage of the evolution, there is no particular dynamical barrier anymore at $\varepsilon=90^\circ$; in the example of Fig.~\ref{fig:fullycoupled_ex1}, we see that the planet's spin can very well become retrograde and reach the current obliquity of Uranus (dotted line).

\begin{figure*}
   \centering
   \includegraphics[width=\textwidth]{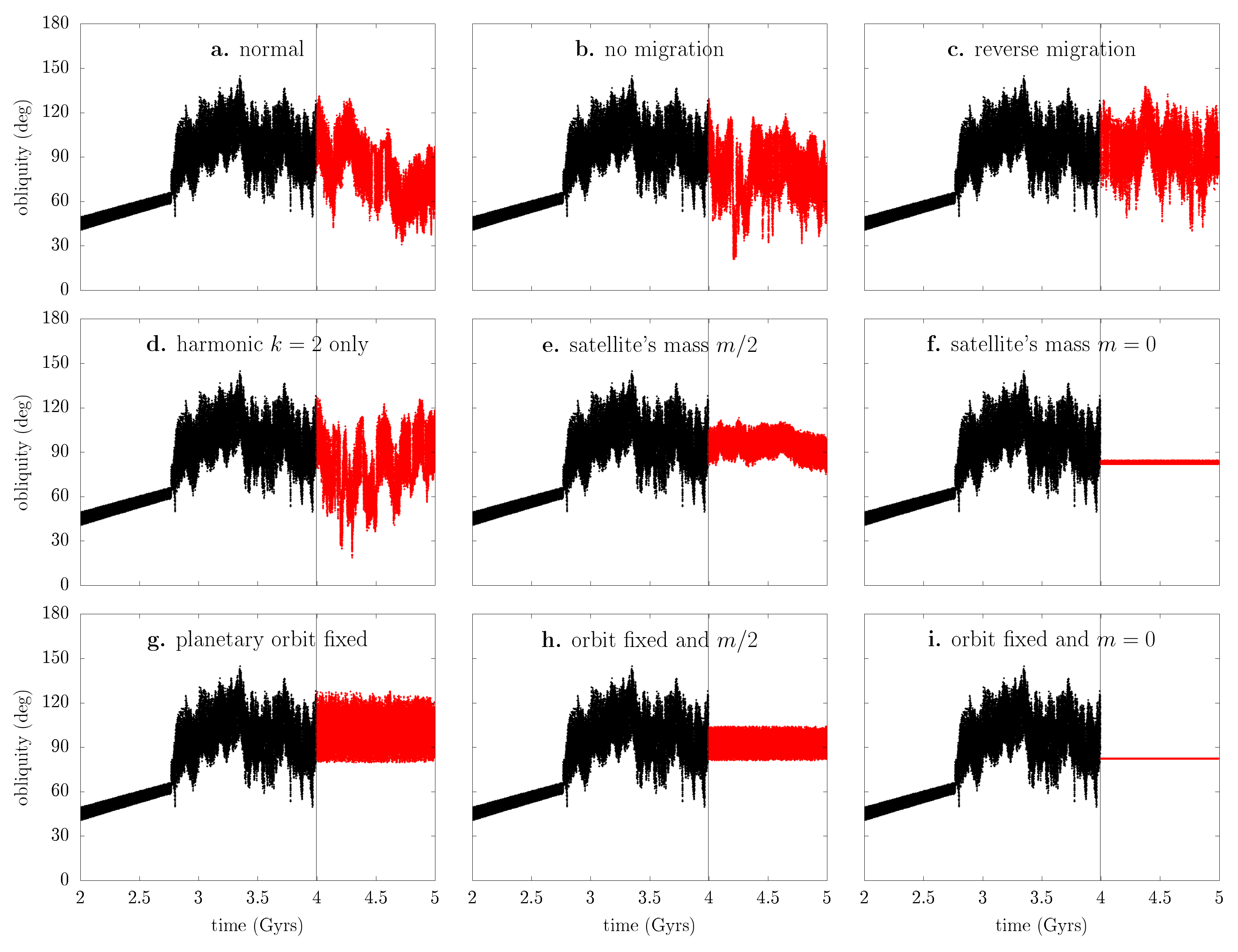}
   \caption{Chaotic diffusion of obliquity for different physical parameters. The black trajectory is the same as in Fig.~\ref{fig:fullycoupled_ex1}. At $t=4$~Gyrs (vertical bar), parameters are instantly changed and the system is integrated for $1$~Gyr more (red part). \emph{Panel a}: no parameter change. \emph{Panel b:} the satellite no longer migrates. \emph{Panel c:} the satellite migrates inwards. \emph{Panel d:} the planet's orbital variations are restricted to the $2$nd harmonic of $\zeta$. \emph{Panel e:} the mass of the satellite is halved. \emph{Panel f:} the mass of the satellite is zero. \emph{Panel g:} the planet's orbit is held fixed. \emph{Panel h:} the planet's orbit is held fixed and the satellite's mass is halved. \emph{Panel i:} the planet's orbit is held fixed and the satellite's mass is zero.}
   \label{fig:chaostest}
\end{figure*}

The cause of the satellite's destabilisation is well understood (see \citealp{Tremaine-etal_2009,Saillenfest-Lari_2021}); the reason for the random walk in obliquity is less obvious. In order to investigate this mechanism further, we continued the numerical integration of Fig.~\ref{fig:fullycoupled_ex1} for one more gigayear with different physical parameters. The results are shown in Fig.~\ref{fig:chaostest}. Because of the strongly chaotic nature of the dynamics, the slightest change in initial condition soon leads to a strong divergence of trajectories. In this regard, the evolutions shown in panels a, b, and c are indistinguishable: from a qualitative point of view, they may just as well be three different realisations of the same chaotic process. Hence, from panels a, b, and c, it seems clear that the chaotic obliquity variations do not depend on the residual migration of the satellite. Indeed, the destabilisation of the system is an irreversible process, and the wild orbital variations of the satellite persist even if its tidal migration stops or is reversed. Changes in the satellite's migration are expected to take place in reality when the satellite becomes unstable; these changes can be due to tidal dissipation within the satellite (not modelled here), to the break of the tidal resonant link (in case of a migration driven by a mechanism similar to that described by \citealp{Fuller-etal_2016}), or to the reversed effect of energy dissipation when the satellite becomes retrograde. The absence of a visible influence of the satellite's migration properties on the chaotic dynamics guarantees that our idealised model still yields a qualitatively relevant description of the dynamics when the system goes deep into the unstable region.

Conversely, panels d and g of Fig.~\ref{fig:chaostest} show that the orbital nodal precession of the planet plays an important role in the chaotic behaviour of the obliquity. If the planet's orbit is held fixed (panel~g), its obliquity still oscillates as a result of the orbital variations of the satellite, but the large-scale obliquity kicks are suppressed. Panel~d shows that one single harmonic of nodal precession is enough to restore the sporadic jumps in obliquity (and term $k=2$ is the largest). Other panels in Fig.~\ref{fig:chaostest} show that the mass of the satellite is directly related to the amplitude of the erratic variations in obliquity: the diffusion is slower for a less massive satellite, and totally suppressed for $m=0$. Panel~i confirms that if the satellite is removed (e.g. because of a collision; see below), the planet's obliquity instantly freezes.

From these comparisons, we deduce that the large-scale chaotic variations in the planet's obliquity are due to a double synergistic destabilisation: \emph{i)} the satellite's orbit is strongly unstable because of the combined attractions of the planet's equatorial bulge and of the star, as previously described by \cite{Tremaine-etal_2009} and \cite{Saillenfest-Lari_2021}; \emph{ii)} by a partial conservation of angular momentum, the wild orbital variations of the satellite produce fast oscillations and some diffusion in the planet's obliquity (direct effect), but also large-amplitude variations in the planet's spin-axis precession frequency. Because of these frequency variations, the planet goes in and out of the nearby secular spin-orbit resonances, resulting in strong additional obliquity kicks. These obliquity kicks reinforce the satellite's instability and make it explore an even wider range of orbital elements. Since all secular spin-orbit resonances converge to the singular point $\mathrm{S}_1$ (see e.g. Fig.~\ref{fig:reswidths}), all available resonances are actually `nearby' and contribute to the obliquity kicks. The largest of them is the $\nu_2$ resonance, that is, a resonance with the nodal orbital precession mode of Uranus itself. Fig.~\ref{fig:chaostest}d shows that it contributes most to the obliquity kicks.

\subsection{Exploration of the parameter space}\label{sec:explor}

The previous example shows that the double destabilisation mechanism, while putting an end to the stable adiabatic drift inside the resonance, is able to pump the obliquity high enough to reproduce the current state of Uranus. We must now investigate the conditions under which the double destabilisation is triggered and estimate its efficiency. To this end, we turn to Monte Carlo experiments.

In our first experiment, the migration timescale of the satellite is set to $\tau=7.5$~Gyrs and the initial obliquity of the planet is $\varepsilon_0=3^\circ$ for all numerical integrations. This puts the system in the parameter region producing a $100\%$-sure capture and tilting in most resonances. As shown in Fig.~\ref{fig:resstats}, the choice of $\varepsilon_0$ is not critical, as the same efficiency can be obtained in a large range of initial obliquities. The initial semi-major axis of the satellite is $a_0=42$~$R_\mathrm{eq}$ (that is, about $0.8\,r_\mathrm{M}$). The satellite is started close to the circular Laplace equilibrium, with an eccentricity $e=10^{-4}$ and an inclination $I_\mathrm{LP}=10^{-4}$~rad with respect to the local Laplace plane. For a given value of the satellite's mass $m$, we draw a random sample of initial conditions for the planet's spin-axis precession angle $\psi\in[0,2\pi)$ and for the satellite's argument of pericentre $\omega_\mathrm{LP}\in[0,2\pi)$ and longitude of ascending node $\Omega_\mathrm{LP}\in[0,2\pi)$ measured with respect to the local Laplace plane. All initial conditions are propagated numerically for $4$~Gyrs using the fully coupled secular model, and the maximum obliquity value reached over the full numerical integration is recorded.

\begin{figure*}
   \includegraphics[width=\textwidth]{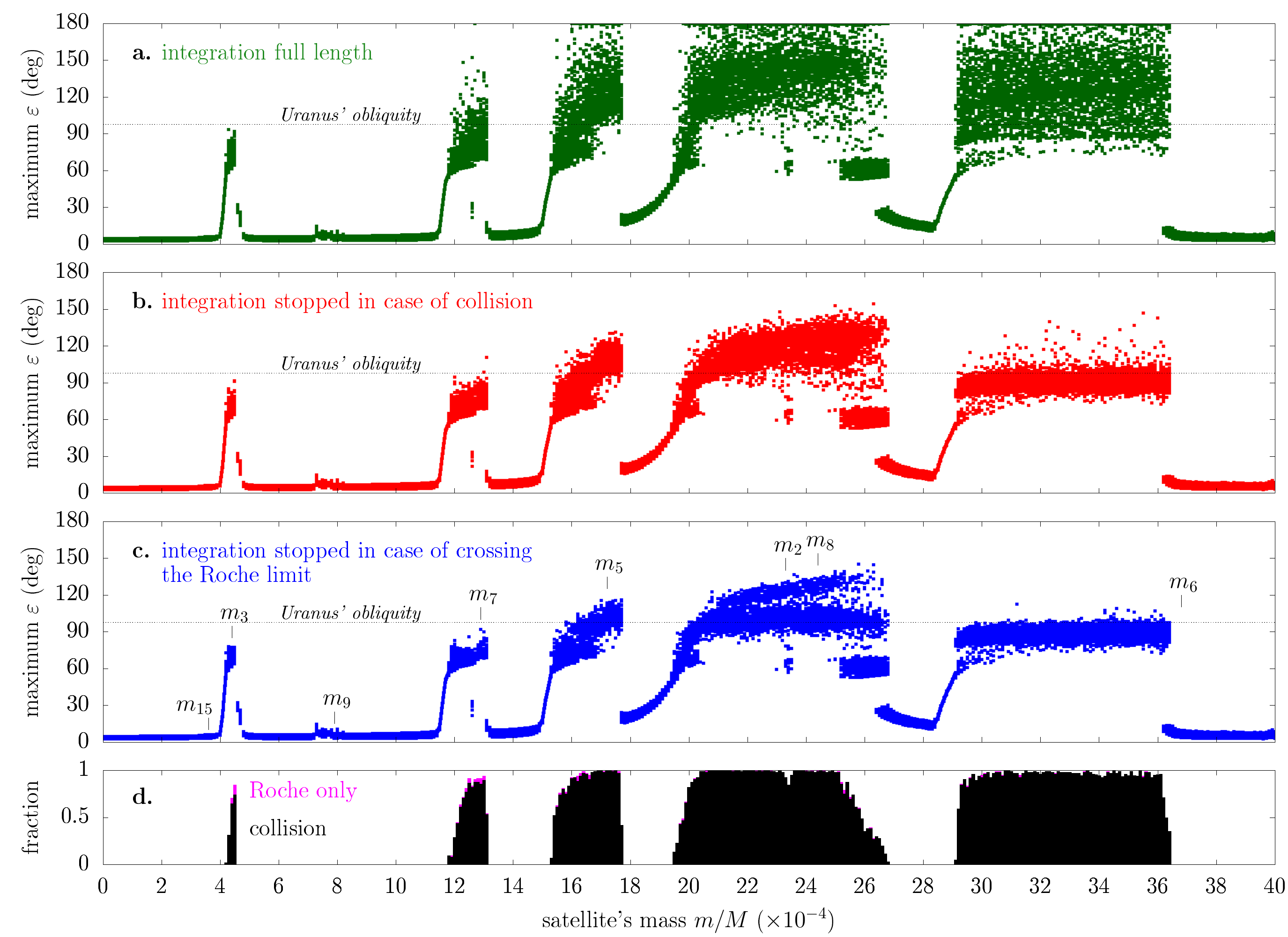}
   \caption{Maximum obliquity reached in the course of numerical integrations using a fully coupled secular model. For each value of the satellite's mass (bottom axis), $96$ numerical integrations are performed with random initial conditions as described in the text. The initial semi-major axis of the satellite is $a_0=42$~$R_\mathrm{eq}$ and it is made migrating according to Eq.~\eqref{eq:asat} with timescale $\tau=7.5$~Gyrs. \emph{Panel~a}: the maximum obliquity reached over a full duration of $4$~Gyrs is shown by a small dot. The horizontal dotted line shows Uranus' current obliquity. \emph{Panel~b}: same as panel~a, except that integrations are stopped if the satellite collides into the planet. If a collision occurs, the dot on the graph shows the maximum obliquity reached before the collision; the time of reaching this maximum may be different from the collision time. \emph{Panel~c}: same as panel~b, except that integrations are stopped if the satellite goes below the Roche limit (taken equal to $2$~$R_\mathrm{eq}$). The mass estimates from Table~\ref{tab:res} are labelled (small vertical lines). \emph{Panel~d:} histogram showing the fraction of trajectories featuring a collision (black). The magenta bars on top show the small fraction of trajectories for which the satellite goes below the Roche limit but does not collide into the planet within the integration timespan.}
   \label{fig:obmaxvsmass}
\end{figure*}

The result is shown in the panel~a of Fig.~\ref{fig:obmaxvsmass}. We can identify the effects of each individual resonance which progressively appears as we take a larger mass for the satellite, and whose bottom extremity sweeps over the satellite's initial location at $a_0\approx 0.8\,r_\mathrm{M}$ (see Fig.~\ref{fig:reswidths} for the geometry of the resonances). Beyond a given mass value, the resonance still exists, but its bottom extremity is closer to the planet than $0.8\,r_\mathrm{M}$, so the system does not encounter it during the outward migration of the satellite. The minimum mass values estimated previously (see Table~\ref{tab:res}) correspond quite accurately to the top of the obliquity peaks; their locations are marked by a small line in panel~c. As expected, resonances $\nu_{15}$ and $\nu_9$ only produce tiny obliquity bumps because they are too thin to allow for an adiabatic capture in less than the age of the Solar System (see Sect.~\ref{sec:res}). We also checked that for a very fast satellite migration ($\tau\lesssim 3$~Gyrs), the peak corresponding to resonance $\nu_7$ is greatly reduced and barely reaches $\varepsilon=30^\circ$ because the probability of adiabatic capture falls to zero (see Fig.~\ref{fig:resstats}). The minimum mass estimate for resonance $\nu_8$ appears quite far from the corresponding obliquity peak in Fig.~\ref{fig:obmaxvsmass}, because resonance $\nu_8$ is very distorted by the large neighbour resonance $\nu_2$. In fact, resonance $\nu_2$ is so large that it almost completely overlaps resonance $\nu_8$; it is also surrounded by several high-order resonances that have a non-negligible width and contribute to the production of chaos (see \citealp{Saillenfest-etal_2019}). By carefully analysing the trajectories in Fig.~\ref{fig:obmaxvsmass}, we found that some of them feature captures in second-order resonances. An example of capture in a resonance with critical angle $\sigma = 2\psi + \phi_2 + \phi_5$ is given in Appendix~\ref{asec:rese2}; this example confirms that the dynamics in the neighbourhood of resonance $\nu_2$ are very rich and not limited to interactions with harmonic $\phi_2$ taken in isolation.

Panel~a of Fig.~\ref{fig:obmaxvsmass} reveals that the double destabilisation mechanism described above can routinely reproduce Uranus' obliquity through a capture into resonances $\nu_7$, $\nu_5$, $\nu_2$, $\nu_8$, or $\nu_6$. The minimum mass of the satellite for which successful trajectories are obtained is $m/M\approx 1.2\times 10^{-3}$ (i.e. $m\approx m_7$). For smaller masses, no trajectory in our sample goes beyond the current obliquity of Uranus. As expected from previous sections, captures into resonance $\nu_3$ for $m/M\approx 4.4\times 10^{-4}$ are perfectly stable and adiabatic, and trajectories do reach the unstable region, but then they barely manage to go over $\varepsilon=90^\circ$. This is because the satellite is too small to produce a strong back reaction on the spin axis when it becomes unstable, and the $\nu_3$ resonance is quite isolated from other strong resonances. Both these aspects contribute to slow down the chaotic diffusion in obliquity, and one would need to greatly expand the integration timespan in order to give enough time for trajectories to diffuse high enough (see Sect.~\ref{sec:doubledestab}).

Moreover, the satellite can actually not go on kicking and jumping all over the place for a long time, because sooner or later it reaches an orbit that makes it collide into the planet. Indeed, our simulations show that the secular destabilisation of the satellite is strong enough to destroy it, similarly to asteroid disruption around white dwarfs \citep{OConnor-etal_2022}. In panel~a of Fig.~\ref{fig:obmaxvsmass}, all integrations are continued for $4$~Gyrs even if the pericentre distance $q$ of the satellite becomes smaller than $1$~$R_\mathrm{eq}$ (this situation is unphysical). For comparison, panel~b shows the same Monte Carlo experiment, but for which integrations are stopped in case of satellite-planet collision (i.e. if ever $q\leqslant 1$~$R_\mathrm{eq}$). The fraction of collisional trajectories is shown in panel~d (black histogram): we see that the transition from $0\%$ to nearly $100\%$ collisions is very steep for all resonances. This means that if the system goes deep in the unstable zone, collision becomes inescapable. Because of the existence of collisional trajectories, the destabilisation of the satellite can be thought of as the start of a countdown that will eventually put an end to the evolution of the system. In order to reproduce the current state of Uranus, we need that the erratic motion of the satellite brings the planet's obliquity high enough before the collision. As shown in panel~b, the maximum obliquity values reached before collision are substantially smaller than what we observed in panel~a. A satellite with mass $m/M\approx 1.2\times 10^{-3}$ now only marginally allows for reaching Uranus' obliquity.

The effect of the satellite's impact on the spin rate and obliquity of the planet can be estimated from the conservation of angular momentum (see Appendix~\ref{asec:impact} and the related discussion in Sect.~\ref{sec:smallestsat} below). For the range of masses considered in Fig.~\ref{fig:obmaxvsmass}, the relative change in spin rate due to the impact is a few percent at most, and the change in spin direction is no larger than a few degrees.

Depending on the cohesive strength of the satellite and the rate of its eccentricity increase, the satellite may not even reach collision but be torn apart by tidal forces farther away from Uranus. Like the amount of tidal dissipation within the satellite (that we chose to ignore for now), the exact Roche limit for the satellite depends on its internal structure and composition and can only be speculative at this stage. For definitiveness, we adopt a limit of $2$~$R_\mathrm{eq}$, which is about halfway between the fluid and rigid Roche limits (assuming that the satellite is homogeneous and has a density similar to Uranus' current regular satellites; see e.g. \citealp{Canup-Esposito_1995,Crida-Charnoz_2012,Hesselbrock-Minton_2019}). In fact, the satellite is not expected to be completely torn apart in only one passage, but rather be progressively eroded and form a torus of material. If the satellite is differentiated, its denser core may survive much longer than its outer layers, possibly down to collision \citep{Canup_2010}. Some fraction of material may also be lost in space (see e.g. \citealp{Malamud-Perets_2020,Brouwers-etal_2022}). In any case, the remaining debris would reorganise into a thin equatorial disc confined below the fluid Roche limit (see e.g. \citealp{Hyodo-etal_2017} and discussions in Sect.~\ref{sec:currsat}). Hence, if the satellite is tidally disrupted, there should actually be a continuous (yet fast) transition regime from the solid satellite, whose influence on the planet's spin axis is strong, to the confined debris disc, whose effect on the planet's spin axis is negligible. Here, we do not model these phenomena, and consider that the satellite instantly disappears when crossing $2$~$R_\mathrm{eq}$. This approximation may seem quite crude but the results obtained are still informative, as they can be compared to the pure collisional case described above: a more realistic behaviour of the system probably lies somewhere between these two extremes. The maximum obliquity reached by the planet before the satellite crosses the Roche limit is shown in panel~c of Fig.~\ref{fig:obmaxvsmass}. Obliquity values are globally lower than in panel~b, but the picture is not very different in a qualitative point of view. Indeed, almost all trajectories that cross the Roche limit reach collision shortly afterwards (see panel~d).

\subsection{Probability of reproducing the current state of Uranus}\label{sec:coupledproba}

\begin{figure*}
   \centering
   \includegraphics[width=\textwidth]{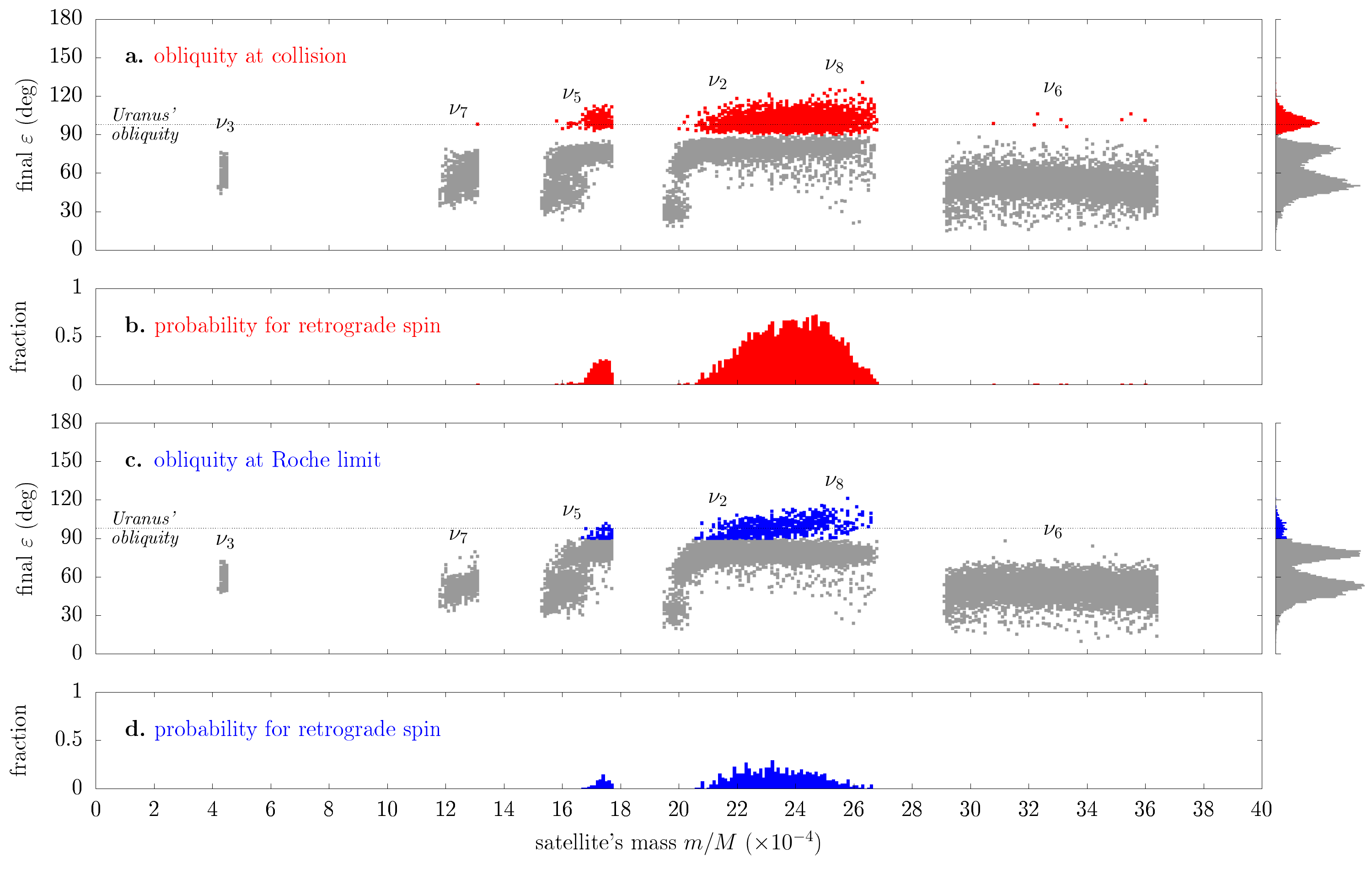}
   \caption{Final value of the planet's obliquity in the simulations of Fig.~\ref{fig:obmaxvsmass}. \emph{Panels a and b:} the satellite is removed when it collides into the planet. Trajectories in which the satellite does not collide within the integration timespan are not shown. The main resonances are labelled. The obliquity histogram is given on the right. Obliquity values larger than $90^\circ$ are highlighted in red; their ratio over all trajectories (including those without collision) is shown in panel~b as a function of mass. \emph{Panels c and d}: the satellite is removed when it goes below the Roche limit. Trajectories in which it never does are not shown. Obliquity values larger than $90^\circ$ are highlighted in blue.}
   \label{fig:obcolvsmass}
\end{figure*}

It is now clear that starting from a small axis tilt, the obliquity of Uranus can be reached through the tidal migration of an ancient satellite, by a capture in secular spin-orbit resonance followed by a global instability. For reproducing the current state of Uranus, however, the planet must not only reach Uranus' obliquity, but stop at its value. In order to estimate the probability of successful trajectories, we now examine the configuration of the system at collision, that is, at the time the satellite is expected to be destructed and leave the planet's obliquity in a fossilised state. Panel~a of Fig.~\ref{fig:obcolvsmass} presents the same Monte Carlo simulations as in Fig.~\ref{fig:obmaxvsmass}, but showing the obliquity of the planet at the time of collision (instead of the maximum obliquity reached). The intervals of mass devoid of data points are regions in which no trajectory in our sample features a collision. Obliquity values are sensibly lower than in Fig.~\ref{fig:obmaxvsmass}b, which means that the satellite generally does not collide when the planet's obliquity is at its maximum, but somewhat later on. We see that collisions occur when the spin axis is either plainly prograde or retrograde, but almost no collision occurs at all while the obliquity is $\varepsilon\approx90^\circ$. The explanation for this bizarre phenomenon is given in Sect.~\ref{sec:keyholes}. The colour code in Fig.~\ref{fig:obcolvsmass} distinguishes trajectories ending with $\varepsilon<90^\circ$ (prograde family) from those ending with $\varepsilon>90^\circ$ (retrograde family). Interestingly, the bulk of the retrograde family is located right at the obliquity value of Uranus (see histogram on the right). For this reason, we define a run as `successful' to reproduce the current state of Uranus when it ends up in the retrograde family.

Panel~b of Fig.~\ref{fig:obcolvsmass} shows the probability of success as a function of mass (i.e. it shows the fraction of trajectories in our sample that feature a collision and for which the planet's obliquity at collision is larger than $90^\circ$). In the most favourable region, probabilities reach $60$ to $70\%$ (the highest peak is $73\%$). In other words, reproducing the current state of Uranus is the most likely outcome in this range of parameters. The resonance responsible for the obliquity increase in this region is $\nu_2$ (partially overlapping with resonance $\nu_8$). This is not a coincidence: resonance $\nu_2$ is the largest secular spin-orbit resonance that appears in the orbital series of Uranus.

Panels~c and d of Fig.~\ref{fig:obcolvsmass} present the same Monte Carlo experiment, but where the satellite is supposed to be instantly destructed when it goes below the Roche limit. The same kind of structure as in panel~a is observed, but the division between the prograde and retrograde families is less marked. The probability to end up retrograde is quite smaller: it reaches $20\%$ to $30\%$ in the most favourable region (the highest peak is~$29\%$). Again, this can be explained by the coupled destabilisation mechanism: as the satellite is removed earlier than in panel~a, the system is offered less time to substantially excite the obliquity.

\begin{figure*}
   \centering
   \includegraphics[width=0.95\textwidth]{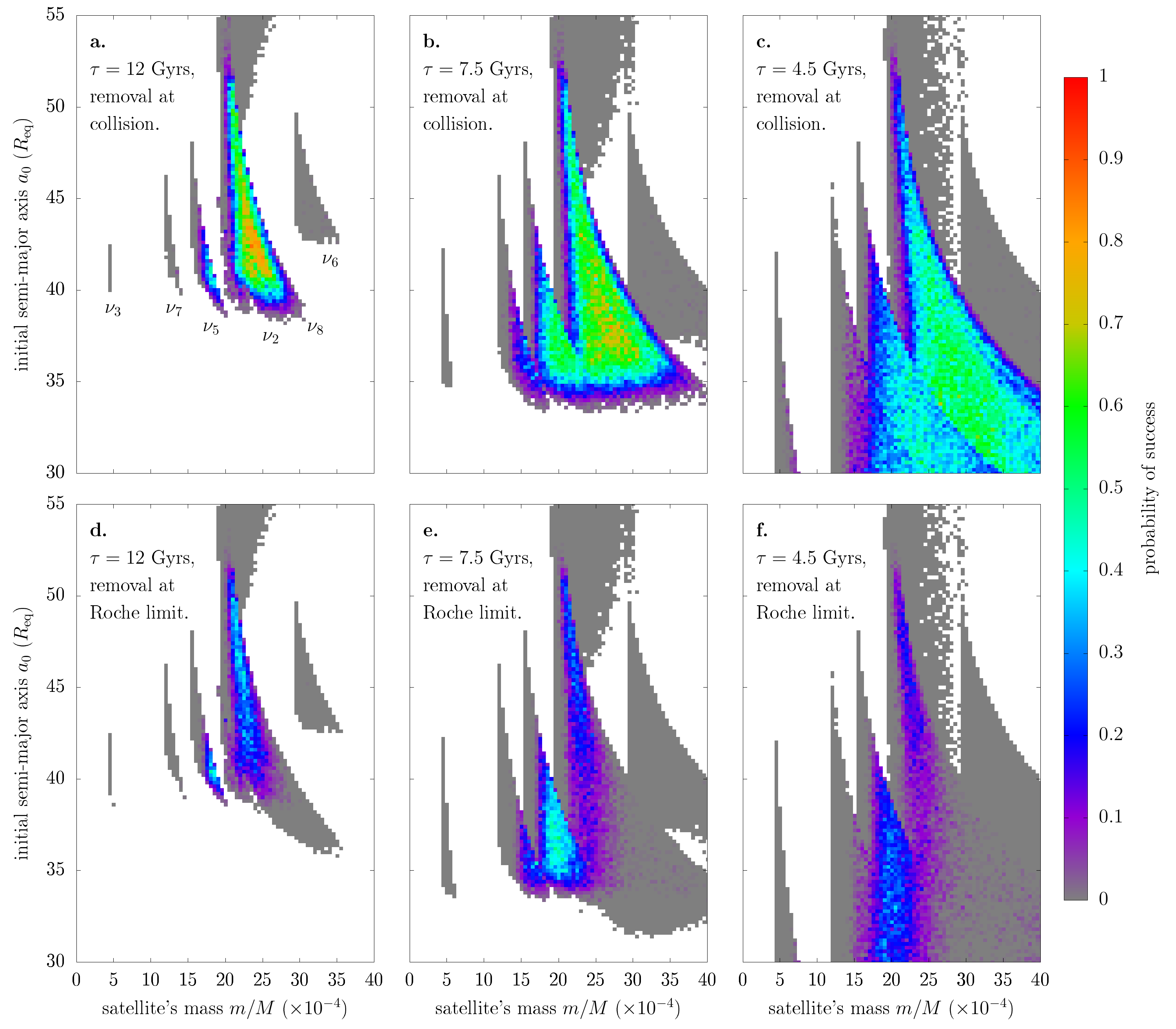}
   \caption{Probability of destructing the satellite and producing a retrograde obliquity as a function of the mass of the satellite (bottom axis), its initial semi-major axis (vertical axis), and its migration timescale (labels). Each pixel features $96$ numerical integrations over $4$~Gyrs with random initial angles (same setting as in Fig.~\ref{fig:obmaxvsmass}). In the upper panels, the satellite is removed when it collides into the planet $(q<R_\mathrm{eq}$). In the lower panels, the satellite is removed when it goes below the Roche limit ($q<2\,R_\mathrm{eq}$). White pixels mean that the satellite survived in all $96$ simulations. Coloured pixels show the fraction of simulations in which the satellite is destructed and the planet's spin axis ends in the retrograde family. Resonances are labelled in panel~a.}
   \label{fig:massmap}
\end{figure*}

Because of the geometry of secular spin-orbit resonances, the intervals of mass showing the maximum success probability in Fig.~\ref{fig:obcolvsmass} are specific to the choice of initial semi-major axis $a_0$ of the satellite. The choice $a_0\approx 0.8\,r_\mathrm{M}$ has been made to roughly minimise the distance over which the satellite must migrate in order to reach the unstable region (see Sect.~\ref{sec:res}). However, similar success probabilities can be obtained if the satellite starts closer to the planet and migrates further out. In this case, the same resonances are reachable up to larger satellite masses. In order to clarify this point, we repeated our Monte Carlo experiment for different initial locations of the satellite. Figure~\ref{fig:massmap} shows the probability of ending in the retrograde family as a function of both the satellite's mass $m$ and its initial distance $a_0$. The effect of each individual resonance is clearly visible. As expected, resonance $\nu_2$ for $m/M\gtrsim 2\times 10^{-3}$ produces the largest fraction of successful trajectories, but resonances $\nu_5$ and $\nu_7$ also generate substantial probabilities of success in particular regions of the parameter space. The range of mass having the highest success ratio increases when we decrease the initial distance of the satellite; this is a direct consequence of the geometry of secular spin-orbit resonances. Below a given initial distance, however, the satellite has not enough time in $4$~Gyrs to reach the unstable region, so the success probability is zero. Larger migration velocities are needed to counteract this problem; they allow one to initialise the satellite closer to the planet (compare panels a, b, and c).

If the satellite is initialised closer to the planet, the system can catch a resonance that brings it to higher obliquities before triggering the instability (see e.g. resonance $\nu_3$ in Fig.~\ref{fig:reswidths}b, to be compared with the same resonance in Fig.~\ref{fig:reswidths}a). This can be a way to increase the success ratio for small satellites. And indeed, because of this property, panel~c in Fig.~\ref{fig:massmap} shows that resonance $\nu_3$ is able to generate a small fraction of successful trajectories if the satellite is started at $a_0\lesssim 35\,R_\mathrm{eq}$ and migrates fast (see the violet shade for $m/M\approx 7\times 10^{-4}$). The success ratios, however, remain very small in resonance $\nu_3$.

\begin{figure*}
   \centering
   \includegraphics[width=0.95\textwidth]{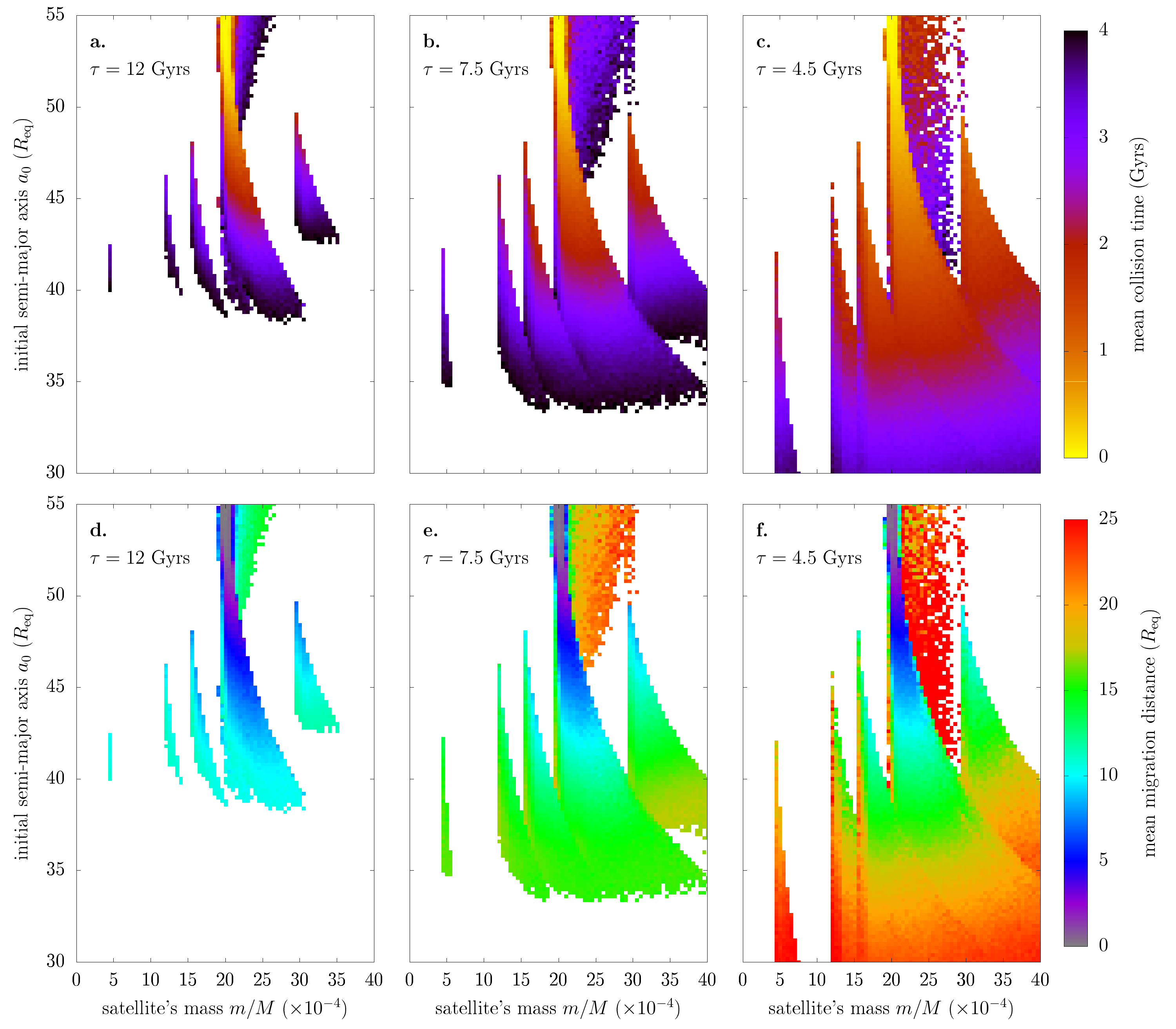}
   \caption{Collision time and migration distance covered by the satellite in the simulations of Fig.~\ref{fig:massmap}. \emph{Top panels:} integration time at collision. \emph{Bottom panels:} distance covered by the satellite before collision. Similar graphs are obtained when considering the time of crossing the Roche limit (with only slightly smaller integration times and migration distances).}
   \label{fig:massmap_time}
\end{figure*}

In summary, Fig.~\ref{fig:massmap} reveals that the regions of the parameter space allowing for reproducing the current state of Uranus are quite vast. Through a capture in resonance $\nu_2$, the probability of obtaining a final retrograde spin axis (with an obliquity similar to Uranus') can be larger than $80\%$. Other promising resonances are $\nu_5$ and $\nu_7$, with success ratios up to $50\%$. Thanks to the large width of resonance $\nu_2$, the tilting mechanism can also operate down to very small migration ranges: this is the case in the uppermost portion of the graphs, for initial semi-major axes $a_0\gtrsim 50~R_\mathrm{eq}$. The time elapsed before collision and the migration range covered by the satellite in our Monte Carlo experiments are shown in Fig.~\ref{fig:massmap_time}. In some narrow range of parameters, we see that the current state of Uranus can be reproduced even if the satellite migrates over distances as small as $2\,R_\mathrm{eq}$. This reduced migration range increases the effective quality factor $Q$ that would be required for Uranus (see Fig.~\ref{fig:k2sQ}).

\subsection{Collision keyholes}\label{sec:keyholes}

\begin{figure}
   \centering
   \includegraphics[width=\columnwidth]{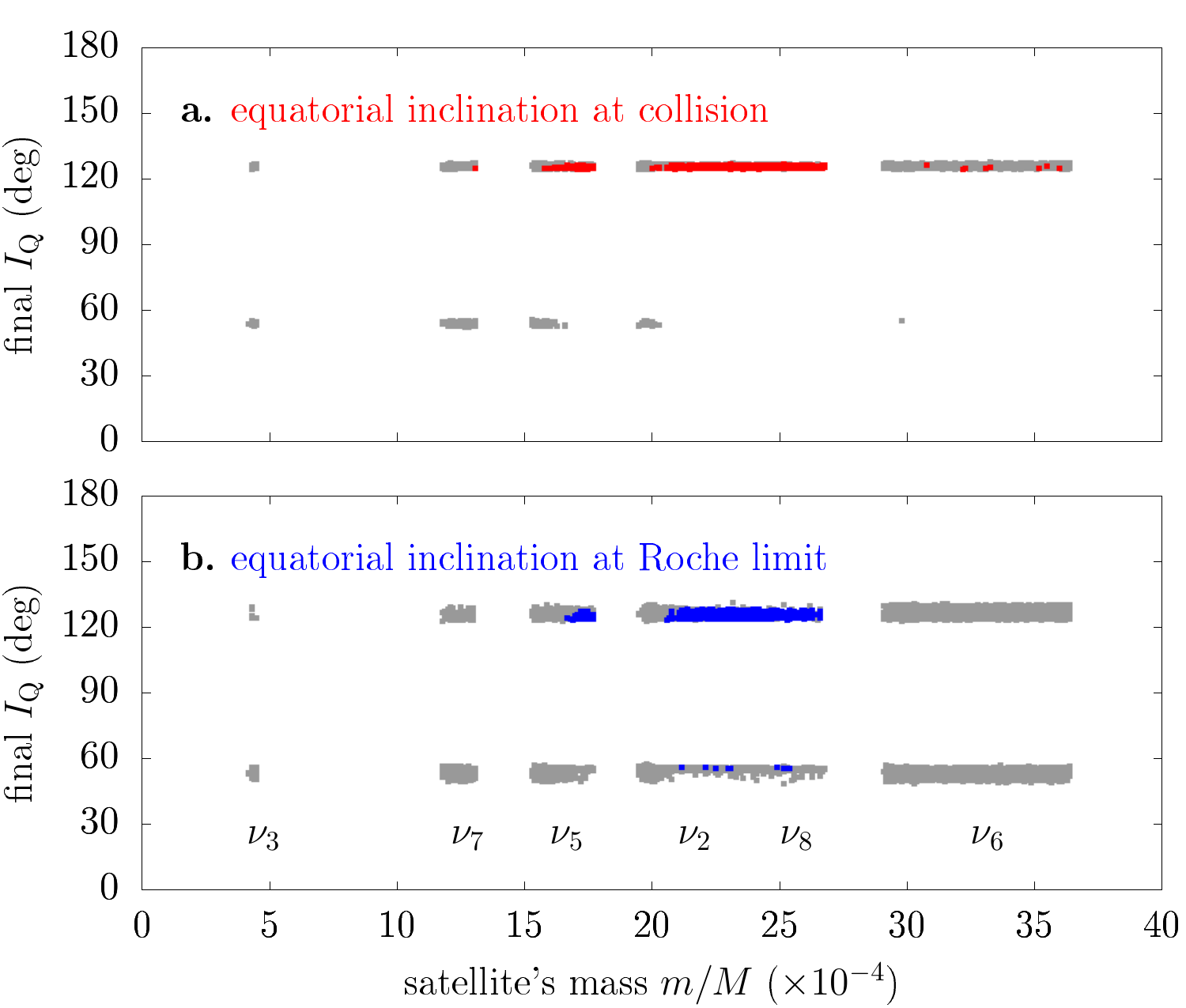}
   \caption{Same as Fig.~\ref{fig:obcolvsmass}, but showing the orbital inclination of the satellite at the time of its destruction. Coloured dots show obliquity values larger than $90^\circ$; grey dots show obliquity values smaller than $90^\circ$ (same as Fig.~\ref{fig:obcolvsmass}).}
   \label{fig:IQcolvsmass}
\end{figure}

The peculiar distribution of obliquity in Fig.~\ref{fig:obcolvsmass}, showing accumulations in a prograde and a retrograde spin `families' and much depletion around $\varepsilon\approx 90^\circ$, deserves a more in-depth analysis. We write $I_\mathrm{Q}$ the orbital inclination of the satellite measured with respect to the planet's equator and $I_\mathrm{C}$ its inclination measured with respect to the planet's orbital plane ($\mathrm{Q}$ and $\mathrm{C}$ stand for `equator' and `ecliptic'). When examining the orbital elements of the satellite at the time of its collision, we find that the values of $I_\mathrm{Q}$ are localised in two very narrow bands that can be visualised in Fig.~\ref{fig:IQcolvsmass}. These two bands are localised at $I_\mathrm{Q}\approx 55^\circ$ and $125^\circ$; we see that they are less narrow when the satellite crosses the Roche limit than when the satellite impacts the planet. In contrast, we find no preferential orientation of the satellite with respect to the planet's orbital plane, with inclinations $I_\mathrm{C}$ covering almost the entire range from $0^\circ$ to $180^\circ$. This seems to indicate that the pathways to collision (`collision keyholes') involve predominantly planet-satellite interactions.

\begin{figure}
   \centering
   \includegraphics[width=\columnwidth]{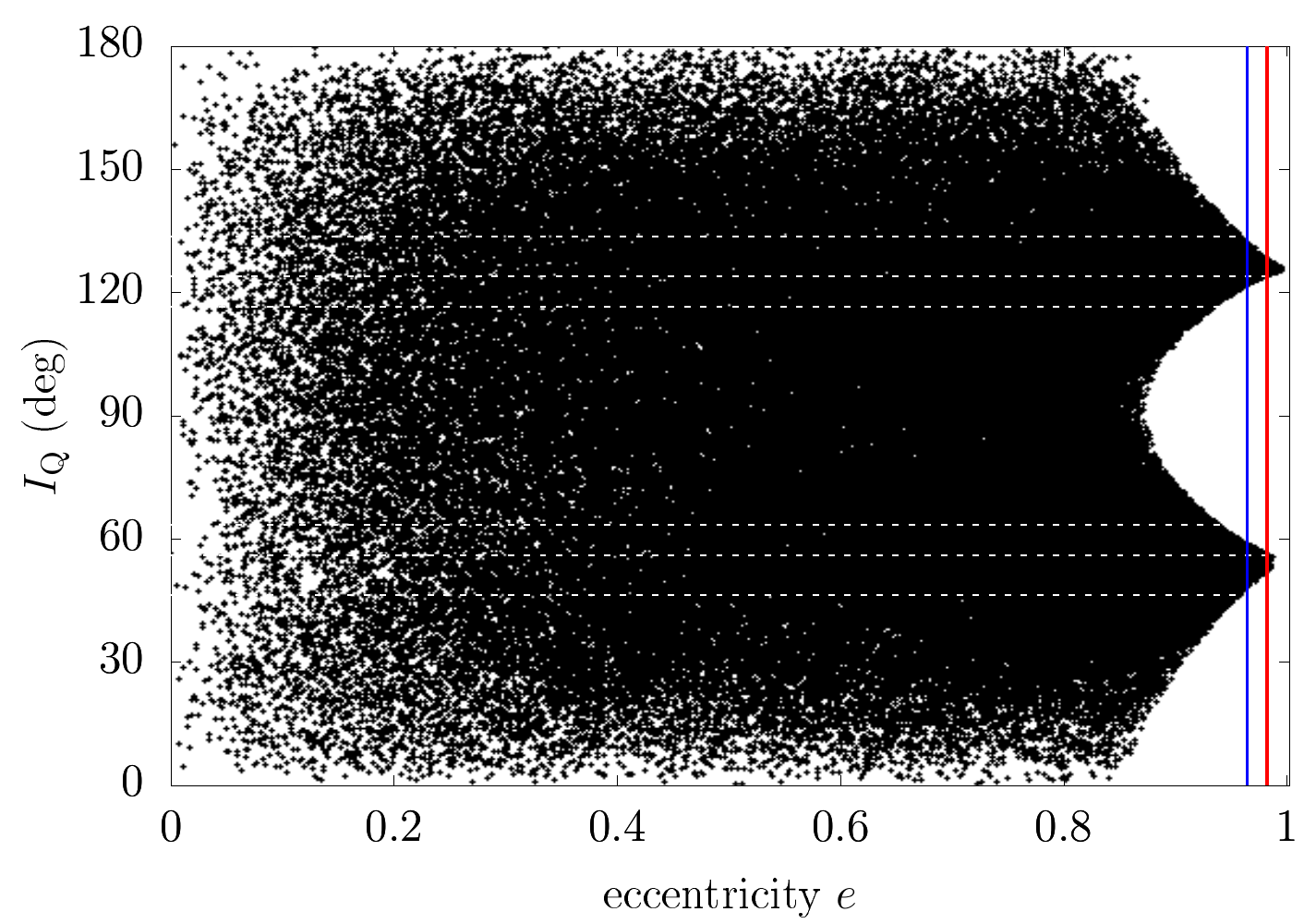}
   \caption{Chaotic region explored by the satellite. The black dots show the track left during $4$~Gyrs by a single numerical integration. The physical parameters are the same as in Fig.~\ref{fig:fullycoupled_ex1}. The blue and red vertical lines show the Roche and collision limits, respectively. The white horizontal dashed lines show the nominal locations of the secular resonances that have the largest width in eccentricity (see text).}
   \label{fig:EcIQ_full}
\end{figure}

Figure~\ref{fig:EcIQ_full} shows an example of the track left by the satellite in the plane $(e,I_\mathrm{Q})$ as the system wanders in the chaotic zone. Even though the obliquity of the planet covers a large range of values during the evolution (see Fig.~\ref{fig:fullycoupled_ex1}), the two collision keyholes are extremely localised in inclination $I_\mathrm{Q}$: we clearly see that no collision can possibly occur outside of $I_\mathrm{Q}\approx 55^\circ$ and $125^\circ$, which explains the existence of the narrow bands in Fig.~\ref{fig:IQcolvsmass}. We performed various numerical tests, and found that the same chaotic structure as in Fig.~\ref{fig:EcIQ_full} is produced even in the most simple case of a massless satellite around a planet with fixed orbit and fixed spin axis. This may not be surprising, because collisions occur on a very short timescale compared to the orbital and spin-axis motions of the planet, and we only consider relatively small satellites here. The nature of the collision keyholes should therefore have a simple dynamical origin, that we investigate here.

The Hamiltonian function describing the secular motion of a massless satellite around a planet with fixed orbit and spin axis can be written
\begin{equation}\label{eq:Hsecsimp}
   \mathcal{H} = k_\mathrm{P}\mathcal{H}_\mathrm{P} + k_\odot\mathcal{H}_\odot\,,
\end{equation}
where the first term comes from the interaction of the satellite with the equatorial bulge of the planet, and the second term contains the orbital perturbation produced by the star. The explicit expression of the constant factors $\{k_\mathrm{P},k_\odot\}$ and of the Hamiltonian functions $\{\mathcal{H}_\mathrm{P},\mathcal{H}_\odot\}$ expanded at quadrupolar order are given in Appendix~\ref{asec:secres}. Because $\mathcal{H}_\mathrm{P}$ contains a factor $(1-e^2)^{-3/2}$, the first term in Eq.~\eqref{eq:Hsecsimp} becomes strongly dominant whenever the eccentricity grows large. Indeed, the contribution of $k_\mathrm{P}\mathcal{H}_\mathrm{P}$ produces a fast precession of the satellite's argument of pericentre $\omega_\mathrm{Q}$ and longitude of ascending node $\Omega_\mathrm{Q}$, at rates
\begin{equation}\label{eq:J2prec}
   \dot{\omega}_\mathrm{Q} = k_\mathrm{P}\frac{5\cos^2I_\mathrm{Q}-1}{\sqrt{\mathcal{G}Ma}(1-e^2)^2}
   \quad;\quad
   \dot{\Omega}_\mathrm{Q} = k_\mathrm{P}\frac{-2\cos I_\mathrm{Q}}{\sqrt{\mathcal{G}Ma}(1-e^2)^2} \,,
\end{equation}
with the result of averaging out the angular dependency in $k_\odot\mathcal{H}_\odot$. As a consequence, the eccentricity and equatorial inclination of the satellite become constants of motion when the eccentricity grows large. Because of this property, the eccentricity of the satellite can never grow arbitrary close to $1$: the contribution from $k_\mathrm{P}\mathcal{H}_\mathrm{P}$ makes it freeze at some point and leaves it with no alternative but decreasing again.

In order to determine the orbital configurations in which the eccentricity can grow the largest, we must focus on the intermediate regime in which $k_\mathrm{P}\mathcal{H}_\mathrm{P}$ is dominant but the contribution from $k_\odot\mathcal{H}_\odot$ is not yet negligible. If we adopt a perturbative approach and consider that the two degrees of freedom are weakly coupled, any combination $\xi = i\omega_\mathrm{Q}+j\Omega_\mathrm{Q}$ can become a resonant angle, where $(i,j)\in\mathbb{Z}^2\setminus\{(0,0)\}$. The possible resonances at first order in the perturbation are given in the left column of Table~\ref{tab:resconst}. Their nominal location is obtained by imposing $\dot{\xi}=0$ from Eq.~\eqref{eq:J2prec}, which results in one specific value for the inclination $I_\mathrm{Q}$, that we note $I_0$.

\begin{table}
   \caption{Properties of the secular resonances appearing in the orbital dynamics of a massless satellite at first order in the solar perturbation.}
   \label{tab:resconst}
   \vspace{-0.7cm}
   \begin{equation*}
      \begin{array}{rrr}
      \hline
      \text{resonance angle} & \text{nominal location} & \text{constant quantity} \\
      \xi & \cos I_0 & K \\
      \hline
      \hline
      \omega_\mathrm{Q}+\Omega_\mathrm{Q} & (\sqrt{6}+1)/5   & \sqrt{1-e^2}(\cos I_\mathrm{Q}-1)   \\
      2\omega_\mathrm{Q}+\Omega_\mathrm{Q}   & (\sqrt{21}+1)/10 & \sqrt{1-e^2}(2\cos I_\mathrm{Q}-1)  \\
      \omega_\mathrm{Q}                     & \sqrt{5}/5       & \sqrt{1-e^2}\cos I_\mathrm{Q}       \\
      2\omega_\mathrm{Q}-\Omega_\mathrm{Q}   & (\sqrt{21}-1)/10 & \sqrt{1-e^2}(2\cos I_\mathrm{Q}+1)  \\
      \omega_\mathrm{Q}-\Omega_\mathrm{Q} & (\sqrt{6}-1)/5   & \sqrt{1-e^2}(\cos I_\mathrm{Q}+1)   \\
      \Omega_\mathrm{Q}                     & 0                & e                                   \\
      \hline
      \end{array}
   \end{equation*}
   \vspace{-0.3cm}
   \tablefoot{The retrograde resonances are obtained by inverting the sign of $\cos I_0$ and of $\Omega_\mathrm{Q}$.}
\end{table}

\begin{figure*}
   \centering
   \includegraphics[width=\textwidth]{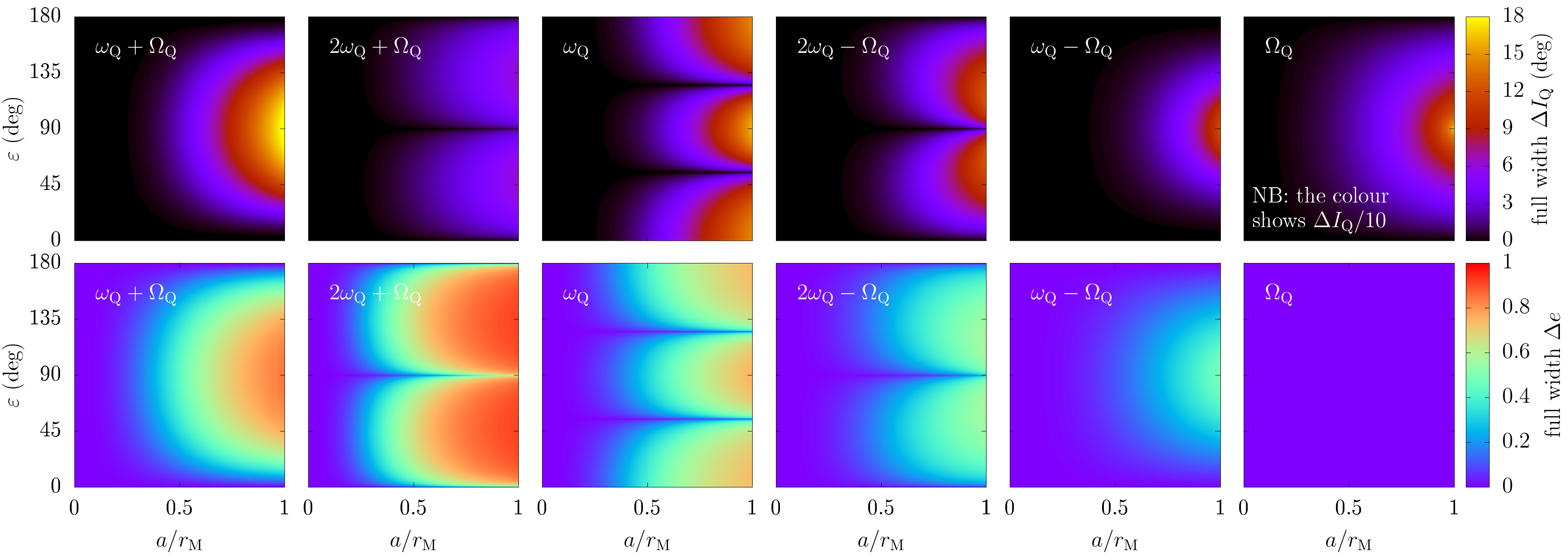}
   \caption{Widths of first-order secular resonances as a function of the parameters. Only prograde resonances are shown. The widths of retrograde resonances are obtained by inverting the sign of $\Omega_\mathrm{Q}$ in the titles. For each resonance, the inclination and eccentricity of the satellite are linked through a constant quantity $K$ given in Table~\ref{tab:resconst}. The fixed value of $K$ used to plot this figure is given in Appendix~\ref{asec:secres}. The colour gradient shows the width of the resonance as measured in inclination (top row) and in eccentricity (bottom row). Because of its overly large $\Delta I_\mathrm{Q}$, the width displayed in the top right panel is shown divided by $10$ (as illustrated in Fig.~\ref{fig:resoverlap}, it reaches $180^\circ$ at $a=r_\mathrm{M}$ and $\varepsilon=90^\circ$).}
   \label{fig:reswidthall}
\end{figure*}

The properties of these resonances can be studied with the method described in Appendix~\ref{asec:secres}. In the vicinity of each resonance, the averaged system admits a constant quantity $K$ that links the eccentricity and the inclination of the satellite, in the same way as the Kozai constant (see e.g. \citealp{Lidov_1962,Kozai_1962,Saillenfest-etal_2016}). The expression of $K$ associated with each resonance is given in the right column of Table~\ref{tab:resconst}. Because of this constant quantity, the resonance width $\Delta I_\mathrm{Q}$ in inclination has an equivalent width $\Delta e$ in eccentricity (except for the `resonance' $\xi=\Omega_\mathrm{Q}$ for which the eccentricity itself is constant; see Table~\ref{tab:resconst}). The widths of each individual resonance is shown in Fig.~\ref{fig:reswidthall} as a function of the distance of the satellite $a$ and obliquity of the planet $\varepsilon$.

\begin{figure*}
   \centering
   \includegraphics[width=\textwidth]{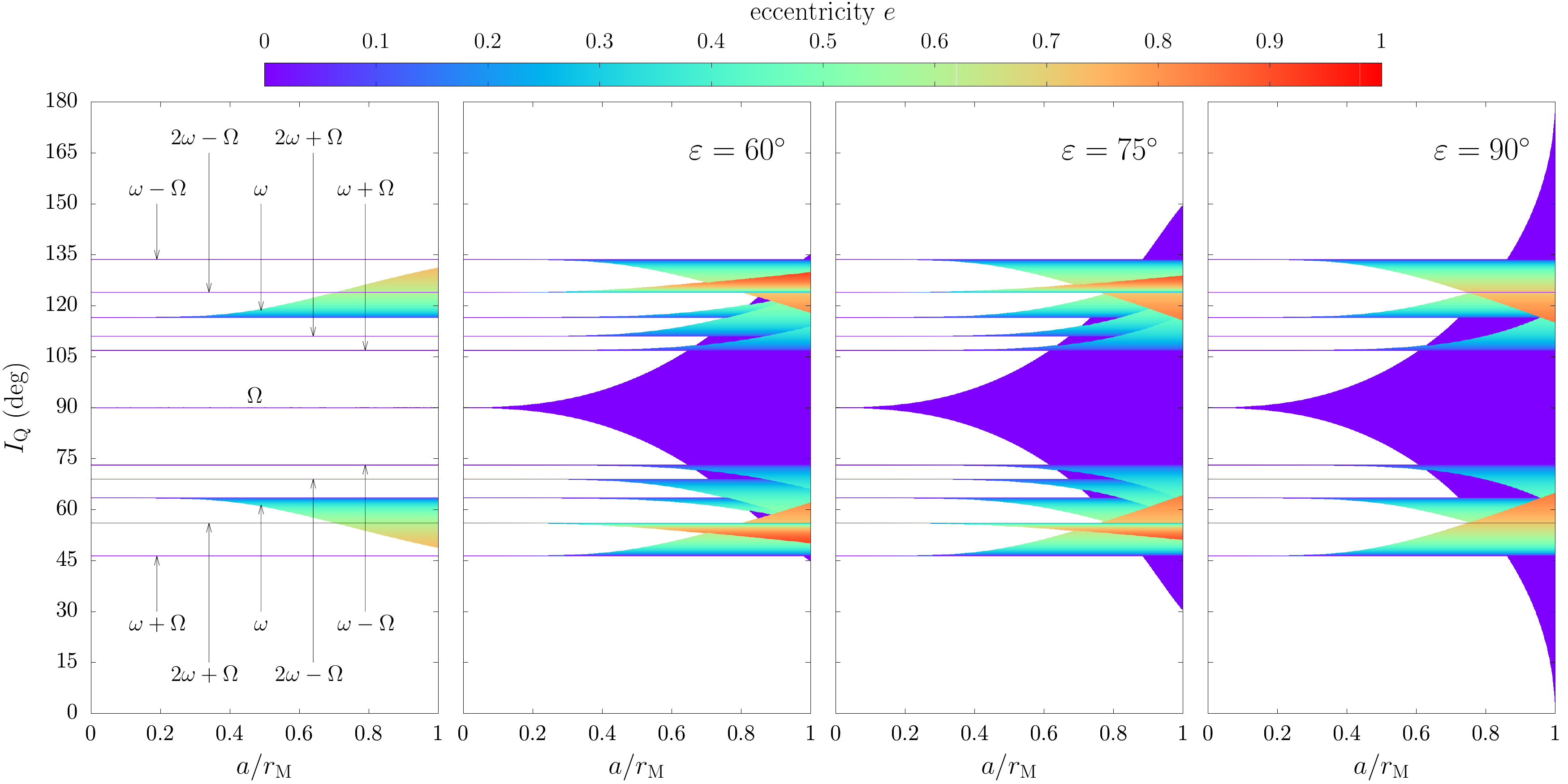}
   \caption{Location and width of first-order secular resonances for the satellite. Each panel is plotted for one fixed value of the planet's obliquity $\varepsilon$ from $0^\circ$ (left) to $90^\circ$ (right). Resonances are labelled in the leftmost panel by their respective critical angle $\xi$ (omitting the $\mathrm{Q}$ indexes). The coloured intervals show the extremal values of equatorial inclination $I_\mathrm{Q}$ spanned by the resonance separatrix. Inside each resonance, a conserved quantity links $I_\mathrm{Q}$ to the eccentricity $e$, as shown by the colour gradient.}
   \label{fig:resoverlap}
\end{figure*}

For a fixed value of obliquity $\varepsilon$, the extension of the resonances can be visualised in the space of the satellite's orbital elements, as illustrated in Fig.~\ref{fig:resoverlap}. As the resonances all feature a distinct constant quantity $K$, they lie in a different space and we cannot directly judge whether they are overlapping or not in the sense of \cite{Chirikov_1979}. However, Fig.~\ref{fig:resoverlap} gives a precise idea of the variations of orbital elements expected for the satellite under the influence of each resonance. For $a\lesssim 0.2\,r_\mathrm{M}$, resonances are small and well separated from each other for any obliquity value. When the satellite is close to $a=r_\mathrm{M}$, on the contrary, and especially if the obliquity is large, resonances allow for extreme variations in eccentricity and inclination\footnote{The same dynamical mechanism has been reported by \cite{Saillenfest-etal_2019b} for trans-Neptunian objects perturbed by the galactic tides.}, in accordance with what we observed in the previous simulations. Moreover, it is clear from Fig.~\ref{fig:resoverlap} that no resonance at all is able to substantially increase the eccentricity over large intervals of equatorial inclination located around $I_\mathrm{Q}=0^\circ$, $90^\circ$, and $180^\circ$. This property explains the peculiar shape of the black region in Fig.~\ref{fig:EcIQ_full}; the two spikes are produced by the two triplets of resonances that have the largest $\Delta e$ (see the horizontal dashed lines).

The resonance that produces by far the largest variations in eccentricity is $2\omega_\mathrm{Q}+\Omega_\mathrm{Q}$ (which becomes $2\omega_\mathrm{Q}-\Omega_\mathrm{Q}$ if the orbit is retrograde). This resonance, however, is very narrow in inclination: a few degrees at most. Its nominal location is $I_\mathrm{Q}\approx 56^\circ$  (or $I_\mathrm{Q}\approx 124^\circ$ for a retrograde orbit), as given in Table~\ref{tab:resconst}; this location closely matches the collision keyholes in Figs.~\ref{fig:IQcolvsmass} and \ref{fig:EcIQ_full}. We deduce that the secular resonance $2\omega_\mathrm{Q}+\Omega_\mathrm{Q}$ (when the orbit is prograde) or $2\omega_\mathrm{Q}-\Omega_\mathrm{Q}$ (when the orbit is retrograde) is the main responsible for the collision of the satellite into the planet. This resonance is well known for artificial satellites; it has been studied in depth by \cite{Daquin-etal_2022} in the more complicated case of lunisolar perturbations. In our case, Figs.~\ref{fig:reswidthall} and \ref{fig:resoverlap} reveal that the widths of this resonance abruptly drop to zero for obliquity values $\varepsilon=0^\circ$, $90^\circ$, and $180^\circ$; this explains why no collision at all occurs around these values of obliquity, as noticed previously in Fig.~\ref{fig:obcolvsmass}. One can verify this property by running a simulation with the planet's obliquity fixed to $\varepsilon=90^\circ$: we obtain a graph similar to Fig.~\ref{fig:EcIQ_full}, but in which the two spikes are smoother, like eroded, and the collision limit is out of reach for the satellite. As shown in Fig.~\ref{fig:reswidthall}, the strongest resonance that persists for $\varepsilon=90^\circ$ is $\xi=\omega_\mathrm{Q}+\Omega_\mathrm{Q}$ located at $I_\mathrm{Q}\approx 46^\circ$ (or equivalently $\omega_\mathrm{Q}-\Omega_\mathrm{Q}$ at $I_\mathrm{Q}\approx 134^\circ$), but it is not quite large enough for the satellite to reach collision.

It should be noted, however, that the simplified model used here to describe the secular resonances predicts an equal probability for collisions at $I_\mathrm{Q}\approx 56^\circ$ and $124^\circ$. Equal probabilities are indeed obtained for small satellites, but more massive satellites happen to collide much more frequently in the retrograde keyhole (see Fig.~\ref{fig:IQcolvsmass}). A careful examination of Fig.~\ref{fig:EcIQ_full} reveals that the retrograde peak is indeed slightly bigger. This subtlety cannot be explained using a restricted model (but the restricted model does provide a qualitative understanding of the dynamics).

\begin{figure}
   \centering
   \includegraphics[width=0.9\columnwidth]{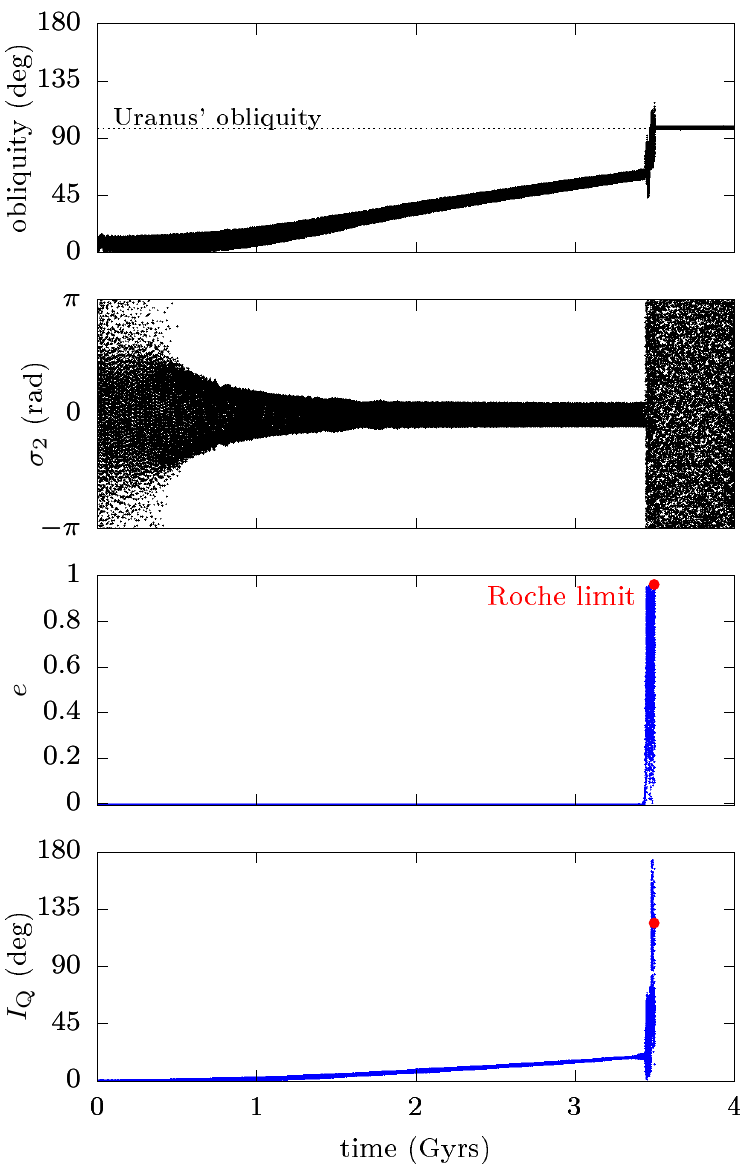}
   \caption{Example of simulation that reproduces the current state of Uranus. We assume that the satellite is instantly removed when it goes below the Roche limit (red point). The mass of the satellite is $m/M=2.2\times 10^{-3}$ and it migrates with a timescale $\tau=7.5$~Gyrs.}
   \label{fig:fullycoupled_ex2}
\end{figure}

\begin{figure}
   \centering
   \includegraphics[width=0.9\columnwidth]{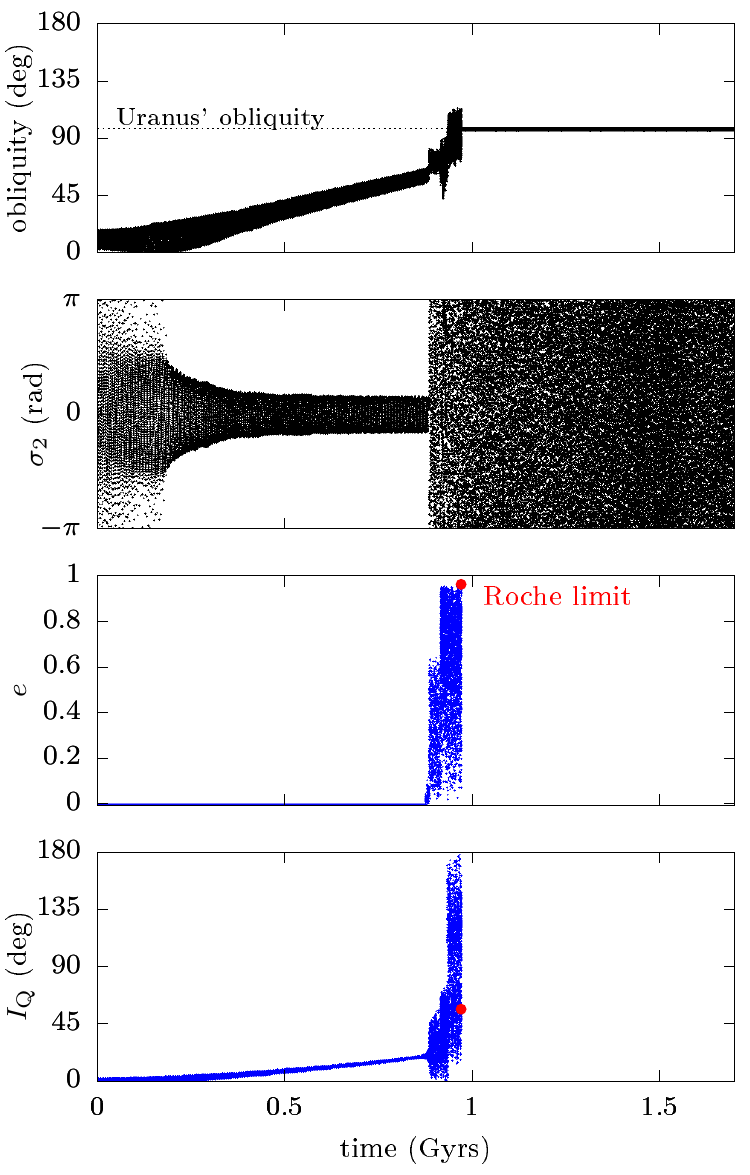}
   \caption{Same as Fig.~\ref{fig:fullycoupled_ex2}, but where the satellite crosses the Roche limit while having a prograde orbit with respect to the planet's equator. The mass of the satellite is $m/M=2\times 10^{-3}$ and it migrates with a timescale $\tau=3$~Gyrs.}
   \label{fig:fullycoupled_ex3}
\end{figure}

Another property of Fig.~\ref{fig:IQcolvsmass} needs to be explained: the extreme majority of satellite destructions (collisions of Roche crossings) that occur when $\varepsilon>90^\circ$ (coloured dots) go through the retrograde keyhole at $I_\mathrm{Q}\approx 124^\circ$ and almost never in the prograde keyhole, whatever the mass of the satellite. In other words, in almost every trajectory that successfully reproduces the current state of Uranus, the ancient satellite collides into Uranus while being retrograde with respect to its equator. This property can be explained by the `countdown to collision' mentioned earlier: As the satellite wanders about in the chaotic zone, it will reach sooner or later a collision keyhole. Because of the partial conservation of the planet-satellite angular momentum, the planet's spin axis diffuses much more efficiently from $\varepsilon<90^\circ$ to $\varepsilon>90^\circ$ (or back again) when the satellite is highly retrograde with respect to its equator (see e.g. Fig.~\ref{fig:fullycoupled_ex1}: the largest obliquity jumps occur when the satellite remains retrograde for a long period of time). During the short interval while $\varepsilon\approx 90^\circ$, collision keyholes are closed and the satellite is temporarily safe from destruction (see above). However, when the obliquity becomes substantially larger than $90^\circ$ and the collision keyholes open again, the satellite has a much larger probability of colliding right away, while still being retrograde with respect to the planet's equator, than going back prograde and then collide. Figures~\ref{fig:fullycoupled_ex2} and \ref{fig:fullycoupled_ex3} present an example of trajectory for both cases. Figure~\ref{fig:fullycoupled_ex2} shows the most common situation, in which the satellite is destructed right away after having become retrograde and driven the obliquity to values larger than $90^\circ$. Figure~\ref{fig:fullycoupled_ex3} shows the very improbable situation (much less than $1\%$ of cases) in which the satellite goes back prograde before being destructed.

One can note that these two trajectories have been obtained using two very different migration timescales; this highlights the large variety of parameters from which the current state of Uranus can be reproduced via the mechanism described here.

\section{Discussions}\label{sec:discuss}

\subsection{Forming Uranus' ancient satellite}\label{sec:form}

In Sect.~\ref{sec:coupledmod}, we have seen that when trying to reproduce the current spin state of Uranus via the migration of a single ancient satellite, large probabilities of success can be obtained for a satellite mass $m/M\approx 1.7\times 10^{-3}$ (i.e. about the mass of Jupiter's moon Ganymede, $m\approx 15\times 10^{22}$\,kg) or larger. In order to achieve the tilting of Uranus in a billion-year timescale without invoking extravagant migration rates, this satellite should be formed during the early stages of the Solar System evolution and start its outward migration at an initial distance $a_0$ ranging from about $40$ to $50\,R_\mathrm{eq}$. We must now discuss whether the existence a such a satellite around Uranus could appear realistic in the context of satellite formation theories.

For postulating that Uranus and Neptune had small initial obliquities, our main hypothesis is that they formed in a similar manner as Jupiter and Saturn, with the infall of gas dominating their final spin state (see Sect.~\ref{sec:intro}). With this hypothesis in mind, it seems natural to consider that a primordial generation of satellites formed around Uranus and Neptune through a similar mechanism as the Galilean satellites around Jupiter or Titan around Saturn. \cite{Mosqueira-Estrada_2003a,Mosqueira-Estrada_2003b} modelled the formation of the main satellites of Jupiter, Saturn, and Uranus in a gaseous circumplanetary disc. According to their work, Uranus today `lacks' a big satellite, that should have been formed at a distance of $57~R_\mathrm{eq}$ and would have been Uranus' analogue of Ganymede or Titan. The authors interpreted this discrepancy as the signature of a substantial inward migration of Uranus' proto-satellites in the gaseous disc; they did not notice that a satellite at this distance around Uranus is unlikely to survive because it is close to the unstable zone (see \citealp{Tremaine-etal_2009,Saillenfest-Lari_2021}). Even though major improvements have been made since then in satellite formation theories (see e.g. \citealp{Batygin-Morbidelli_2020}), a large distant satellite does seem to lack in Uranus' system when compared to the systems of Jupiter and Saturn. This missing satellite may resemble what we are looking for, but some caution is needed. On the one hand, it is now admitted that planets and satellites substantially migrated and reorganised after their formation (see e.g. \citealp{Tsiganis-etal_2005,Nesvorny-Morbidelli_2012,Lainey-etal_2009,Lainey-etal_2020,Lari-etal_2020}). Hence, trying to adjust formation scenarios to today's locations of the planets and their satellites is probably inadequate. On the other hand, we assumed so far that at the time of Uranus' tilting, the hypothetical ancient satellite was not affected by the attraction of additional close-in satellites. To lowest order, the gravitational attraction of inner moons is equivalent to increasing the planet's $J_2$ \citep{Tremaine-etal_2009}. If we take the current satellites of Uranus into account at their current locations, the effective value of $J_2$ felt by the hypothetical distant satellite would be multiplied by a factor $k\approx 5.4$. As a result, the Laplace radius is multiplied by $k^{1/5}\approx 1.4$ (see Eq.~\ref{eq:rM}) and $r_\mathrm{M}$ becomes $75\,R_\mathrm{eq}$ instead of $53\,R_\mathrm{eq}$. This new value still does not seem too unrealistic, but it does not match any more that proposed by \cite{Mosqueira-Estrada_2003a,Mosqueira-Estrada_2003b}. In order to avoid this increase in distance for the hypothetical ancient satellite, the current moons of Uranus should either have been formed after the tilting (see Sect.~\ref{sec:currsat} below), or have been located closer to Uranus than they are today.

The total satellites-to-planet mass ratios for Jupiter, Saturn, and Uranus are today a few times $10^{-4}$. \cite{Canup-Ward_2006} argued that such a similar mass ratio can be explained naturally if the satellites of all giant planets formed in a similar way within a circumplanetary disc of gas and dust, with multiple generations of satellites being formed and lost by migrating through the disc. Under the widely accepted assumption that Triton is a captured object, \cite{Rufu-Canup_2017} then showed that a primordial satellite system with mass ratio roughly equal to $10^{-4}$ or less is also expected to have existed around Neptune. This general picture seems to contradict the existence of a primordial satellite with mass $m/M\gtrsim 10^{-3}$ around Uranus, as proposed here. However, even though the typical ratio of $10^{-4}$ was found by \cite{Canup-Ward_2006} to only weakly depend on the poorly known parameters involved, we have no guarantee that satellites around Jupiter, Saturn, and Uranus did form in roughly similar external conditions. Mass ratios larger than $10^{-3}$ can actually be obtained from the formulas of \cite{Canup-Ward_2006} by slightly tweaking the poorly known parameters. Moreover, the assumption that the regular satellites of all four giant planets formed through the same process is itself debated. Circumplanetary discs around the Solar System giant planets probably had much more diverse histories than initially envisioned by \cite{Canup-Ward_2006}, and additional phenomena, as the truncation of the circumplanetary disc due to the planet's magnetosphere, are expected to have played an important role (see e.g. \citealp{Sasaki-etal_2010,Batygin-Morbidelli_2020}).

Recently, \cite{Szulagyi-etal_2018} breathed new life into the possibility that the satellites of Uranus and Neptune formed early in a gaseous circumplanetary disc. The range of masses that they obtain in the framework of their population synthesis appears in line with our proposed ancient satellite. Yet, in this paper, the formation of the disc relies on a heat sink forcing the gas to not exceed a temperature of $100$~K in the neighbourhood of the planet, which may not be realistic. Besides, the satellite population obtained strongly depends on the flux of mass injected within the disc during planetary formation, which is an unknown quantity. For these reasons, it seems to us that the mass distribution obtained by \cite{Szulagyi-etal_2018} cannot be used as an argument in favour of given mass values in our scenario: using a higher mass flux would increase the mass of the satellites obtained, and vice versa. What we conclude from \cite{Szulagyi-etal_2018} and previous works is that satellites as massive as what we need here can be formed routinely from circumplanetary discs that appear `realistic' for Uranus and Neptune. As a general rule, very big satellites may also exist around giant planets (with masses as large as $m/M\approx 10^{-2}$), but they are probably not a generic outcome of formation processes within gaseous circumplanetary discs. Recent 3D hydrodynamic simulations seem to suggest that, unlike Jupiter and Saturn, planets similar to Uranus and Neptune do not develop significantly massive discs unless the equation of state of the gas is taken to be isothermal (see \citealp{Fung-etal_2019}). For this reason, the formation of a massive satellite around Uranus may have been unlikely.

Unusually big satellites may also be formed during the protoplanetary disc phase as captured coorbital proto-cores \citep{Hansen_2019}. However, those are expected to have large orbital inclinations, with most likely values near $180^\circ$, in contrast with the scenario proposed here in which the satellite starts close to its local (prograde) Laplace plane. Hence the formation scenario of \cite{Hansen_2019} could be invoked in this context only if it is associated with an efficient mechanism of inclination damping. If instead the satellite remains highly retrograde, then tidal dissipation within the planet would result in an inward (rather than outward) migration. As discussed in Sect.~\ref{sec:smallestsat}, the mechanism for tilting Uranus can work just as well, or even better, if the satellite migrates inwards, however there is no evidence showing that a powerful inward migration of retrograde satellites is possible. The capture of a large body during the planetesimal-driven planetary migration is also a possibility for Uranus to have acquired a big ancient satellite (see e.g. \citealp{Agnor-Hamilton_2006,Li-Christou_2020}); however, captured objects typically have large eccentricities and inclinations \citep{Nesvorny-etal_2007}, in contrast to the satellite needed here for the tilting of Uranus.

\subsection{Smallest ancient satellite allowing for tilting Uranus}\label{sec:smallestsat}

Even though a mass ratio of $m/M\approx 1.7\times 10^{-3}$ does not appear absurdly high in the context of satellite formation theories, smaller satellites are generally thought to be more likely for a primordial formation in a gaseous circumplanetary disc. For comparison, the mass ratios for the Galilean satellites range from about $3$ to $8\times 10^{-5}$, and the mass ratio for Titan is about $2\times 10^{-4}$. The possibility for Uranus and Neptune to have had a massive circumplanetary disc during the late stages of their formation is itself much debated (see e.g. \citealp{Szulagyi-etal_2018,Reinhardt-etal_2020} and Sect.~\ref{sec:form} above). As the formation of small satellites requires less material and imposes weaker constraints on their hypothetical gaseous birth disc, it appears natural to try to minimise as much as possible the mass required for the ancient satellite in our scenario.

Because of the characteristics of Uranus' orbital motion, we can be assured from strong dynamical grounds that no tilting can be achieved for a single ancient satellite with mass smaller than $m/M\approx 4.4\times 10^{-4}$ (see Sect.~\ref{sec:tilting}). With this minimum mass, a satellite allows for a capture of Uranus in secular spin-orbit resonance with $\nu_3$ and adiabatic tilting up to large obliquity values. When the system reaches the unstable zone and the satellite goes wild, however, a mass of $m/M\approx 4.4\times 10^{-4}$ is insufficient to produce a strong back reaction on the planet, and therefore the satellite is generally destructed before the planet's spin axis has time to diffuse much higher up (see Sect.~\ref{sec:coupledmod}). As a result, the planet's obliquity is left fossilised to values that are still close to the boundary of the unstable zone (i.e. from about $75^\circ$ to $80^\circ$ using the parameters of Fig.~\ref{fig:obcolvsmass}), and far from Uranus' current value. We can think of different ways to solve this problem and increase the success rate for small satellites.

The first and most obvious possibility would be that the ancient satellite was not alone around Uranus. As explained above, the addition of close-in satellites increases the effective $J_2$ of Uranus (see e.g. \citealp{Tremaine-etal_2009}). This increase pushes away the Laplace radius of our hypothetical ancient satellite, but it also decreases the mass that it needs to have in order for Uranus to reach a given secular spin-orbit resonance. If Uranus' $J_2$ is increased by a factor $k$ by a set of inner satellites, then $r_\mathrm{M}$ is enlarged by a factor $k^{1/5}$ and the mass needed to reach a given resonance is divided by $k^{2/5}$ (see equations in Sect.~\ref{sec:mec}). It is difficult to judge a priori which satellite configuration would be most appropriate here, and how would inner satellites behave when their distant neighbour becomes unstable. To get a rough idea of the effect of inner satellites, we note that the current satellite system of Uranus would push the critical radius $r_\mathrm{M}$ to $75\,R_\mathrm{eq}$ and divide the mass needed for the ancient satellite by roughly a factor $2$. In particular, the strong resonance $\nu_2$ showing the highest success probabilities in our scenario would be reachable by an ancient satellite with mass $m/M\approx 10^{-3}$ instead of $2\times 10^{-3}$. This might increase the plausibility of our scenario. However a set of inner satellites would probably not help to increase the success probabilities obtained in other resonances, because the issue does not come from the tilting phase (resonance $\nu_3$ is perfect for that) but from the destabilisation phase (during which small satellites do not produce a strong enough back reaction on the planet's obliquity). A variant of this possibility would be to have a succession of several migrating satellites that reach the unstable region one after another: the first one would produce the resonance capture and tilting to about $80^\circ$, and the other ones would just reach the unstable region (with the obliquity being already high) and produce some chaotic obliquity diffusion before being disrupted. If they sum up, each of these limited obliquity changes may allow to reach the current obliquity of Uranus. However, in any case, the exact behaviour of a set of satellites is hard to predict and one would need to explore these scenarios self-consistently using numerical simulations.

The second possibility would be to delay the instability of the satellite such that the obliquity can climb higher up inside the resonance before the destabilisation phase is triggered. As the instability originates from an eccentricity increase (see \citealp{Tremaine-etal_2009,Saillenfest-Lari_2021}), such a delay could be produced through tidal dissipation inside the satellite, whose main effect is to damp eccentricity. Yet, dissipation within the satellite should not be too strong either, otherwise the instability could be completely suppressed, or the migration of the satellite could be reversed before the instability has been properly triggered. As mentioned earlier, the instability is so strong that it is not expected to be suppressed for realistic dissipation parameters of the satellite (see Fig.~\ref{fig:ecc_increase} and the discussion above). However, the rate of energy dissipation within the satellite depends on its interior properties which can only be speculative at this stage. Because of the wide range of possible behaviours that can be obtained when varying the dissipation rate, we leave the exploration of this possibility for future works.

The third possibility would be to follow a secular spin-orbit resonance that is connected higher up to the unstable region, so that the instability is triggered at a larger value of obliquity. The different panels in Fig.~\ref{fig:reswidths} show that a given resonance can reach the unstable zone at different obliquity values, depending on the exact mass of the satellite. This change in the resonance shape is the reason why resonances in Fig.~\ref{fig:massmap} show a higher success ratio for smaller $a_0$ and larger mass (i.e. in the lower end of the coloured bands). For resonance $\nu_3$, however, a non negligible fraction of successful trajectories can be obtained only for $a_0\lesssim 33~R_\mathrm{eq}$ (see panels~c and f of Fig.~\ref{fig:massmap}). Such small values of $a_0$ require a very fast tidal migration in order for the system to reach the unstable zone in less than the age of the Solar System. For even smaller initial semi-major axes, the migration rate of the satellite would need to be so large that adiabatic capture in resonance $\nu_3$ would be endangered (see Fig.~\ref{fig:resstats}). This third solution in order to increase the success rate of small satellites is therefore not conclusive. Yet, as shown in Fig.~\ref{fig:reswidths}, very late instabilities would be produced for a satellite migrating inwards from an initial semi-major axis $a_0>r_\mathrm{M}$. In order for the satellite to migrate inwards, it should be retrograde with respect to the planet's spin motion. Retrograde massive satellites with $I_\mathrm{Q}\approx 180^\circ$ can be formed through the mechanism described by \cite{Hansen_2019}. However, it is not clear whether retrograde satellites could be able to trigger tidal dissipation mechanisms that are as efficient as those produced by prograde satellites (e.g. by a process analogous to that described by \citealp{Fuller-etal_2016}). This possibility is therefore quite speculative at this stage.

The fourth possibility would be to invoke an additional ingredient to produce the final missing tilt of about $10^\circ$ to $20^\circ$ at the end of the adiabatic drift in resonance $\nu_3$. The first idea that comes to mind is to take into account the effect of the satellite collision in the final obliquity value of the planet. The satellite destabilisation may also trigger a chain reaction involving other satellites and lead to several impacts. Assuming a perfect merger, the effect of a satellite collision into the planet can be estimated from the conservation of their total angular momentum. Indeed, the spin-axis motion of the planet and the destabilisation of its satellite are due to the secular effect of external perturbers (namely, the Sun and other planets); as such, external perturbations act on long timescales (at least tens of thousands of years; see Table~1 of \citealp{Saillenfest-Lari_2021}). During the very limited time interval of a collision, the planet and its satellite can therefore be considered as an isolated system. The details of computations are given in Appendix~\ref{asec:impact}. Considering that the ancient satellite is destabilised by the mechanism described above, the maximum obliquity change of the planet due to the impact of its satellite is, in radians,
\begin{equation}\label{eq:satimpact}
   \Delta\varepsilon \approx 35 \frac{m}{M}\,.
\end{equation}
This means that in order to produce a $10^\circ$ extra tilt, the ancient satellite must have at least a mass $m/M \approx 5\times 10^{-3}$, that is, bigger than Mercury. In order to produce $15^\circ$, it must have the size of Mars. These very large masses do not seem realistic in the context of satellite formation, and they clearly do not help to increase the success rate of small satellites. The effects of a chain reaction involving several satellites are difficult to predict a priori, but because of the conservation of total angular momentum, the total mass of the satellites needs to be quite large anyway, and in the same order of magnitude as in Eq.~\eqref{eq:satimpact}.

\begin{figure}
	\centering
	\includegraphics[width=\columnwidth]{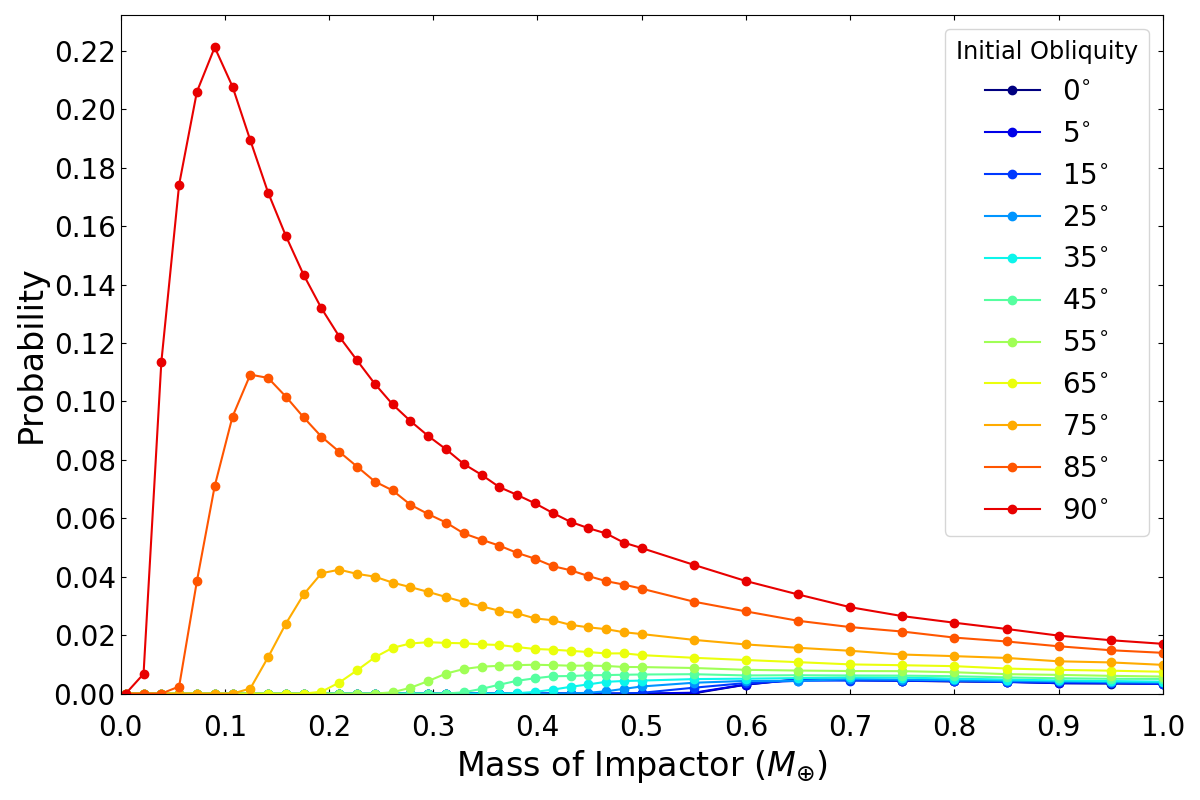}
	\caption{Probability of reproducing Uranus's current spin state as a function of the mass of a single impact strike. The planet is initialised with the current spin rate of Uranus, considered to be primordial. The mass of the impactor is given in Earth masses ($M_\oplus$).}
	\label{fig:impacts}
\end{figure}

Alternatively, given the small size of the obliquity gap to be filled, one can think of a subsequent impact with a rogue minor planet, once the tilting up to $75^\circ$ to $80^\circ$ has been achieved. We consider here the scenario of a small late impact that only marginally alters the spin rate and obliquity (quite differently from giant impact scenarios mentioned in Sect.~\ref{sec:intro}, for which the impact itself fully settles the spin rate and obliquity of the planet). Since collisions capable of such tilting can still vary a planet's spin rate by as much as a factor of two \citep{Rogoszinski-Hamilton_2020}, the most likely size of impactors must be determined with the condition that it should not significantly change the planet's spin rate. We accomplish this by using the code developed by \cite{Rogoszinski-Hamilton_2020,Rogoszinski-Hamilton_2021} to calculate a planet's final spin state after collisions. The planet grows by summing the angular momenta of impactors and the planet, and the planet's final tilt and spin are extracted from its resulting angular momentum vector. For simplicity, we assume that the impactors originate from within the Solar System, they travel on trajectories that are parallel to the plane of the planet's orbit, and all their mass is absorbed upon impact. These approximations are sufficient because we assume small orbital inclinations, and less than $0.1\,M_{\oplus}$ of debris is expected to be ejected from the system after an Earth-mass strike \citep{Kegerreis-etal_2018,Kegerreis-etal_2019}. The impactor hits at a random location on the planet's surface and over all possible spin-axis precession phases, so we run this code for a half million iterations to generate probability distributions of the planet's spin state. Finally, since the impactors are likely to travel on low-eccentricity orbits prograde to Uranus', we follow the approach of \cite{Hamilton-Burns_1994} and sample the relative speed of impactors between $0$ and $0.4$ times the Keplerian speed of a circular orbit at the distance of Uranus ($6.8$~km\,s$^{-1}$). The values obtained are much less than Uranus' escape speed ($21.4$~km\,s$^{-1}$), so we include gravitational focusing (i.e. trajectories are aimed closer to the planet's centre). Spin and obliquity must be handled simultaneously: we calculate the likelihood of generating Uranus' current spin state by measuring the fraction of instances that are within both $\pm 5^\circ$ of Uranus current obliquity and $\pm 10\%$ of its current spin rate. Figure~\ref{fig:impacts} shows the results obtained for initial obliquities between $0^\circ$ and $90^\circ$ and impactor masses from $0$ to $1\,M_{\oplus}$. The most likely impactor that can generate Uranus' current spin state from an initial tilt of $85^\circ$ has a mass of about $0.1\,M_{\oplus}$. Again, this is close to the size of Mars. Even though pebble accretion models posit a surplus of Mars-sized cores in the formation region of the giant planets (see e.g. \citealp{Levison-etal_2015}), and especially beyond Saturn \citep{Izidoro-etal_2015}, these planetesimals are thought to have been accreted and scattered quickly during planetary formation. After billions of years of evolution, as required by our tilting mechanism, the structure of the Solar System was most likely very similar to what it is now. The most massive objects that may possibly impact Uranus today are trans-Neptunian dwarf planets, whose small masses lead to a negligible probability of producing the small missing tilt that we are looking for (Eris is $0.003\,M_\oplus$; see Fig.~\ref{fig:impacts}). Hence, even though planetesimal impacts probably happened during the late stages of planetary formation and possibly produced a small initial obliquity excitation (which would not be in contradiction with our tilting scenario; see Fig.~\ref{fig:resstats}), impacts of external bodies do not allow for decreasing the mass of the satellite needed to tilt Uranus in our scenario.

In conclusion of this subsection, it seems difficult to reproduce the current spin state of Uranus if its ancient satellite is single and has a mass smaller than $m/M\approx 1.7\times 10^{-3}$. Taking into account a system of additional inner satellites resembling the current ones would allow to decrease this minimum limit by roughly a factor $2$. Below $m/M\approx 10^{-3}$, however, our simulations show that satellites hardly manage to excite the planet's obliquity high enough. In order to increase further the success rate of small satellites, another possibility would be to include the effects of tidal energy dissipation within the satellites. This could help to decrease the minimum mass for a single satellite down to $m/M\approx 4.4\times 10^{-4}$ at most.

\subsection{Forming Uranus' current satellites in a debris disc}\label{sec:currsat}

Many previous works have been devoted to the formation of Uranus' current satellites in the context of the giant impact hypothesis (see e.g. \citealp{Morbidelli-etal_2012,Ida-etal_2020,Rufu-Canup_2022,Salmon-Canup_2022,Woo-etal_2022}). If our new tilting scenario is correct, then it should also not be in contradiction with the existence of Uranus' current satellite system. As discussed above, strong dynamical arguments show that Uranus cannot be tilted through the scenario considered here if its ancient satellite is single and has a mass smaller than $m/M\approx 4.4\times 10^{-4}$. This minimum mass is about four times the total mass of Uranus' current satellite system. Assuming that the ancient satellite goes below the Roche limit during the final instability (see Sect.~\ref{sec:coupledmod}), then it would be torn apart into a large amount of debris. The remaining pieces of material would then collide with each other and rapidly reorganise into an equatorial disc confined inside the Roche limit (see \citealp{Morbidelli-etal_2012,Hyodo-etal_2017}).

Since the current regular satellites of Uranus seem to have been formed in such a tidal disc \citep{Crida-Charnoz_2012,Hesselbrock-Minton_2019}, it is tempting to link them to the disruption of our hypothetical ancient massive satellite. If the ancient satellite had mass $m/M=4.4\times 10^{-4}$ and was initially the only regular satellite of Uranus, then about $75\%$ of its mass should have fallen onto the planet or been ejected, while the remaining $25\%$ served as building blocks for the current satellite system of Uranus. These proportions are very similar to those obtained by \cite{Hesselbrock-Minton_2017} in their simulations of the formation of Phobos from a debris disc. \cite{Hesselbrock-Minton_2017} show that the precise formation efficiency of the new generation of satellites depends on the exact radius at which the ancient satellite is torn apart (which itself depends on its cohesive strength): if the ancient satellite is destructed closer to the planet, then the viscous spreading of the debris disc results in more material to be lost by falling onto the planet. Hence, in the context of the tilting mechanism described above, the minimum mass $m/M\approx 4.4\times 10^{-4}$ for the ancient satellite is compatible with the formation of all current regular satellites of Uranus from the debris disc, but more massive ancient satellites are not ruled out.

When the disc spreads beyond the fluid Roche limit, the pieces of debris aggregate into new satellites through the mechanism described by \cite{Crida-Charnoz_2012}. For a disc mass $m_\mathrm{disc}$, the characteristic lifetime of the disc $T_\mathrm{disc}=m_\mathrm{disc}/\dot{m}_\mathrm{disc}$ is
\begin{equation*}
   T_\mathrm{disc}\approx 0.0425\left(\frac{M}{m_\mathrm{disc}}\right)^2T_\mathrm{Roche}\,,
\end{equation*}
where $T_\mathrm{Roche}$ is the orbital period at the fluid Roche limit. Assuming that the disc contains just enough material to form all the current regular satellites of Uranus, one obtains a characteristic lifetime of about $4000$~years. This timescale is reduced to $200$~years if the disc has an initial mass $m_\mathrm{disc}/M\approx 4.4\times 10^{-4}$. The formation of the new generation of satellites is therefore extremely fast at the beginning, and it slows down asymptotically as $m_\mathrm{disc}\propto t^{-1/2}$. When the disc mass has decreased below a given threshold, other phenomena can take over: \cite{Estrada-Durisen_2021} show that micro-meteoroid bombardment can slow down ring particles and force small-mass rings to drift inwards and eventually vanish relatively quickly. These short evolution timescales would explain why no massive disc remains around Uranus today, even in the hypothesis that the destruction of its ancient satellite (and the end of the tilting mechanism) occurred recently in the history of the Solar System, perhaps a few hundreds of million years ago. The current inner sparse ring system of Uranus may be a remnant of the old massive debris disc, and the innermost satellites of Uranus may be evolved residuals of the last generation of satellites formed from the disc. These innermost satellites are known to be unstable on short timescales (see e.g. \citealp{Duncan-Lissauer_1997,French-Showalter_2012,French-etal_2015}), and it is probable that they continuously form new ring particles and new small satellites through mutual collisions and reaccretion \citep{Tiscareno-etal_2013}.

After their formation in the debris disc, the regular satellites of Uranus must have migrated up to their current location. Oberon is the farthest regular satellite of Uranus and it is located today at a semi-major axis of $a\approx 23$~$R_\mathrm{eq}$. This distance is to be compared to the synchronous orbit around Uranus, which lies somewhat below $3.5$~$R_\mathrm{eq}$ (the migration from the fluid Roche limit at $2.5$~$R_\mathrm{eq}$ to the synchronous orbit is produced through Lindblad torques from the disc while it is massive enough; see \citealp{Hesselbrock-Minton_2019}).

The current migration rate of Uranus' satellites is essentially unconstrained today from direct observations \citep{Lainey_2016}, and even though theoretical arguments can be used to give bounds to the unknown parameters \citep{Tittemore-Wisdom_1990,Cuk-etal_2020}, our lack of knowledge about the dissipation mechanisms at play in the interior of Uranus leaves room for a large range of possible scenarios. For instance, previous arguments based on mean-motion resonance encounters may be invalidated if satellites are undergoing a mechanism similar to the `tidal resonance locking' of \cite{Fuller-etal_2016}, for which the effective quality factor $Q$ of the planet evolves over time in such a way that semi-major axis ratios are constant \citep{Crida_2020}. In fact, as shown by \cite{Lainey-etal_2020}, the migration history of satellites around giant planets can result to be spectacularly different from what one would expect from constant-$Q$ models. Based on the observed migration rate of Saturn's satellites, and assuming that the tidal mechanism of \cite{Fuller-etal_2016} was not triggered long after their formation, then Titan, for instance, would have migrated across a radial distance of $10$ radii of Saturn in $4$~Gyrs, that is, more than $600\,000$~km. In contrast, if we suppose that Oberon was formed from a debris disc at the Roche limit of Uranus, then it would need to have migrated over about $20$ radii of Uranus before today, which are about $500\,000$~km. These large numbers may seem extraordinary high when viewed in the context of previous works. For instance, \cite{Morbidelli-etal_2012} argued that a tidal disc around Uranus cannot have been the birth place of its current satellites because they are too far away\footnote{A fast satellite migration is also discarded by \cite{Rufu-Canup_2022}, even though it could be a promising solution to their alternative version of the coaccretion + giant impact model; see their final discussion.}. However, these distances are in line with the measurement of the current orbital expansion of Saturn's satellites \citep{Lainey-etal_2020}. In any case, if we assume that Uranus had a distant ancient satellite subject to substantial tidal migration (which is our basic hypothesis throughout this paper), then some mechanism of efficient energy dissipation must exist in the interior of Uranus. Regardless of the precise nature of this mechanism, the dissipation efficiency needed to move Oberon to its current location (measured by the equivalent constant parameter $k_2/Q$; see Sect.~\ref{sec:res}) is smaller than that needed to move the distant satellite over $10$~$R_\mathrm{eq}$, so if we assume that the latter is possible, then the former should follow.

Yet, unless the migration rate of Oberon is much larger than that of Titan (which cannot be firmly ruled out yet, but may seem unlikely), we cannot expect it to have migrated up its current position in less than a couple billion years. This means that the end of the tilting mechanism of Uranus and the destruction of its ancient satellite cannot be arbitrarily recent if one wants to explain the formation of its current satellites in the debris disc formed. For instance, one could imagine that the migration of the ancient satellite and tilting of Uranus took $2$~Gyrs; then it was followed by the formation of its current satellites in the debris disc; and then the migration of its current satellites from the synchronous orbit to their current location took an extra $2$~Gyrs. Of course, the exact timing of these two phases is highly speculative.

The creation of a tidal disc that would be suitable for the formation of the current satellites of Uranus is itself not straightforward. Section~\ref{sec:keyholes} reveals that, for dynamical reasons, the massive ancient satellite is much more likely to collide into Uranus through the retrograde keyhole, whereas the current satellites of Uranus are prograde with respect to its spin axis. Yet, several factors could lead to the formation of a prograde tidal disc even if the satellite collides on a retrograde orbit. For instance, the satellite is not expected to be destructed at once but to form a torus of debris, and all pieces of debris still orbit in the highly unstable region before they are damped to the equator plane, so whether the final tidal disc should be prograde or retrograde is perhaps not as obvious as it may appear. Impacts of debris into the planet could also lead to the production of prograde ejecta.

Instead of being directly torn apart by tides, the destabilisation of the ancient satellite may also have triggered a collision cascade in a pre-existing system of inner moons. In this case, a large portion of the total satellite mass would need to be ejected or collide into the planet anyway, in order to reproduce the current mass of Uranus' system. Furthermore, dramatic satellite collisions beyond the Roche limit are not expected to produce a tidal disc, but to lead to a rapid reaccretion of debris into new moons without substantial spreading \citep{Hyodo-Charnoz_2017}. This means that if satellite-satellite collisions occurred far from the Roche limit, then the current satellites of Uranus would need to be the remnants of the collision cascade, which seems to be in contradiction with their spacing and mass distribution that closely match those expected from formation in a tidal disc \citep{Crida-Charnoz_2012}. Given the large eccentricity and inclination of the massive ancient satellite when it is destabilised, one would also expect a substantial orbital excitation from such large satellite-satellite collisions. Even though eccentricities can be quickly damped through tidal dissipation, this is not the case of inclinations, which have a much longer damping timescale. Today, only Miranda has a substantial orbital inclination, and its origin is well explained as a consequence of a past $5$:$3$ resonance between Ariel and Umbriel \citep{Cuk-etal_2020}. Previous studies obtained similar inclination excitations via the evolution of Miranda through the $3$:$1$ mean-motion resonance with Umbriel \citep{Tittemore-Wisdom_1989,Malhotra-Dermott_1990,Verheylewegen-etal_2013}. Depending on the actual migration rate of the satellites, this inclination increase may be quite recent in both scenarios, and trying to link it to the mechanism discussed here would be doubtful. A way to protect the current satellites of Uranus from the destabilising action of their distant neighbour would be to place them closer to the planet at the time of the tilting, because this would lock them more tightly to Uranus' equator. Hence, independently of whether the current satellites of Uranus formed before or after the tilting, invoking a substantial outward migration for them appears inescapable.

\section{Summary and conclusion}\label{sec:ccl}

Giant planets are thought to form in a protoplanetary disc made of gas and dust. During their last formation stages, gas accretion dominates their final spin state and gives them a primordial near-zero obliquity. The similarities between the four giant planets of the Solar System -- their spin rates in particular -- suggest such a common formation mechanism, and may disfavour stochastic processes, like giant impacts, as being a main formation ingredient. In this view, the small to moderate obliquity values of Jupiter ($3^\circ$), Saturn ($27^\circ$), and Neptune ($30^\circ$) can be explained by post-formation events (see e.g. \citealp{Ward-Hamilton_2004,Hamilton-Ward_2004,Ward-Canup_2006,Vokrouhlicky-Nesvorny_2015,Rogoszinski-Hamilton_2020,Saillenfest-etal_2021a}). Explaining the extreme obliquity of Uranus ($98^\circ$) is more challenging. Tilting Uranus without affecting much its spin rate would require a slow process involving the spin-axis precession motion. Such a process hardly fits the timespan offered by some given stage of the early Solar System evolution (e.g. dissipation of the gas disc, or late planetary migration). This is why previous studies investigating the tilting of Uranus as a post-formation event faced a timescale issue \citep{Boue-Laskar_2010,Quillen-etal_2018,Rogoszinski-Hamilton_2021}.

In contrast, the tidal migration of satellites is an everlasting process that can slowly change the orbit of natural satellites over billions of years. Massive satellites are known to affect the spin-axis motion of planets (see e.g. \citealp{Tremaine_1991,French-etal_1993,Boue-Laskar_2006}), and over the lifetime of the Solar System the migration of the main satellites of Jupiter and Saturn can potentially produce dramatic obliquity changes to their host planet \citep{Saillenfest-etal_2020,Saillenfest-etal_2021b}. In this article, we examined whether this mechanism could also apply to Uranus and possibly explain its extreme obliquity.

The mechanism considered here involves secular spin-orbit resonances, that is, resonances between the precession of the spin axis of the planet and some harmonics appearing in its orbital precession. It is made of two different phases as satellites migrate: \emph{i)} a capture in resonance and steady obliquity increase, and \emph{ii)} a violent destabilisation of the satellite when the obliquity exceeds $70^\circ$ to $80^\circ$ \citep{Saillenfest-Lari_2021}. Both phases can be used to determine the properties of the satellite that would be needed to reproduce the current spin state of Uranus.

\emph{Phase~1:} As we are studying a very slow process and the orbit of Uranus has been stable for billions of years, we need to consider the current orbital dynamics of Uranus as a forcing to its spin-axis motion. This puts strong constraints on the possible secular spin-orbit resonances involved. The orbital forcing term that is closest to Uranus' spin-axis precession rate has frequency $\nu_{15}=g_7-g_8+s_7$ (where $g_j$ and $s_j$ are the apsidal and nodal precession modes of the Solar System planets numbered from $1$ for Mercury to $8$ for Neptune). In order to produce a resonance with $\nu_{15}$, Uranus must have had a satellite with minimum mass $m/M=3.5\times 10^{-4}$ (assuming that Uranus had a single satellite at that time); however, this resonance is too weak to allow for an adiabatic capture and tilting in less than the age of the Solar System. The second closest forcing term has frequency $\nu_3=s_8$. Triggering a resonance with $\nu_3$ and tilting Uranus up to a large obliquity would require a satellite with minimum mass $m/M=4.4\times 10^{-4}$, that is, $m\approx 4\times 10^{22}$\,kg (smaller than Jupiter's moon Europa). Simulations show that there is a $100\%$ probability for Uranus to be captured and tilted in this resonance in a wide interval of satellite migration rates and for primordial obliquities $\varepsilon$ between $0^\circ$ and $20^\circ$. In order to complete the tilting phase, the satellite must migrate over a range of about $10$ radii of Uranus during its history; covering this distance in less than the age of the Solar System would require the satellite to migrate at least $6$~cm\,yr$^{-1}$ in average. This migration rate is orders of magnitude higher than what can be inferred from historical models of tidal dissipation within Uranus, but quite comparable to the $11$~cm\,yr$^{-1}$ measured for Titan (see \citealp{Lainey-etal_2020}). A mechanism similar to that described by \cite{Fuller-etal_2016} could be a viable explanation for a fast satellite expansion around Uranus.

A mass of $m/M=4.4\times 10^{-4}$ is roughly four times the total mass of the current satellites of Uranus; this means that the satellite involved does not exist any more and must have been destructed during the second phase of the evolution. More massive ancient satellites allow for reaching other resonances. The strongest resonance is $\nu_2=s_7$ which can be reached for a satellite mass $m/M=2.2\times 10^{-3}$. Other promising resonance candidates are $\nu_7=g_5-g_7+s_7$ and $\nu_5=-g_5+g_6+s_6$, reachable for masses $m/M=1.3\times 10^{-3}$ and $1.7\times 10^{-3}$, respectively (i.e. between the masses of Jupiter's moons Callisto and Ganymede). These mass estimates hold for a single ancient satellite. If Uranus had a system of additional inner moons at that time, then the mass needed would be reduced. For instance, if we place the hypothetical ancient satellite together with Uranus' current moon system, all mass estimates given above should be divided by roughly a factor two.

\emph{Phase~2:} We explored the destabilisation phase using a coupled secular model including altogether the satellite's orbit and the planet's spin-axis dynamics. When the system reaches the unstable zone, the satellite's orbit can be excited very quickly and reach almost any value of eccentricity and inclination. If the satellite is massive enough, its wild orbital variations produce a back reaction on the planet's spin axis: the planet sweeps over various secular spin-orbit resonances and undergoes erratic obliquity kicks. These kicks allow the planet to go much deeper in the unstable region and reach obliquity values similar to Uranus' today. Sooner or later, the satellite reaches the Roche limit (or collides into the planet), which puts an end to the chaotic dynamics and fossilises the planet's spin in its last state. Satellite collisions occur through the action of the secular resonance $2\omega+\Omega$ located at equatorial inclination $I_\mathrm{Q}\approx 56^\circ$, and to its symmetric counterpart $2\omega-\Omega$ at $I_\mathrm{Q}\approx 124^\circ$. This resonance is well known for artificial satellites (see e.g. \citealp{Daquin-etal_2022}). Since it vanishes for a planet obliquity $\varepsilon=90^\circ$, satellite collisions can only occur when the planet's obliquity is substantially larger than $90^\circ$ (as Uranus today), or substantially smaller than $90^\circ$. Interestingly, the current state of Uranus is near the maximum of our histograms in the $\varepsilon>90^\circ$ side. This may be in favour of the mechanism proposed here.

Larger resonances and more massive satellites produce stronger obliquity kicks. Our simulations show that for $m/M=4.4\times 10^{-4}$, the satellite generally collides into the planet before the obliquity reaches $80^\circ$. Hence, it seems hard to reproduce the current state of Uranus through a capture in resonance $\nu_3$, even if this resonance demonstrates a great capture and tilting efficiency. We discussed other phenomena that may produce the missing extra tilt for such small ancient satellites (e.g. a subsequent collision with a minor planet) but found none satisfactory. More massive satellites are much more successful in reproducing the current spin state of Uranus. If the satellite has a mass $m/M$ between $2$ and $3\times 10^{-3}$, the probability of success through a capture in resonance $\nu_2$ can be as large as $86\%$. For slightly smaller masses ($m/M\approx 1.5$ to $2\times 10^{-3}$), resonances $\nu_5$ and $\nu_7$ also demonstrate large success ratios, that reach $50\%$ in some ranges of parameters. As before, these mass estimates can be reduced if we assume that Uranus already had a set of additional inner moons at that time; their survival, however, is far from being guaranteed.

Resonance $\nu_2$ is special, because it is a secular spin-orbit resonance with Uranus' own nodal precession mode $s_7$. Due to its large width, it produces large-amplitude oscillations of the obliquity that can destabilise the satellite very early on, after only a short passage through Phase~1. Moreover, resonance $\nu_2$ produces very strong kicks that can quickly pump the obliquity up to Uranus' current value. For this reason, our simulations show that as long as the satellite is initialised at the right distance ($a_0\approx 50\,R_\mathrm{eq}$), the tilting mechanism via resonance $\nu_2$ can operate down to very small migration ranges (only a few $R_\mathrm{eq}$, instead of $10\,R_\mathrm{eq}$ in the general case). When studying the formation of the main satellites of Jupiter, Saturn, and Uranus in gaseous circumplanetary discs, \cite{Mosqueira-Estrada_2003a,Mosqueira-Estrada_2003b} found that a distance of about $50\,R_\mathrm{eq}$ is precisely where one would expect a massive satellite to form around Uranus. The current main satellites of Uranus (Miranda, Ariel, Umbriel, Titania, and Oberon) are much further in and they may seem peculiarly small as compared to other satellites of the giant planets. From these argument, one may argue at least that forming a satellite with mass ratio $m/M\approx 2\times 10^{-3}$ at a distance of $a_0\approx 50\,R_\mathrm{eq}$ does not seem too unreasonable in the context of satellite formation theories.

Any successful tilting scenario must not only reproduce the current spin state of Uranus, but also allow for the existence of its observed satellite system. Today's main satellites of Uranus have prograde orbits, and they show signs of having been formed in a tidal disc \citep{Crida-Charnoz_2012}. We discussed the possibility that they formed from the debris disc produced by the destruction of Uranus' hypothetical ancient satellite, after the latter went below the Roche limit. The formation of the new generation of satellites is expected to be very fast compared to other dynamical processes at play. Then, the newborn satellites would need to migrate outwards from the synchronous orbit ($a\approx 3.4\,R_\mathrm{eq}$) up to their current distance (Oberon is at $a\approx 23\,R_\mathrm{eq}$). Such a large migration range appears unrealistic in the context of traditional tidal models, but it could be achievable on a billion-year timescale if a mechanism similar to the `tidal lock' recently proposed for Titan is also at play for the satellites of Uranus \citep{Fuller-etal_2016,Lainey-etal_2020}.Yet, if this is the case, the satellites would have no particular reason to follow the peculiar distribution described by \cite{Crida-Charnoz_2012} as an evidence of their origin in a tidal disc. The possible relationship between our proposed ancient satellite and the current ones is therefore not a simple question.

Moreover, in our subset of simulations that successfully reproduce the spin state of Uranus, the disruption of the ancient satellite almost always occur when the latter is retrograde ($I_\mathrm{Q}\approx 124^\circ$). Therefore, our simplified secular model does not provide a straightforward way to produce a prograde tidal disc from its debris, and one would need to model the disruption process in more details to assess the properties of the disc -- including tidal dissipation within the satellite, orbital and collisional evolution of the debris, and the possibility of additional ejecta from the planet.

These issues are solved if one considers that the current satellites of Uranus are primordial and formed in the same gaseous disc as our hypothetical ancient satellite (i.e. following the scenario of \citealp{Mosqueira-Estrada_2003a,Mosqueira-Estrada_2003b}). It is not clear, however, whether the current satellites would have survived during the violent unstable phase of their massive neighbour, and whether they would have kept their orbits as unexcited as they are today. Eccentricities can be subsequently damped by tidal dissipation, but inclinations have much longer damping timescales. A possible solution to this problem would be to consider that the current satellites were closer to Uranus during the tilting and more tightly locked to its equatorial plane, and that they substantially migrated afterwards. As these scenarios involve several moons, one would need to investigate them using $N$-body simulations to get definite answers.

Our results are based on numerical explorations performed with secular models. Being fast and valid for arbitrary values of eccentricity and inclination, these models allow us to explore a vast range of parameters. During the first phase of the evolution, we are guaranteed that they give a qualitatively accurate picture. During the second phase, on the contrary, it may seem questionable to average the dynamics over the orbit of the satellite even when its eccentricity grows close to $1$ and its precession becomes very fast. More work is now needed to determine the exact ultimate fate of the satellite and the properties of the tidal disc that it would create. In this extreme dynamical regime, unaveraged numerical integrations are probably required.

We did not investigate yet the effects of tidal dissipation within the satellite. Once again, tidal dissipation would affect the second phase of the evolution, when the satellite's eccentricity strongly increases. The exact outcome of the destabilisation process is hard to predict a priori because it involves three competing effects: tidal dissipation within the planet (which produces the outward satellite migration), dynamical destabilisation (which increases the eccentricity), and tidal dissipation within the satellite (which circularises the orbit and produces an inward migration). The net effect on the satellite strongly depends on the relative timescales of these three competing effects, which are related to the specific mechanisms of energy dissipation at play. As dissipation processes inside ice giant planets are essentially unknown today, and the internal composition of Uranus' ancient satellite is highly speculative, we did not try to obtain a definite answer here. However, our results can serve as a robust starting point for future experiments: According to the level of tidal dissipation within the satellite, we can expect three different behaviours: \emph{i)} If dissipation is exceptionally strong, the destabilisation of the satellite would be inhibited and our scenario would fail to reproduce the spin state of Uranus; \emph{ii)} If dissipation is moderate, the destabilisation may be delayed but not suppressed, and this could increase the fraction of successful runs for a single small satellite, down to a mass $m/M\approx 4.4\times 10^{-4}$; \emph{iii)} If dissipation is small, the picture outlined here would remain unchanged.

In this article, we have settled the basic ingredients that would allow for reproducing the current state of Uranus via the migration of an ancient satellite. This scenario leaves room for refinements. One promising variant would be to invoke two satellites: a big distant one (as that described here), plus a small one located much closer to Uranus. In this configuration, the inner satellite would be much more prone to tidal migration than the outer one; and while the inner satellite migrates outwards over a few $R_\mathrm{eq}$, the outer one would feel a change in the effective $J_2$ parameter of Uranus, which would increase its characteristic radius $r_\mathrm{M}$. As a result, the whole tilting mechanism would be triggered all the same, even though the distant big satellite would not need to migrate at all. This variant scenario exemplifies how it would be possible to relax constraints both on the large energy dissipation required in the interior of Uranus (as the migration of close-in satellites requires much less energy dissipation that the migration of distant ones), and on the large mass needed for the ancient satellite (as the combined action of several moderately massive satellites can produce the same effect). The exploration of this and other possible variant scenarios are left for future works.

Despite the unanswered questions mentioned above, this new picture for the tilting of Uranus appears quite promising to us. To our knowledge, this is the first time that a single mechanism is able to both tilt Uranus (phase~1) and fossilise its spin axis in its final state (phase~2) without invoking a giant impact or other external phenomena. The bulk of our successful runs peaks at Uranus' location, which appears as a natural outcome of the dynamics. This picture also seems appealing as a generic phenomenon: Jupiter today is about to begin the tilting phase \citep{Saillenfest-etal_2020}, Saturn may be halfway in \citep{Saillenfest-etal_2021b}, and Uranus would have completed the final stage, with the destruction of its satellite.

Confronting this picture to observations is not an easy task. Part of the answer would be given by a measure of the tidal migration rate of Uranus' current satellites. A high rate would indicate that they formed substantially closer to Uranus: this would give the possibility that they formed from the debris of the ancient satellite, or that they were protected from its wild destabilisation phase.

\begin{acknowledgements}
   We thank the anonymous referee for his/her thorough analysis of our manuscript and inspiring comments about this new tilting mechanism.
\end{acknowledgements}

\bibliographystyle{aa}
\bibliography{uranusspin}
\appendix

\section{Tilting Uranus during the planetary migration}\label{asec:migtilt}
If we restrict the precession of the planet's orbital plane to a single dominant harmonic with amplitude $S_0=\sin(I_0/2)$ and angle $\phi_0$ evolving at frequency $\nu_0$ (see Eq.~\ref{eq:qprep}), the Hamiltonian function governing the secular spin-axis dynamics of the planet is simplified into
\begin{equation}
   \begin{aligned}
      \mathcal{H} &= -\frac{\alpha}{2}\frac{X^2}{(1-e_\odot^2)^{3/2}} + \nu_0\Phi_0 \\
      &+ \nu_0(1-\cos I_0)X - \nu_0\sin I_0\sqrt{1-X^2}\cos(\psi+\phi_0)\,,
   \end{aligned}
\end{equation}
where $\Phi_0$ is the momentum conjugate to $\phi_0$, added to obtain an autonomous Hamiltonian system. In this expression, $\alpha$ is the precession constant, $e_\odot$ is the eccentricity of the planet, $X$ is the cosine of obliquity, and $\psi$ is the precession angle (see e.g. \citealp{Saillenfest-etal_2019}). Passing to the resonant canonical coordinates $X$ and $-\sigma = -\psi-\phi_0$ and dropping unnecessary constant parts, the Hamiltonian function can be rewritten
\begin{equation}\label{eq:Coltop}
   \mathcal{H} = -\frac{1}{2}X^2 + \gamma X + \beta\sqrt{1-X^2}\cos\sigma\,,
\end{equation}
using the rescaled timescale $\mathrm{d}\tau = p\mathrm{d}t$, where
\begin{equation}\label{eq:gambet}
   \gamma = -\frac{\nu_0\cos I_0}{p}\,,
   \quad\quad
   \beta = -\frac{\nu_0\sin I_0}{p}\,,
\end{equation}
and $p=\alpha (1-e_\odot^2)^{-3/2}$ is the characteristic spin-axis precession frequency of the planet. Equation~\eqref{eq:Coltop} is the Hamiltonian of Colombo's top problem, which has been widely studied \citep{Colombo_1966,Henrard-Murigande_1987,Saillenfest-etal_2019,Haponiak-etal_2020,Su-Lai_2020,Su-Lai_2022}. Its equilibrium points are generally called the `Cassini states', following \cite{Peale_1969}.

\begin{figure}
   \includegraphics[width=\columnwidth]{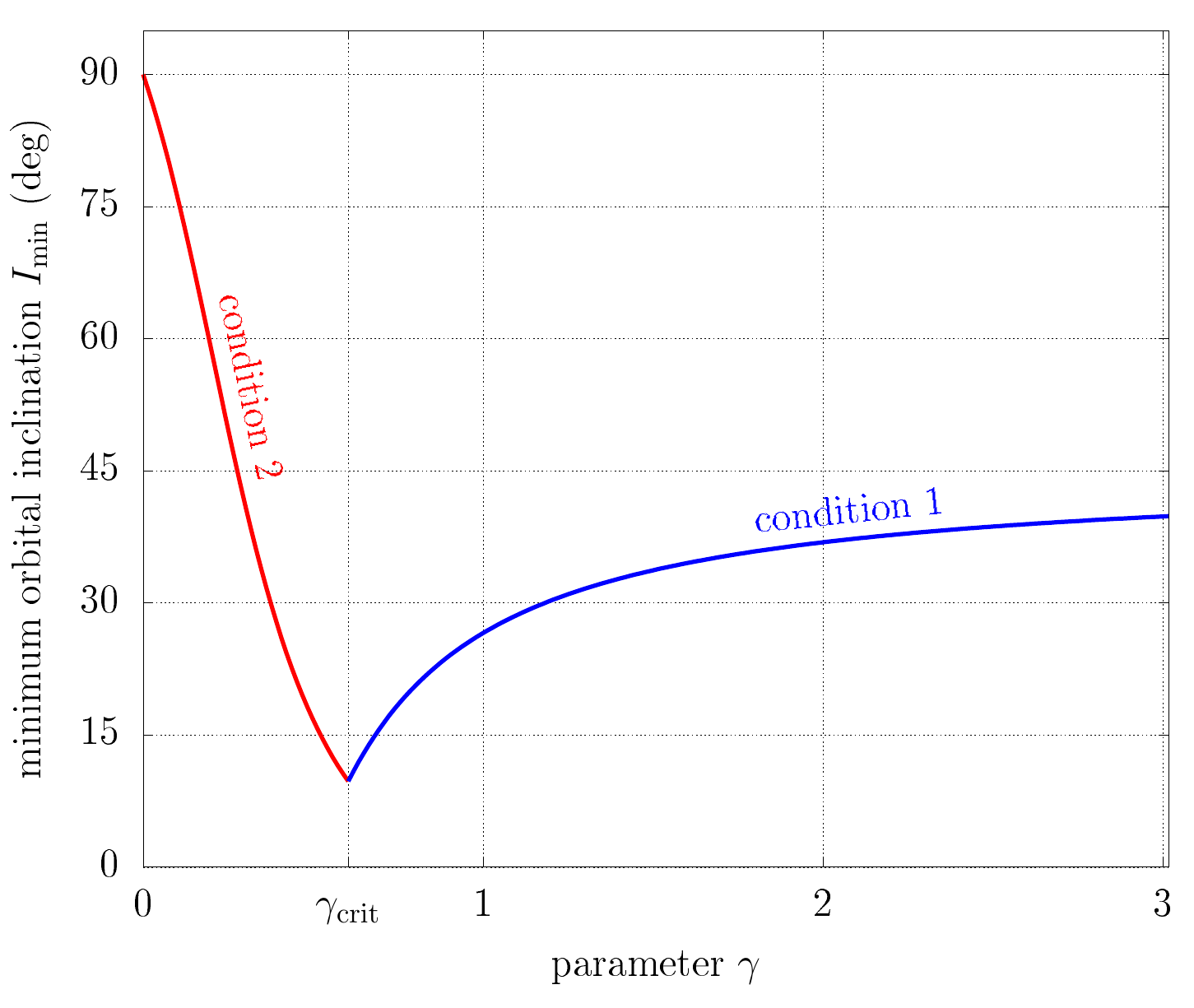}
   \caption{Minimum orbital inclination of a planet such that its obliquity can evolve from $0^\circ$ to more than $90^\circ$ over a single libration in a secular spin-orbit resonance. Conditions~1  and 2 are given in Eqs.~\eqref{eq:cond1} and \eqref{eq:cond2}, respectively. The minimum inclination is reached at $\gamma_\mathrm{crit}=(1+\sqrt{2})/4$ and its value is given by $\tan I_\mathrm{min} = 3-2\sqrt{2}$.}
   \label{fig:Imin}
\end{figure}

Since mutual planetary perturbations dominantly produce a retrograde nodal precession, we have $\nu_0<0$, so that $\gamma$ and $\beta$ are both positive with these notations. Under the hypothesis that Uranus' obliquity is small initially, the fastest tilting is obtained when Uranus acquires its current obliquity during a single cycle of $\sigma$ (`resonance kick'). Assuming that the planet's orbit is stable enough to consider that $\gamma$ and $\beta$ do not vary much over this cycle, the existence of such an extreme trajectory puts strong constraints on the parameters. The existence of a trajectory connecting $\varepsilon=0^\circ$ to more than $\varepsilon=90^\circ$ first requires that
\begin{equation}\label{eq:cond1}
   \beta\geqslant|\gamma-1/2|\,,
\end{equation}
which is the condition for which a trajectory passing at $X=0$ has the same Hamiltonian value has a trajectory passing at $X=1$ (see Eq.~\ref{eq:Coltop}). Moreover, in order for these two points to belong to the same trajectory, the separatrix surrounding Cassini state~2 must either contain the North pole of the sphere or be inexistent. As shown by \cite{Saillenfest-etal_2019}, this condition can be written
\begin{equation}\label{eq:cond2}
   8\beta^2 \geqslant 1-20\gamma-8\gamma^2+(1+8\gamma)^{3/2}\,.
\end{equation}
The two curves defined by Eqs.~\eqref{eq:cond1} and \eqref{eq:cond2} connect at $\gamma = \gamma_\mathrm{crit}$, where $\gamma_\mathrm{crit} = (1+\sqrt{2})/4$. For $\gamma<\gamma_\mathrm{crit}$, Eq.~\eqref{eq:cond2} is the more stringent condition, whereas for $\gamma>\gamma_\mathrm{crit}$, Eq.~\eqref{eq:cond1} is the more stringent condition. According to Eq.~\eqref{eq:gambet}, $\gamma$ and $\beta$ are linked through $\beta=\gamma\tan I_0$. Therefore, the two previous conditions can be formulated as a minimum value for $I_0$ instead of a minimum value for $\beta$. The minimum inclination of the planet allowing for its tilting from $0^\circ$ to more than $90^\circ$ as a function of $\gamma$ is shown in Fig.~\ref{fig:Imin}. The global minimum of this function is located at $\gamma=\gamma_\mathrm{crit}$, and its value is given by $\tan I_\mathrm{min} = 3-2\sqrt{2}$, that is, $I_\mathrm{min}\approx 9.74^\circ$.

\begin{figure}
   \centering
   \includegraphics[width=0.9\columnwidth]{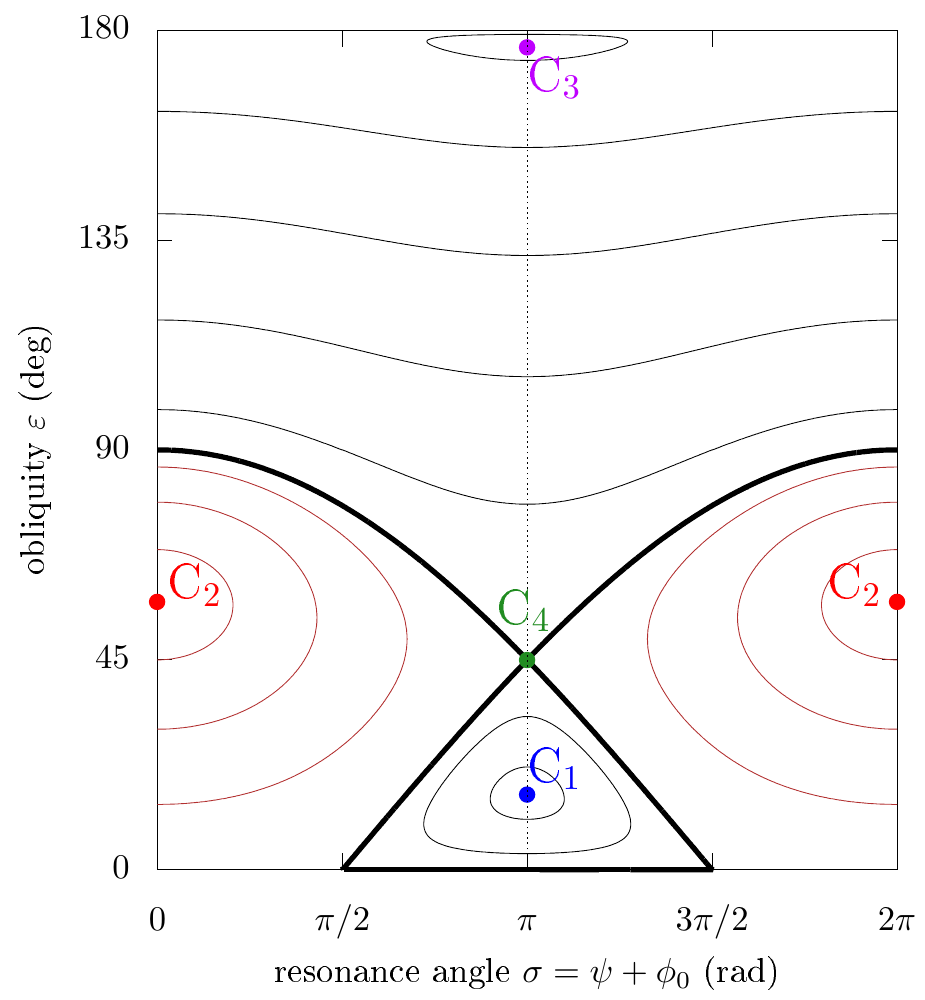}
   \includegraphics[width=0.8\columnwidth]{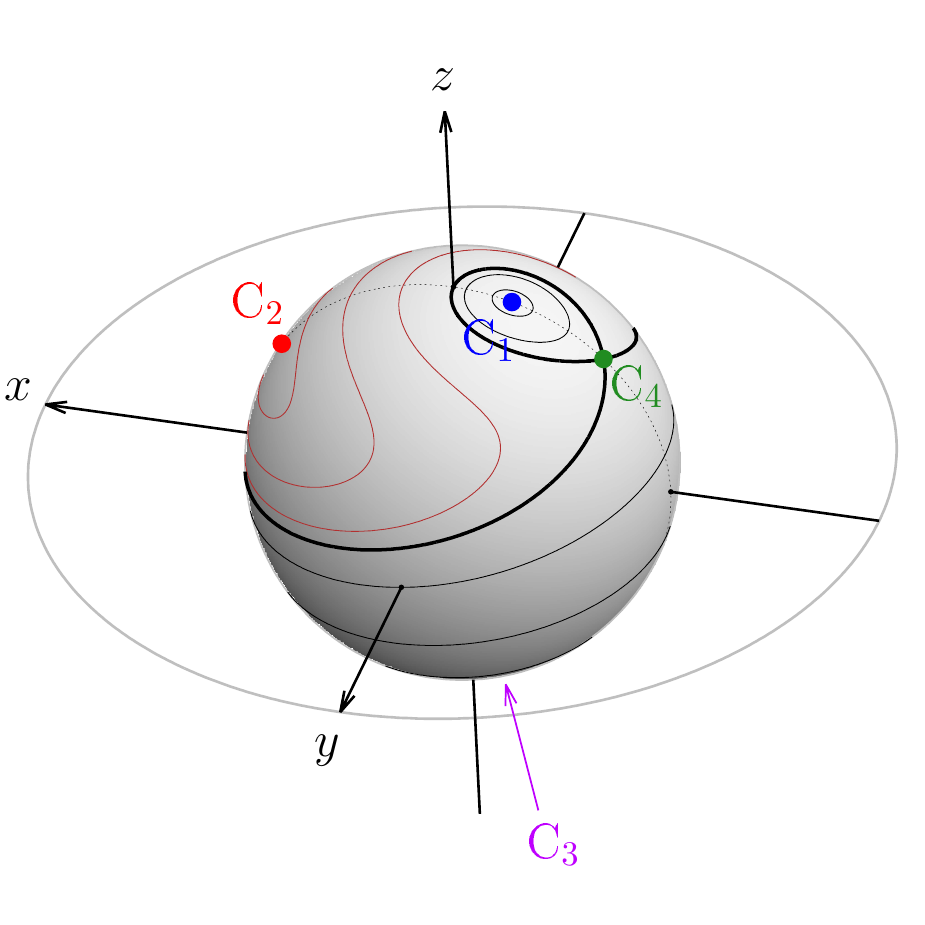}
   \caption{Phase portrait of Colombo's top Hamiltonian when the planet has the minimum orbital inclination required for going from an obliquity $\varepsilon=0^\circ$ to $\varepsilon\geqslant 90^\circ$ over a single libration. The top and bottom panels show the same phase portrait using two sets of coordinates. In the bottom panel, the obliquity $\varepsilon$ is the tilt from the $z$-axis and the resonance angle $\sigma$ is the polar angle measured in the $xy$-plane. The parameters are $\gamma=(\sqrt{2}+1)/4$ and $\beta=(\sqrt{2}-1)/4$, respectively. The Cassini states are labelled and showed with coloured dots. The thick black level is the separatrix of the resonance.}
   \label{fig:HnivImin}
\end{figure}

The phase portrait of the Hamiltonian at the critical point is shown in Fig.~\ref{fig:HnivImin}. We see that it corresponds to the case where the separatrix enclosing Cassini state~2 goes exactly from $\varepsilon=0^\circ$ to $\varepsilon=90^\circ$.

The problem is the same if Uranus' spin axis resonates with its own orbital precession (i.e. $\nu_0=s_7$) or with a harmonic stemming from an other planet (e.g. Neptune, as investigated by \citealp{Rogoszinski-Hamilton_2021}, that is, $\nu_0=s_8$): this harmonic must anyway fulfil the conditions described above, which means that the planet must have a large inclination in order to allow for a fast tilting. Imposing small inclination values, instead, means that Uranus cannot acquire a large obliquity during a single libration, and that we need an additional drift of the resonance centre (i.e. a variation of $\gamma$ and $\beta$). As discussed in Sect.~\ref{sec:intro}, this alternative strongly increases the tilting timescale, because the drift must be adiabatic enough in order for Uranus to remain inside the resonance and gradually follow its drift.

\section{Orbital solution for Uranus}\label{asec:QPS}

The secular orbital solution of \cite{Laskar_1990} is obtained by multiplying the normalised proper modes $z_i^\bullet$ and $\zeta_i^\bullet$ (Tables~VI and VII of \citealp{Laskar_1990}) by  the matrix $\tilde{S}$ corresponding to the linear part of the solution (Table~V of \citealp{Laskar_1990}). In the series obtained, the terms with the same combination of frequencies are then merged together, resulting in 56 terms in eccentricity and 60 terms in inclination. This forms the secular part of the orbital solution of Uranus, which is what is required by our averaged models.

The orbital solution is expressed in the variables $z$ and $\zeta$ as described in Eqs.~\eqref{eq:qprep} and \eqref{eq:munu}. In Tables~\ref{tab:z} and \ref{tab:zeta}, we give all terms of the solution in the J2000 ecliptic and equinox reference frame.

Because of the chaotic dynamics of the Solar System \citep{Laskar_1989,Laskar_1990}, the fundamental frequencies related to the terrestrial planets (e.g. $s_1$, $s_2$, and $\gamma$ appearing in Table~\ref{tab:zetashort}) could vary noticeably over billions of years \citep{Hoang-etal_2021}. However, they only marginally contribute to Uranus' orbital solution, and none of them takes part in the resonances studied in this article. We can therefore safely consider that this orbital solution is valid on a billion-year timescale, at least in a qualitative point of view.

\begin{table}
   \caption{Quasi-periodic decomposition of Uranus' eccentricity and longitude of perihelion (variable $z$).}
   \label{tab:z}
   \vspace{-0.7cm}
   \small
   \begin{equation*}
      \begin{array}{rrrr}
      \hline
      \hline
      k & \mu_k\ (''\,\text{yr}^{-1}) & E_k\times 10^9 & \theta_k^{(0)}\ (^\text{o}) \\
      \hline
       1 &   4.24882 &  37351497 & 210.67 \\
       2 &   3.08952 &  29110713 & 121.36 \\
       3 &   0.66708 &   1640609 &  73.98 \\
       4 &  28.22069 &   1540208 & 308.11 \\
       5 &   1.93168 &    423140 & 208.85 \\
       6 &   5.40817 &    411981 & 301.79 \\
       7 &   2.97706 &     75968 & 306.81 \\
       8 &  27.06140 &     53352 & 218.76 \\
       9 &   4.89647 &     47827 & 115.82 \\
      10 &  52.19257 &     39627 &  45.83 \\
      11 & -20.88236 &     35859 & 203.93 \\
      12 &   1.82121 &     28918 & 154.92 \\
      13 &   0.77840 &     28310 &  65.10 \\
      14 & -19.72306 &     27851 & 293.24 \\
      15 &  28.86795 &     10506 &  32.64 \\
      16 &  27.57346 &      9228 &  43.74 \\
      17 &  -0.49216 &      8225 & 164.74 \\
      18 &  29.37998 &      7279 &  38.31 \\
      19 &  53.35188 &      7234 & 314.93 \\
      20 &   3.60029 &      4594 & 301.39 \\
      21 & -56.90922 &      4109 & 224.11 \\
      22 &   5.59644 &      3787 & 110.35 \\
      23 &   7.45592 &      3521 & 200.24 \\
      24 &  17.91550 &      2648 & 335.34 \\
      25 &  76.16447 &      2387 & 143.03 \\
      26 &   5.47449 &      1426 & 275.01 \\
      27 &  51.03334 &      1325 & 316.29 \\
      28 &   5.71670 &      1082 & 120.52 \\
      29 &   5.35823 &       582 &  94.89 \\
      30 &   6.93423 &       536 & 168.90 \\
      31 &   7.05595 &       423 & 358.70 \\
      32 &  17.36469 &       409 & 303.95 \\
      33 &   7.34103 &       297 & 207.85 \\
      34 &   4.36906 &       221 & 220.83 \\
      35 &   5.99227 &       214 & 113.56 \\
      36 &   7.57299 &       207 &  11.47 \\
      37 &   5.65485 &       206 &  39.22 \\
      38 &   5.23841 &       171 & 272.97 \\
      39 &  17.08266 &       114 & 359.38 \\
      40 &  16.81285 &        98 &  93.77 \\
      41 &   6.82468 &        97 & 194.53 \\
      42 &   7.20563 &        61 & 143.91 \\
      43 &  17.63081 &        42 &  10.70 \\
      44 &  19.01870 &        41 &  39.75 \\
      45 &  17.15752 &        39 & 145.02 \\
      46 &   7.71663 &        38 &  93.52 \\
      47 &  17.81084 &        33 & 238.56 \\
      48 &  18.18553 &        30 & 237.28 \\
      49 &  17.72293 &        25 & 228.46 \\
      50 &  16.52731 &        24 & 311.91 \\
      51 &  18.01611 &        21 & 224.83 \\
      52 &  17.47683 &        21 &  80.26 \\
      53 &  17.55234 &        17 &  17.65 \\
      54 &  18.46794 &        15 & 183.15 \\
      55 &  16.26122 &        15 & 238.89 \\
      56 &  18.08627 &        12 & 176.17 \\
      \hline
      \end{array}
   \end{equation*}
   \vspace{-0.5cm}
   \tablefoot{This solution has been directly obtained from \cite{Laskar_1990} as explained in the text. The phases $\theta_k^{(0)}$ are given at time J2000.}
\end{table}

\begin{table}[h!]
   \caption{Quasi-periodic decomposition of Uranus' inclination and longitude of ascending node (variable $\zeta$).}
   \label{tab:zeta}
   \vspace{-0.7cm}
   \small
   \begin{equation*}
      \begin{array}{rrrr}
      \hline
      \hline
      k & \nu_k\ (''\,\text{yr}^{-1}) & S_k\times 10^9 & \phi_k^{(0)}\ (^\text{o}) \\
      \hline
       1 &   0.00000 &  13773646 & 107.59 \\
       2 &  -3.00557 &   8871413 & 320.33 \\
       3 &  -0.69189 &    563042 & 203.96 \\
       4 & -26.33023 &    347710 & 307.29 \\
       5 &  -2.35835 &    299979 & 224.75 \\
       6 &  -4.16482 &    187859 & 231.66 \\
       7 &  -1.84625 &    182575 & 224.56 \\
       8 &  -3.11725 &     59252 & 146.97 \\
       9 &  -1.19906 &     25881 & 313.99 \\
      10 &  11.50319 &     18941 & 101.01 \\
      11 &  10.34389 &     11930 &  11.68 \\
      12 & -26.97744 &     10362 & 225.10 \\
      13 &  -5.61755 &     10270 & 348.70 \\
      14 &  20.96631 &      7346 & 237.78 \\
      15 &  -0.58033 &      5474 & 197.32 \\
      16 &  -5.50098 &      3662 & 342.89 \\
      17 & -50.30212 &      2748 &  29.83 \\
      18 &  -7.07963 &      2372 & 273.81 \\
      19 &   0.46547 &      1575 & 106.88 \\
      20 &  82.77163 &      1514 & 308.95 \\
      21 &  -5.85017 &      1238 & 165.47 \\
      22 &  -7.19493 &      1177 & 105.12 \\
      23 &  -6.96094 &      1120 &  97.96 \\
      24 & -17.74818 &      1109 & 303.28 \\
      25 &  -5.21610 &       927 &  18.91 \\
      26 &  -5.37178 &       922 &  35.48 \\
      27 &  -5.10025 &       903 & 195.38 \\
      28 &  58.80017 &       887 &  32.90 \\
      29 &  -6.84091 &       826 & 109.96 \\
      30 &  -7.33264 &       810 & 196.75 \\
      31 &  34.82788 &       701 & 114.12 \\
      32 &   0.57829 &       545 & 283.72 \\
      33 &  -5.96899 &       529 & 350.64 \\
      34 & -18.85115 &       463 & 240.24 \\
      35 & -27.48935 &       449 &  38.53 \\
      36 & -25.17116 &       425 &  35.94 \\
      37 &  -6.15490 &       383 &  89.77 \\
      38 &   9.18847 &       310 & 181.15 \\
      39 & -28.13656 &       282 & 134.07 \\
      40 &  -6.73842 &       265 &  44.50 \\
      41 &  -7.40536 &       245 & 233.35 \\
      42 &  -7.48780 &       228 &  47.95 \\
      43 &  -6.56016 &       214 & 303.47 \\
      44 &  -8.42342 &       178 & 211.21 \\
      45 &  18.14984 &       139 & 111.19 \\
      46 & -19.40256 &       122 &  28.16 \\
      47 & -17.19656 &       119 & 153.99 \\
      48 & -18.01114 &        84 &  62.09 \\
      49 & -17.66094 &        80 & 318.93 \\
      50 & -17.83857 &        65 & 109.13 \\
      51 & -17.54636 &        58 &  66.71 \\
      52 & -17.94404 &        41 &  32.26 \\
      53 & -18.59563 &        37 & 278.11 \\
      54 & -19.13075 &        15 & 125.90 \\
      55 & -18.30007 &        11 &  78.29 \\
      56 & -18.97001 &         6 & 253.36 \\
      57 & -18.69743 &         3 &  41.70 \\
      58 & -18.77933 &         3 &  42.83 \\
      59 & -18.22681 &         3 & 226.30 \\
      60 & -19.06544 &         3 & 230.21 \\
      \hline
      \end{array}
   \end{equation*}
   \vspace{-0.5cm}
   \tablefoot{This solution has been directly obtained from \cite{Laskar_1990} as explained in the text. The phases $\phi_k^{(0)}$ are given at time J2000.}
\end{table}

\section{Coupled model for the satellite and the spin axis of its host planet}\label{asec:selfcons}

In Sect.~\ref{sec:coupledmod}, we investigate the effect of the satellite destabilisation on the coupled dynamics of the satellite's orbit and the planet's spin axis. This investigation requires a self-consistent coupled model that also incorporates the orbital variations of the planet due to perturbations from other planets in the system, in order for the secular spin-orbit resonances to also be present. In this section, we describe the model that we use for this purpose.

\subsection{Unaveraged Hamiltonian function}

Our setting is similar to that of \cite{Correia-etal_2011}. We consider three bodies with masses $m_i$ and positions $\mathbf{x}_i$ measured in an inertial reference frame. The momenta conjugate to the positions $\mathbf{x}_i$ are $\mathbf{X}_i=m_i\dot{\mathbf{x}}_i$. We use the index $0$ for the planet, index $1$ for the satellite, and index $2$ for the star. The planet is assumed to be an extended rigid body. The position $\mathbf{x}_0$ of the planet is that of its centre of mass. We write $(\mathbf{I},\mathbf{J},\mathbf{K})$ the basis vectors associated with the principal axes of inertia of the planet; the unitary vectors $(\mathbf{I},\mathbf{J},\mathbf{K})$ are attached to the rigid planet and rotate with it. In this system of coordinates, the inertia tensor $\mathcal{I}$ of the planet is diagonal and writes:
\begin{equation}
   \mathcal{I}=
   \begin{pmatrix}
      A & 0 & 0 \\
      0 & B & 0 \\
      0 & 0 & C
   \end{pmatrix}
   _{(\mathbf{I},\mathbf{J},\mathbf{K})}
   \quad\text{where }A\leqslant B\leqslant C\,.
\end{equation}
Following \cite{Boue-Laskar_2006}, the Hamiltonian function describing the dynamics of the system, as built from its total energy, can be written
\begin{equation}
   \mathcal{H} = \mathcal{H}_\mathrm{E} + \mathcal{H}_\mathrm{N} + \mathcal{H}_\mathrm{I}\,,
\end{equation}
where $\mathcal{H}_\mathrm{E}$ corresponds to the Eulerian free rigid rotation of the planet, $\mathcal{H}_\mathrm{N}$ describes the Newtonian attraction of three mass points, and $\mathcal{H}_\mathrm{I}$ contains the interactions of the non-spherical component of the planet with the satellite and the star mass points. The Eulerian part $\mathcal{H}_\mathrm{E}$ is
\begin{equation}
   \mathcal{H}_\mathrm{E} = \frac{1}{2}\mathbf{G}^\mathrm{T}\mathcal{I}^{-1}\mathbf{G} = \frac{(\mathbf{G}\cdot\mathbf{I})^2}{2A} + \frac{(\mathbf{G}\cdot\mathbf{J})^2}{2B} + \frac{(\mathbf{G}\cdot\mathbf{K})^2}{2C}\,,
\end{equation}
where $\mathbf{G}$ is the rotational angular momentum of the planet. The Newtonian part $\mathcal{H}_\mathrm{N}$ can be written
\begin{equation}
   \mathcal{H}_\mathrm{N} = \sum_{i=0}^2\frac{\|\mathbf{X}_i\|^2}{2m_i} - \sum_{0\leqslant i<j\leqslant 2}\frac{\mathcal{G}m_im_j}{\|\mathbf{x}_i-\mathbf{x}_j\|}\,,
\end{equation}
where $\mathcal{G}$ is the gravitational constant. Because of the hierarchical nature of the system, we use the following set of Jacobi coordinates:
\begin{equation}
   \left\{
   \begin{aligned}
      \mathbf{r}_0 &= \frac{m_0\mathbf{x}_0 + m_1\mathbf{x}_1 + m_2\mathbf{x}_2}{m_0+m_1+m_2}\,,\\
      \mathbf{r}_1 &= \mathbf{x}_1-\mathbf{x}_0\,, \\
      \mathbf{r}_2 &= \mathbf{x}_2 - \frac{m_0\mathbf{x}_0 + m_1\mathbf{x}_1}{m_0+m_1}\,,
   \end{aligned}
   \right.
\end{equation}
with conjugate momenta
\begin{equation}
   \left\{
   \begin{aligned}
      \mathbf{p}_0 &= \beta_0\dot{\mathbf{r}}_0 = \mathbf{X}_0 + \mathbf{X}_1 + \mathbf{X}_2\,, \\
      \mathbf{p}_1 &= \beta_1\dot{\mathbf{r}}_1 = \mathbf{X}_1 - m_1\frac{\mathbf{X}_0 + \mathbf{X}_1}{m_0+m_1}\,, \\
      \mathbf{p}_2 &= \beta_2\dot{\mathbf{r}}_2 = \mathbf{X}_2 - m_2\frac{\mathbf{X}_0 + \mathbf{X}_1 + \mathbf{X}_2}{m_0 + m_1 + m_2}\,,
   \end{aligned}
   \right.
\end{equation}
where the reduced masses are defined as
\begin{equation}
   \left\{
   \begin{aligned}
      \beta_0 &= m_0 + m_1 + m_2\,, \\
      \beta_1 &= \frac{m_0m_1}{m_0+m_1}\,, \\
      \beta_2 &= \frac{(m_0+m_1)m_2}{m_0+m_1+m_2}\,.
   \end{aligned}
   \right.
\end{equation}
The vector $\mathbf{r}_0$ describes the position of the barycentre of the whole system in the inertial reference frame, the vector $\mathbf{r}_1$ is the position of the satellite with respect to the planet, and the vector $\mathbf{r}_2$ is the position of the star with respect to the barycentre of the planet and its satellite. In this system of coordinates, the Newtonian part of the Hamiltonian function can be rewritten
\begin{equation}\label{eq:HN}
   \mathcal{H}_\mathrm{N} = \frac{\|\mathbf{p}_0\|^2}{2\beta_0} + \mathcal{H}_\mathrm{K} + \mathcal{H}_\mathrm{M}\,,
\end{equation}
where $\mathcal{H}_\mathrm{K}$ is the Hamiltonian describing two decoupled Keplerian motions:
\begin{equation}
   \mathcal{H}_\mathrm{K} = \sum_{i=1}^2\left(\frac{\|\mathbf{p}_i\|^2}{2\beta_i} - \frac{\mu_i\beta_i}{\|\mathbf{r}_i\|}\right) = -\frac{\mu_1\beta_1}{2a_1} -\frac{\mu_2\beta_2}{2a_2}\,,
\end{equation}
in which we define $\mu_1=\mathcal{G}(m_0+m_1)$ and $\mu_2 = \mathcal{G}(m_0+m_1+m_2)$, and $a_i$ are the associated semi-major axes. The Hamiltonian function $\mathcal{H}_\mathrm{M}$ contains the mutual Newtonian perturbations:
\begin{equation}
   \mathcal{H}_\mathrm{M} = \left(\frac{\mathcal{G}(m_0+m_1)m_2}{\|\mathbf{r}_2\|} - \frac{\mathcal{G}m_0m_2}{\|\mathbf{x}_2-\mathbf{x}_0\|}\right) - \frac{\mathcal{G}m_1m_2}{\|\mathbf{x}_2-\mathbf{x}_1\|}\,.
\end{equation}
We take the hierarchical nature of system into account by noting that
\begin{equation}
   \begin{aligned}
      \mathbf{x}_2-\mathbf{x}_0 &= \frac{m_1}{m_0+m_1}\mathbf{r}_1 + \mathbf{r}_2\,, \\
      \mathbf{x}_2-\mathbf{x}_1 &= -\frac{m_0}{m_0+m_1}\mathbf{r}_1 + \mathbf{r}_2\,,
   \end{aligned}
\end{equation}
where $\|\mathbf{r}_1\|\ll\|\mathbf{r}_2\|$, and by performing a multipolar expansion of $\mathcal{H}_\mathrm{M}$ using the Legendre polynomials. This gives
\begin{equation}
   \mathcal{H}_\mathrm{M} = \frac{\mathcal{G}\beta_1m_2}{2r_2^3}\left(r_1^2 - 3\frac{(\mathbf{r}_1\cdot\mathbf{r}_2)^2}{r_2^2}\right) + \mathcal{O}\left(\left[\frac{r_1}{r_2}\right]^3\right)\,,
\end{equation}
in which we write $r_i\equiv \|\mathbf{r}_i\|$.

The interaction part $\mathcal{H}_\mathrm{I}$ is obtained from the gravitational potential of a mass element within the planet interacting with the outer two point-mass bodies. Writing $\mathbf{y}$ the position of the mass element measured with respect to the planet's barycentre and $\mathbf{y}_i\equiv \mathbf{x}_i - \mathbf{x}_0$, we again take advantage of the hierarchical nature of the system by noting that $\|\mathbf{y}\|\ll\|\mathbf{y}_i\|\,\forall i=1,2$ and that the planet is almost spherical. By using a multipolar development of $1/\|\mathbf{y}_i-\mathbf{y}\|$ and integrating over the planet's volume, we get
\begin{equation}
   \begin{aligned}
      \mathcal{H}_\mathrm{I} &= \sum_{i=1}^2\Bigg[\frac{\mathcal{G}m_i}{2\|\mathbf{y}_i\|^5}\bigg[(2A-B-C)\left(\mathbf{y}_i\cdot\mathbf{I}\right)^2
      + (2B-A-C)\left(\mathbf{y}_i\cdot\mathbf{J}\right)^2\\
      &+ (2C-A-B)\left(\mathbf{y}_i\cdot\mathbf{K}\right)^2\bigg] + \mathcal{O}\left(\left[\frac{R_\mathrm{eq}}{\|\mathbf{y}_i\|}\right]^3\right)\Bigg]\,,
   \end{aligned}
\end{equation}
where $R_\mathrm{eq}$ is the equatorial (i.e. largest) radius of the planet. Replacing $\mathbf{y}_1$ and $\mathbf{y}_2$ by their expressions as a function of $\mathbf{r}_1$ and $\mathbf{r}_2$, we finally obtain
\begin{equation}
   \begin{aligned}
      \mathcal{H}_\mathrm{I} &= \sum_{i=1}^2\frac{\mathcal{G}m_i}{2r_i^5}\bigg[(2A-B-C)\left(\mathbf{r}_i\cdot\mathbf{I}\right)^2
      + (2B-A-C)\left(\mathbf{r}_i\cdot\mathbf{J}\right)^2\\
      &+ (2C-A-B)\left(\mathbf{r}_i\cdot\mathbf{K}\right)^2\bigg]\\
      &+ \mathcal{O}\left(\left[\frac{R_\mathrm{eq}}{r_1}\right]^3\right)+ \mathcal{O}\left(\left[\frac{R_\mathrm{eq}}{r_2}\right]^3\right)+ \mathcal{O}\left(\left[\frac{r_1}{r_2}\right]^3\right)+ \mathcal{O}\left(\frac{r_1}{r_2}\left[\frac{R_\mathrm{eq}}{r_2}\right]^2\right)\,.
   \end{aligned}
\end{equation}
We notice that $\mathbf{r}_0$ does not appear in the Hamiltonian function, thanks to the reduction of the total barycentre. As a result, the isolated constant term $\|\mathbf{p}_0\|^2/(2\beta_0)$ in Eq.~\eqref{eq:HN} can be dropped from the Hamiltonian function.

\subsection{Secular system}

The Hamiltonian function obtained above can be decomposed into
\begin{equation}
   \mathcal{H} = \mathcal{H}_0 + \epsilon\mathcal{H}_1\,,
\end{equation}
where $\mathcal{H}_0 = \mathcal{H}_\mathrm{E} + \mathcal{H}_\mathrm{K}$ is the dominant integrable part and $\epsilon\mathcal{H}_1 = \mathcal{H}_\mathrm{M} + \mathcal{H}_\mathrm{I}$ is a perturbation (we introduce a factor $\epsilon\ll 1$ to emphasise the smallness of the perturbation). Assuming that there is no resonance involving the mean motions and/or the planet's spin rate, the long-term dynamics of the system at first order in $\epsilon$ is given by the classic averaging method.

In order to average $\epsilon\mathcal{H}_1$ over the fast angles of the rotational motion of the planet, one would need to express $\mathcal{H}_\mathrm{E}$ in action-angle coordinates. Yet, because of the planet's relaxation to hydrostatic equilibrium, we know that the planet's free rotational motion is close to that of an axisymmetric ellipsoid, for which the Andoyer angles $g$ and $\ell$ circulate in an independent manner (i.e. the motion is composed of a main rotation about the constant rotational angular momentum $\mathbf{G}$ and a slower rotation about the principal axis $\mathbf{K}$). Assuming that none of the two corresponding fast frequencies\footnote{$g$ is associated with the proper rotation (or spin motion), and $\ell$ with the Eulerian free nutation, also called Chandler wobble (from \citealp{Chandler_1891} who observed it for the Earth). The period of the free wobble is expected to be a week or so for rigid Jupiter and Saturn, and about $50$~days for Uranus and Neptune (as computed from their spin rates of moments of inertia).} is involved in a resonance, we can therefore bypass the standard procedure and average the Hamiltonian function over $g$ and $\ell$ separately (see \citealp{Boue-Laskar_2006,Vaillant-etal_2019}). Expressing the vectors $\mathbf{I}$, $\mathbf{J}$, and $\mathbf{K}$ in terms of the Andoyer angles, we obtain
\begin{equation}
   \begin{aligned}
      \left<(\mathbf{r}_i\cdot\mathbf{I})^2\right>_{g,\ell} &= \left<(\mathbf{r}_i\cdot\mathbf{J})^2\right>_{g,\ell}\\
      &= \frac{1+\cos^2J}{4}r_i^2 - \frac{1}{2}\left(1-\frac{3}{2}\sin^2J\right)(\mathbf{r}_i\cdot\mathbf{w})^2\,, \\
      \left<(\mathbf{r}_i\cdot\mathbf{K})^2\right>_{g,\ell} &= \frac{\sin^2J}{2}r_i^2 + \left(1-\frac{3}{2}\sin^2J\right)(\mathbf{r}_i\cdot\mathbf{w})^2\,,
   \end{aligned}
\end{equation}
where $\mathbf{w}=\mathbf{G}/G$ and $J$ is the angle between $\mathbf{G}$ and $\mathbf{K}$, such that
\begin{equation}
   \left<\mathcal{H}_\mathrm{I}\right>_{g,\ell} = -\frac{2C-A-B}{2}\left(1-\frac{3}{2}\sin^2J\right)\sum_{i=1}^2\frac{\mathcal{G}m_i}{2r_i^3}\bigg[1 - 3\frac{(\mathbf{r}_i\cdot\mathbf{w})^2}{r_i^2}\bigg]\,.
\end{equation}
Upon averaging, $J$ becomes a constant angle, and the planet's rotational dynamics is reduced to the orientation of the unitary angular momentum vector $\mathbf{w}$.

It remains to average the perturbation Hamiltonian over the fast orbital angles, which are the mean anomalies $M_1$ and $M_2$. The required integrals are performed using the classical formulas (see e.g. Appendix~A of \citealp{Boue-Laskar_2006}). Dropping the constant parts, the secular Hamiltonian function is finally
\begin{equation}
   \bar{\mathcal{H}} = \bar{\mathcal{H}}_\mathrm{M} + \bar{\mathcal{H}}_\mathrm{I}\,,
\end{equation}
where
\begin{equation}
   \bar{\mathcal{H}}_\mathrm{M} = \frac{\gamma}{6}\,\bigg[1 - 6e_1^2 - 3(1-e_1^2)(\mathbf{k}_1\cdot\mathbf{k}_2)^2 + 15(\mathbf{e}_1\cdot\mathbf{k}_2)^2\bigg]\,,
\end{equation}
and
\begin{equation}
   \begin{aligned}
      \bar{\mathcal{H}}_\mathrm{I} &= \sum_{i=1}^2\frac{\alpha_i}{6}\,\bigg[1 - 3(\mathbf{k}_i\cdot\mathbf{w})^2\bigg]\,, \\
   \end{aligned}
\end{equation}
in which we have defined
\begin{equation}
   \gamma = \frac{3\mathcal{G}\beta_1m_2a_1^2}{4a_2^3(1-e_2^2)^{3/2}}\,,
\end{equation}
and
\begin{equation}\label{eq:alpi}
   \alpha_i = \frac{3\mathcal{G}m_0m_iR_\mathrm{eq}^2J_2}{2a_i^3(1-e_i^2)^{3/2}}\left(1-\frac{3}{2}\sin^2J\right)\,,
\end{equation}
where
\begin{equation}
   J_2 = \frac{2C-A-B}{2m_0R_\mathrm{eq}^2}\,.
\end{equation}
In these expressions, $\mathbf{e}_1$ and $\mathbf{e}_2$ are the eccentricity vectors of the satellite around the planet and of the planet-satellite barycentre around the star\footnote{As only the norm of $\mathbf{e}_2$ appears at quadrupolar order, it does not matter whether we consider the orbit of the star around the planet-satellite barycentre or the contrary.} (i.e. the vector pointing to the pericentre and whose norm is the eccentricity), and $\mathbf{k}_1$ and $\mathbf{k}_2$ are the unitary angular momentum vectors of the two orbits.

Following \cite{Correia-etal_2011}, we express the equations of motion in terms of the (non-canonical) vectorial elements $\mathbf{G}$, $\mathbf{e}_1$, and $\mathbf{G}_1$, where $\mathbf{G}_1$ is the orbital angular momentum of the satellite:
\begin{equation}
   \mathbf{G}_1 = \beta_1\sqrt{\mu_1a_1(1-e_1^2)}\,\mathbf{k}_1 \,.
\end{equation}
We obtain
\begin{equation}
   \begin{aligned}
      \dot{\mathbf{G}} &= -\sum_{i=1}^2\alpha_i\,(\mathbf{k}_i\cdot\mathbf{w})(\mathbf{k}_i\times\mathbf{w}) \,,\\
      %%%%%%%%%%%%%%%%%%%%%%%%%%%%%%%%%%%%
      \dot{\mathbf{e}}_1 &= \frac{\alpha_1}{2\|\mathbf{G}_1\|}\bigg[2(\mathbf{k}_1\cdot\mathbf{w})(\mathbf{e}_1\times\mathbf{w}) - \Big(1-5(\mathbf{k}_1\cdot\mathbf{w})^2\Big)(\mathbf{k}_1\times\mathbf{e}_1)\bigg] \\
      &+ \frac{\gamma\, (1-e_1^2)}{\|\mathbf{G}_1\|}\bigg[2(\mathbf{k}_1\times\mathbf{e}_1) + (\mathbf{k}_1\cdot\mathbf{k}_2)(\mathbf{e}_1\times\mathbf{k}_2) \\
      &\hspace{5cm}
      - 5(\mathbf{e}_1\cdot\mathbf{k}_2)(\mathbf{k}_1\times\mathbf{k}_2)\bigg]\,,\\
      %%%%%%%%%%%%%%%%%%%%%%%%%%%%%%%%%%%%
      \dot{\mathbf{G}}_1 &= \alpha_1\,(\mathbf{k}_1\cdot\mathbf{w})(\mathbf{k}_1\times\mathbf{w}) + \gamma\,(1-e_1^2)(\mathbf{k}_1\cdot\mathbf{k}_2)(\mathbf{k}_1\times\mathbf{k}_2) \\
      &- 5\gamma\,(\mathbf{e}_1\cdot\mathbf{k}_2)(\mathbf{e}_1\times\mathbf{k}_2) \,.
   \end{aligned}
\end{equation}
The elements $\mathbf{e}_2$ and $\mathbf{k}_2$ of the planet-satellite barycentre around the star require a specific treatment. Indeed, the three-body system that we consider here is actually not isolated, but subject to perturbations from additional planets. Their direct interactions with the satellite's orbit and the planet's equatorial bulge are negligible; however, they are the main contributors to the secular motion of the planet-satellite barycentre around the star. This contribution is essential because it gives rise to the secular spin-orbit resonances. For this reason, we do not integrate the temporal evolution of $\mathbf{e}_2$ and $\mathbf{k}_2$ but take them as known functions of time obtained from earlier works. Given the small size of the satellites considered in this article, we can safely assume that its influence did not alter noticeably the motion of the barycentre of Uranus and its satellite system around the sun. Hence, we compute $\mathbf{e}_2$ and $\mathbf{k}_2$ from the secular orbital solution of \cite{Laskar_1990} described in Appendix~\ref{asec:QPS}. The conversion from the variables $z$ and $\zeta$ to the vectors $\mathbf{e}_2$ and $\mathbf{k}_2$ is straightforward. This solution acts as an indirect forcing term to the dynamics of the satellite and the spin axis of the planet.

The constant value of Uranus' rotational angular momentum $G$ is quite uncertain, as both its spin rate $\omega$ and moment of inertia are poorly known (see the discussion in Sect~\ref{sec:mass}). For this reason, it is enough to use the approximate expression for a principal axis rotator $G\approx C\omega=\lambda\omega\,m_0R_\mathrm{eq}^2$, where $\lambda$ is the normalised polar moment of inertia. We adopt the nominal value of $\lambda\omega$ described in Sect.~\ref{sec:mass}. We also neglect the $\sin^2J$ term in Eq.~\eqref{eq:alpi} which is not known for Uranus but expected to be extremely small\footnote{The relaxation of planets to hydrostatic equilibrium tends to reduce the angle $J$, and specific internal processes are needed to maintain a non-zero value. The mean value of $J$ is measured to be about $7\times 10^{-7}$~rad for the Earth \citep{Chandler_1891} and $3\times 10^{-8}$~rad for Mars \citep{Konopliv-etal_2020}. The Chandler wobble has not been detected yet for gaseous planets, but from the latest Juno data, the constant $J$ for Jupiter is constrained to be smaller than $10^{-6}$~rad and compatible with zero \citep{Iess-etal_2018,Serra-etal_2019,Durante-etal_2020}. Similar constraints have been obtained for Saturn \citep{Iess-etal_2019}.}. The migration of the satellite is modelled by applying a slow drift in $a_1$ in the equations of motion according to the migration law in Eq.~\eqref{eq:asat}. As discussed in Sect.~\ref{sec:proba}, we neglect the simultaneous decrease in Uranus' spin rate $\omega$, whose contribution is smaller than our uncertainty on the value of $G$.

\section{High-order secular spin-orbit resonances}\label{asec:rese2}

For a given planet, the strongest secular spin-orbit resonances are of first order in the planet's orbital inclination (see e.g. \citealp{Saillenfest-etal_2019}). First-order resonances have resonant angles of the form $\sigma_k = \psi+\phi_k$, where $\psi$ is the spin-axis precession angle and $\phi_k$ is the circulating angle of the $k$th harmonic in the planet's nodal precession spectrum (see Eq.~\ref{eq:qprep}). The secular spin-axis dynamics, however, is not restricted to first-order resonances. At second order in inclination, resonances involve two different harmonics with resonant angles of the form $\sigma_{j,k} = 2\psi+\phi_j+\phi_k$. At order three, there exist five different kinds of resonances whose properties are described by \cite{Saillenfest-etal_2019}.

\begin{figure}[h!]
   \centering
   \includegraphics[width=0.9\columnwidth]{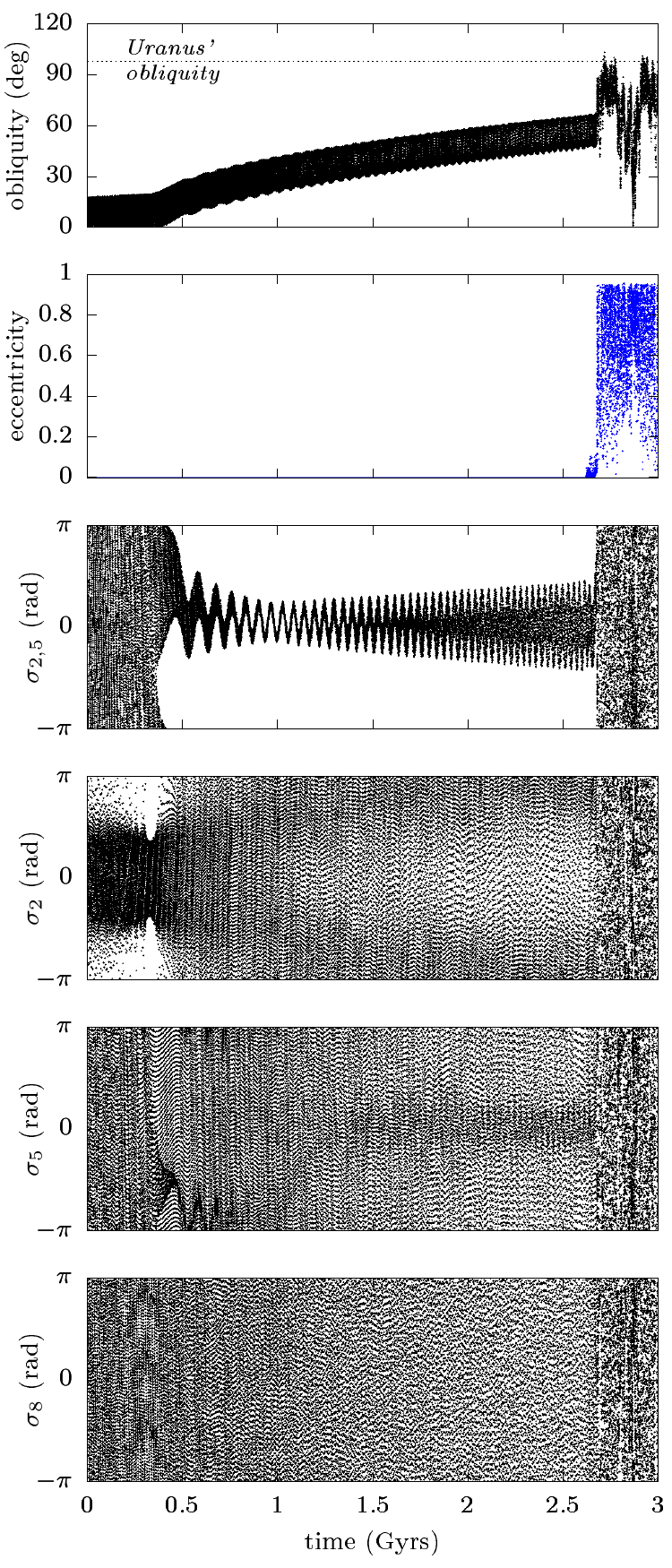}
   \caption{Example of stable capture in a secular spin-orbit resonance of second order. Black and blue are used for quantities related to the planet's spin axis and satellite's orbit, respectively. The numerical integration is performed with the fully coupled secular model presented in Sect.~\ref{sec:coupledmod}. The mass of the satellite is $m/M=2\times 10^{-3}$. The satellite is initialised close to the local circular Laplace equilibrium at $a=42$~$R_\mathrm{eq}$ and it is made migrating outwards over $\Delta a=9$~$R_\mathrm{eq}$.}
   \label{fig:resphi2phi5}
\end{figure}

In the case of Uranus on its current orbit, resonances are rare and isolated from each other. Yet, harmonic $k=2$ has quite a large amplitude (see Table~\ref{tab:zetashort}), and its frequency is not far from those of harmonics $k=5$ and $8$. This suggests that high-order resonances could play a substantial role in the dynamics in the neighbourhood of resonance $\sigma_2$. And indeed, we found that some fraction of the trajectories shown in Fig.~\ref{fig:obmaxvsmass} for satellite masses in the neighbourhood of $m=m_2$ is produced by stable captures in second-order resonances. Figure~\ref{fig:resphi2phi5} shows an example of such trajectories: $\sigma_2=\psi+\phi_2$ first starts to oscillate, but it is soon replaced by $\sigma_{2,5}=2\psi+\phi_2+\phi_5$, while both $\sigma_2$ and $\sigma_5$ circulate. The nearby resonance $\sigma_8$ does not play any role in this example.

\section{Secular resonances of a massless satellite}\label{asec:secres}

In Sect.~\ref{sec:keyholes}, we investigate the origin of the peculiar dynamical pathways through which the satellite can collide into the planet. To this end, we simplify the problem and consider the dynamics of a massless satellite around a planet having a frozen orbit and spin-axis orientation. The secular Hamiltonian function of the satellite can be written $\mathcal{H}=k_\mathrm{P}\mathcal{H}_\mathrm{P} + k_\odot\mathcal{H}_\odot$. Expanding both contributions to quadrupole order, \cite{Saillenfest-Lari_2021} define the two constant coefficients as
\begin{equation}
   k_\mathrm{P} = \frac{3}{4}\frac{\mathcal{G}M}{a}J_2\frac{R_\mathrm{eq}^2}{a^2}
   \quad\text{and}\quad
   k_\odot = \frac{3}{8}\frac{\mathcal{G}m_\odot}{a_\odot(1-e_\odot^2)^{3/2}}\frac{ a^2}{a_\odot^2} \,,
\end{equation}
and the two Hamiltonian functions as
\begin{equation}\label{eq:HP}
   \mathcal{H}_\mathrm{P} = \frac{1-3\cos^2I_\mathrm{Q}}{3(1-e^2)^{3/2}} \,,
\end{equation}
and
\begin{equation}\label{eq:Hs}
   \begin{aligned}
      \mathcal{H}_\odot &= - \frac{1}{8}\bigg[
        8e^2 + 2(3e^2 + 2)(2C^2\cos^2I_\mathrm{Q} + S^2\sin^2I_\mathrm{Q} - 2) \\
      &\quad\quad+ 8CS(3e^2 + 2)\cos I_\mathrm{Q}\sin I_\mathrm{Q}\cos\Omega_\mathrm{Q} \\
      &\quad\quad+ 5S^2e^2(\cos I_\mathrm{Q} + 1)^2\cos(2\omega_\mathrm{Q} + 2\Omega_\mathrm{Q}) \\
      &\quad\quad- 20CSe^2(\cos I_\mathrm{Q} + 1)\sin I_\mathrm{Q}\cos(2\omega_\mathrm{Q} + \Omega_\mathrm{Q}) \\
      &\quad\quad+ 10(3C^2 - 1)e^2\sin^2I_\mathrm{Q}\cos(2\omega_\mathrm{Q}) \\
      &\quad\quad- 20CSe^2(\cos I_\mathrm{Q} - 1)\sin I_\mathrm{Q}\cos(2\omega_\mathrm{Q} - \Omega_\mathrm{Q}) \\
      &\quad\quad+ 5S^2e^2(\cos I_\mathrm{Q} - 1)^2\cos(2\omega_\mathrm{Q} - 2\Omega_\mathrm{Q}) \\
      &\quad\quad+ 2S^2(3e^2 + 2)\sin^2I_\mathrm{Q}\cos(2\Omega_\mathrm{Q})
      \bigg]\,.
   \end{aligned}
\end{equation}
In this expression, $C=\cos\varepsilon$ and $S=\sin\varepsilon$ are constant, and the orbital angles of the satellite $(I_\mathrm{Q},\omega_\mathrm{Q},\Omega_\mathrm{Q})$ are measured with respect to the equator and equinox of the planet. All other quantities that appear in the Hamiltonian function are described in the main text.

Because of the factor $(1-e^2)^{-3/2}$ in Eq.~\eqref{eq:HP}, $k_\mathrm{P}\mathcal{H}_\mathrm{P}$ is the dominant contribution to the dynamics whenever the eccentricity grows large. We consider the intermediate regime in which $k_\mathrm{P}\mathcal{H}_\mathrm{P}$ is the dominant integrable part of the Hamiltonian and $k_\odot\mathcal{H}_\odot$ is a small perturbation. Written in terms of the Delaunay elements, $k_\mathrm{P}\mathcal{H}_\mathrm{P}$ is expressed in action-angle coordinates, so we can directly apply a perturbative method. At first order in the perturbation, the possible resonances are those directly appearing in Eq.~\eqref{eq:Hs}; for each of them, the resonance angle $\xi$ is a linear combination of $\omega_\mathrm{Q}$ and $\Omega_\mathrm{Q}$. These resonances are listed in Table~\ref{tab:resconst} of the main text.

Using a perturbative approach, we can study the orbital dynamics of the satellite in the vicinity of each resonance. The method is the following (see e.g. \citealp{Saillenfest-etal_2019b}, \citealp{Talu-etal_2021}): The angle $\xi$ is first taken as coordinate by a canonical change of variables $(\omega_\mathrm{Q},\Omega_\mathrm{Q})\rightarrow(\xi,\gamma)$. The conjugate momenta of $\xi $ and $\gamma$ are written $\Xi$ and $\Gamma$. In the vicinity of the resonance, $\xi$ is a slow angle and $\gamma$ is a fast angle. At first order in the perturbation, the long-term dynamics is described by averaging the Hamiltonian function over $\gamma$. The Hamiltonian system obtained is independent of $\gamma$, which means that $\Gamma$ is a constant of motion and can be used as parameter. We end up with a one-degree-of-freedom system and any possible trajectory for the satellite can be represented as a level curve of the Hamiltonian function. In practice, $\Gamma$ is rewritten as a constant quantity $K$, whose expression for each resonance is given in Table~\ref{tab:resconst}. By virtue of the constant nature of $K$, it is equivalent to represent the level curves of the Hamiltonian function in the $(e,\xi)$ plane or in the $(I_\mathrm{Q},\xi)$ plane.

\begin{figure}
   \centering
   \includegraphics[width=0.8\columnwidth]{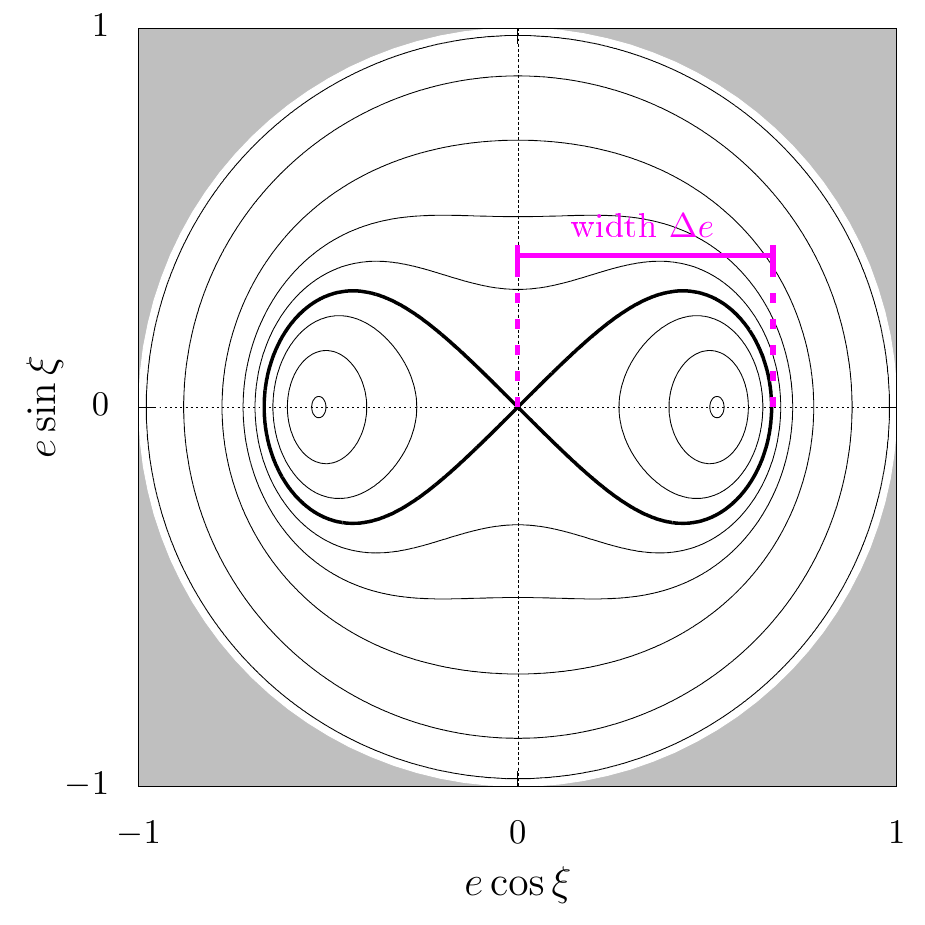}
   \caption{Geometry of the phase portrait describing the orbital dynamics of a satellite in the vicinity of one of the secular resonances listed in Table~\ref{tab:resconst} (except the last one). The eccentricity $e$ and equatorial inclination $I_\mathrm{Q}$ of the satellite are linked through a constant quantity $K$. The origin $e=0$ of the graph corresponds to the nominal inclination $I_0$ of the resonance.}
   \label{fig:phaseRes}
\end{figure}

When it emerges from chaos and reaches the weakly perturbed regime, the satellite can have any value of $K$. For definitiveness, we consider here for each resonance the value of $K$ such that $e=0$ for $I_\mathrm{Q}=I_0$, where $I_0$ is the nominal inclination of the resonance (see Table~\ref{tab:resconst}). Even though this represents a loss of generality, we will see that studying just one value of $K$ is enough to understand the correct hierarchy between the various resonances. The case $\xi=\Omega_\mathrm{Q}$ is special, because the constant quantity is the eccentricity $e$ itself. By putting $e=0$, we obtain the Hamiltonian function that describes the circular Laplace equilibria (see \citealp{Tremaine-etal_2009}). The `resonance' in this case is the libration island surrounding the orthogonal equilibrium (called Laplace state P$_2$ by \citealp{Saillenfest-Lari_2021}), for which the minimum and maximum value of $I_\mathrm{Q}$ along the separatrix are given by
\begin{equation}
   \cos^2I_\mathrm{Q} = \frac{1+u-\sqrt{1+u^2+2u\cos(2\varepsilon)}}{2u}\,,
\end{equation}
where $u=r_\mathrm{M}^5/a^5$. All other resonances in Table~\ref{tab:resconst} produce very similar phase portraits. Since some resonances are very large, we do not use the pendulum approximation but solve for the exact resonant Hamiltonian function. The structure of the phase space is illustrated in Fig.~\ref{fig:phaseRes}. The width $\Delta e$ of the separatrix gives the maximum eccentricity reachable by the satellite inside the resonance. It is linked to an analogous width $\Delta I_\mathrm{Q}$ through the constant quantity $K$. The widths of each resonance as a function of the parameters are shown in Figs.~\ref{fig:reswidthall} and \ref{fig:resoverlap} of the main text.

\section{Effect of a satellite collision on the obliquity}\label{asec:impact}
   In Sect.~\ref{sec:proba}, we observe that large regions of the parameter space lead to a resonance capture of Uranus. As the satellite continues to migrate outwards, the system converges towards a configuration for which the satellite is highly unstable. The effects of this destabilisation are investigated in Sect.~\ref{sec:coupledmod}; they can lead to a direct collision of the satellite into the planet. In this section, we investigate what would be the effect of a satellite-planet collision on the final spin rate and obliquity of the planet.
   
   As explained in Sect.~\ref{sec:smallestsat}, the total angular momentum $\mathbf{L}$ of the planet and its satellite can be considered constant during the collision. The constant vector $\mathbf{L}$ can be written
   \begin{equation}\label{eq:Ltot}
      \mathbf{L} = \mathbf{L}_\mathrm{P} + \mathbf{L}_\mathrm{S}\,,
   \end{equation}
   where $\mathbf{L}_\mathrm{P}$ is the contribution of the planet's spin and $\mathbf{L}_\mathrm{S}$ is the contribution of the satellite's orbit. After the removal of the satellite, the obliquity of Uranus is not expected to vary at all due to external perturbations because Uranus is released far from any kind of spin-orbit resonance. The magnitude of $\mathbf{L}$ can therefore be computed using the current parameters of Uranus. A set of inner satellites can also be included; however, the current satellites of Uranus only contribute to $1\%$ of $L=\|\mathbf{L}\|$, which is much smaller than our uncertainty on Uranus' moment of inertia. To leading order, the relative change in the planet's spin rate is
   \begin{equation}
      \frac{\Delta \omega}{\omega}\approx\frac{L - L_\mathrm{P}}{L_\mathrm{P}}\approx\frac{L_\mathrm{S}}{L_\mathrm{P}}\cos I_\mathrm{Q}\,,
   \end{equation}
   where $L_\mathrm{P}=\|\mathbf{L}_\mathrm{P}\|$, $L_\mathrm{S}=\|\mathbf{L}_\mathrm{S}\|$, and $I_\mathrm{Q}$ is the orbital inclination of the satellite with respect to the planet's equator at the collision time (i.e. the angle between $\mathbf{L}_\mathrm{P}$ and $\mathbf{L}_\mathrm{S}$). Using the gyroscopic approximation for the spin of the planet, the ratio $L_\mathrm{S}/L_\mathrm{P}$ can be written as
   \begin{equation}\label{eq:LSLP}
      \frac{L_\mathrm{S}}{L_\mathrm{P}} = \frac{m}{M}\frac{n}{\omega}\frac{a^2}{R_\mathrm{eq}^2}\frac{\sqrt{1-e^2}}{\lambda}\,,
   \end{equation}
   where $n$ and $e$ are the satellite's mean motion and eccentricity, and other quantities are described in Sect.~\ref{sec:mec}. The collision is expected to occur when the semi-major axis of the satellite is $a\approx r_\mathrm{M}$ as defined in Eq.~\eqref{eq:rM}. Moreover, the collision condition requires that $a(1-e)\lesssim R_\mathrm{eq}$ which gives a minimum bound for the eccentricity value at the collision time. When using in Eq.~\eqref{eq:LSLP} the physical parameters of Uranus, we obtain a value of about $35\,m/M$, where the leading factor varies from $32$ to $39$ when assuming a $10\%$ uncertainty on the value of $\omega\lambda$ as in Sect.~\ref{sec:mass}. The order of magnitude for the relative change in the planet's spin rate is therefore
   \begin{equation}
      \frac{\Delta \omega}{\omega}\approx 35\frac{m}{M}\cos I_\mathrm{Q}\,.
   \end{equation}
   The variation in the planet's spin-axis orientation due to the impact is quantified by the angle $\varphi$ between $\mathbf{L}$ and $\mathbf{L}_\mathrm{P}$. From Eq.~\eqref{eq:Ltot}, it can be expressed as
   \begin{equation}
      \sin\varphi = \frac{L_\mathrm{S}}{L}\sin I_\mathrm{Q}\,,
   \end{equation}
   that is,  
   \begin{equation}
      \sin\varphi\approx 35\frac{m}{M}\sin I_\mathrm{Q}\,.
   \end{equation}
   Under favourable longitude of node of the satellite at the time of collision, the spin-axis deviation $\varphi$ directly equals the obliquity variation $\Delta\varepsilon$. As shown in Sect.~\ref{sec:keyholes}, satellite collisions only occur when $I_\mathrm{Q}\approx 55^\circ$ or $I_\mathrm{Q}\approx 125^\circ$. Therefore, for the mass range discussed throughout this article, the change in spin rate due to the satellite's impact is a few percent at most, and the change in obliquity is no larger than a few degrees.
   
   The maximum spin-axis deviation $\varphi$ would be obtained for a grazing impact with $I_\mathrm{Q}=90^\circ$. Hence, an upper bound for the obliquity change in radians due to the satellite collision is $\Delta\varepsilon\approx 35\,m/M$.

\end{document}